\documentclass[a4paper,11pt]{article}
\pdfoutput=1 

\usepackage{jheppub} 

\usepackage[T1]{fontenc} 
\usepackage[table]{xcolor}
\usepackage[caption=false]{subfig}
\usepackage{epsfig}
\usepackage{colortbl}
\usepackage[T1]{fontenc} 
\usepackage[compat=1.1.0]{tikz-feynman}
\usepackage{adjustbox}
\usepackage{multirow}
\usepackage{slashed}
\usepackage{lipsum}
\usepackage{chngcntr}
\usepackage{bigints}
\usepackage[normalem]{ulem}
\usepackage{amsmath}
\usepackage{textcomp}
\usepackage{cleveref}
\usepackage{lscape}

\usepackage{float}
\usepackage{pstricks}
\usepackage[toc,page]{appendix}
\usepackage{pifont}
\usepackage{braket}
\usepackage{ragged2e}
\usepackage{slashed}
\usepackage{rotating}
\usepackage{tablefootnote}

\usepackage{hyperref}
\usepackage{tikz}
\usepackage{tikz-feynman}
\usetikzlibrary{arrows}
\usepackage{caption} 
\crefname{figure}{fig\,.}{figs\,.} 
\crefname{equation}{eq\,.}{eqs\,.} 
\newcommand{\stkout}[1]{\ifmmode\text{\sout{\ensuremath{#1}}}\else\sout{#1}\fi}
\definecolor{calpolypomonagreen}{rgb}{0.12, 0.3, 0.17}
\newcommand{\bea}{\begin{eqnarray}}
\newcommand{\eea}{\end{eqnarray}}

\usepackage{xcolor}

\title{Correlated study on some $B_{c}\rightarrow\text{ }P$ and $B_{c}\rightarrow\text{ }S$ wave channels in light of new inputs}
	
	\author[a]{Utsab Dey,}
	\author[a]{Soumitra Nandi.}
	
	\affiliation[a]{Department of Physics, Indian Institute of Technology Guwahati,\\North Guwahati, Assam-781039, India,}
	
	\emailAdd{utsab\_dey@iitg.ac.in}
	\emailAdd{soumitra.nandi@iitg.ac.in}

\abstract{
This study investigates the decay modes of the $B_c$ meson, focusing on semileptonic and nonleptonic decay into S and P wave charmonia. The primary objective is to extract the shape parameter of the $B_{c}$ meson distribution amplitude through a data-driven approach, utilizing the lattice results on  $B_{c}\rightarrow\eta_{c},J/\psi$ semileptonic form factors and yielding an estimate of ${\omega_B}_c = 0.998(34)~GeV$ for the same. We use the form factors derived from the modified perturbative QCD framework in the analysis. This result and various other inputs on the radiative decays of the P wave charmonia enable us to estimate the $q^2$ shapes of the $B_{c}\rightarrow\chi_{c0},\chi_{c1}$ and $h_{c}$ form factors using pole expansion parametrization. Using these results, we have obtained predictions of LFUV observables $R(\chi_{c0})=0.169(11)$, $R(\chi_{c1})=0.126(2)$ and $R(h_{c})=0.113(3)$. Finally, we have presented predictions for branching ratios of some nonleptonic decay modes of the $B_{c}$ meson into S and P wave charmonia in the same modified perturbative QCD framework.
}


\begin{document}		
\maketitle
\flushbottom	
	
\section{Introduction}
\label{sec:Introduction}

The $B_c$-meson is a heavy quarkonium with mixed-heavy flavors, which could be useful for studying heavy-quark dynamics. At the same time, it is stable against strong and electromagnetic interactions as it lies below the $B\bar{D}$ threshold and can only decay through weak interactions. It is an ideal system for studying weak decays of both the heavy quarks. This possibility offers a promising opportunity to study various nonleptonic and semileptonic weak decays of heavy mesons. Similar to $B\to D^{(*)}\ell^-\bar{\nu}$ (with $\ell = e, \mu, \tau$) decays the semileptonic decays of $B_c$ meson to the $S$ and $P$-wave charmoniums will be useful to extract the Cabibbo-Kobayashi-Maskawa (CKM) matrix element $|V_{cb}|$ and in the indirect search of the new interactions beyond the standard model (BSM). The modes with light leptons are less sensitive to BSM physics and, hence, could be used to extract $|V_{cb}|$ while the mode with $\ell =\tau$ is expected to help probe BSM scenarios. We define observables like the ratios of the decay rates 
\begin{equation}
R(P (V)) = \frac{\Gamma(B_c^-\to P (V)\tau^- \bar{\nu})}{\Gamma(B_c^-\to P(V) \mu^- \bar{\nu} )}.
\end{equation}
These observables are expected to be potentially sensitive to BSM interactions. At the present level of accuracy, we have observed deviations in the measured values of $R(D^{(*)})$ as compared to the respective SM predictions \cite{hflavnew,HFLAV:2022esi,Ray:2023xjn}. The measurement of the observables in the $B_c$ decays will be important to gain complementary phenomenological information. Such studies can help improve our understanding of the nature of the anomalous results seen in B-meson decays. Moreover, any BSM physics altering these modes' results should be affected and constrained by other $b \to c$ transitions. 

 The LHCb and CMS collaborations have measured this ratio in $B_c^- \to J/\psi \ell^-\bar{\nu}$ decays which are given as follows \cite{LHCb:2017vlu}:
 \begin{equation}
 R(J/\psi) = \frac{\Gamma(B_c^- \to J/\psi \tau^-\bar{\nu})}{\Gamma(B_c^- \to J/\psi \mu^-\bar{\nu})} = 0.71 \pm 0.17 (stat) \pm 0.18 (syst) ~~~~~\text{(LHCb),}
 \label{eq:Rjpsilhcb}
 \end{equation}
  \begin{equation}
  R(J/\psi)= 0.17^{+0.18}_{-0.17} (stat.)^{+0.21}_{-0.22} (syst.)^{+0.19}_{-0.18} (theo)= 0.17\pm 0.33 ~~~ \text{( CMS)}.
  \label{eq:CMSres}
  \end{equation}
 Both the measured values have significant errors and are marginally consistent with each other at their 1$\sigma$ uncertainties. We must wait for more precise data to look for possible new physics (NP) effects. However, precise predictions of all the related observables in the SM are equally important. A model-independent approach regarding the form factors \cite{Cohen:2019zev} leads to the SM prediction of $0.25(3)$. The HPQCD lattice collaboration has recently extracted the $B_c \to J/\psi$ form-factors over the full kinematically allowed region \cite{Harrison:2020gvo}. They have predicted $R(J/\psi) = 0.2582 (38)$ \cite{Harrison:2020nrv}, which is so far the most precise prediction and in tension with the LHCb result given above but in agreement with the measured value at the CMS. So far, no inputs on the form factors of other $B_c \to S$ or $B_c \to P$ wave charmoniums from the lattice are available. Neither data on the corresponding semileptonic or non-leptonic rates is available. Several QCD models exist in the literature, and based on the modelling of the form factors, the value of $R(J/\psi)$ lies in the range $[0,0.48]$ \cite{Anisimov:1998uk,Kiselev:2002vz,Ivanov:2006ni,Hernandez:2006gt,Wang:2012lrc,Rui:2017pre,Rui:2018kqr,PhysRevD.97.113001,Hu:2019bdf}.   
 
 The form factors in $B_c \to J/\psi$, $B_c \to \eta_c$, $B_c\to \chi_{c0}$, $B_c \to \chi_{c1}$ and $B_c \to h_c$ transitions have been calculated in the perturbative QCD (PQCD) framework \cite{Wang:2012lrc,Rui:2017pre,Rui:2018kqr,PhysRevD.97.113001,Hu:2019bdf}. Apart from the perturbatively calculable hard functions, these form factors depend on the non-perturbative wave functions of $B_c$ and other mesons involved in the respective processes. Therefore, estimates of the relevant wave functions should be made to obtain the form factors in the respective decays. In this analysis, we constrain these wave functions using the lattice inputs on $B_c \to J/\psi$ form factors \cite{Harrison:2020gvo} and the available data on the respective radiative decays of the corresponding charmonium states. In addition, we have used the inputs on $B_c\to \eta_c$ form factors, which we have extracted using the available information on $B_c \to J/\psi$ form factors from the lattice in combination with the heavy-quark-spin-symmetry (HQSS), the method is similar to the one used in ref. \cite{Biswas:2023bqz}. In the earlier perturbative QCD analyses \cite{Wang:2012lrc,Rui:2017pre,Rui:2018kqr,PhysRevD.97.113001,Hu:2019bdf}, model-dependent inputs were used to obtain the respective wave functions.
 
 To obtain the semileptonic rates, we need to know the shape of the respective form factors in the kinematically allowed $q^2$ (lepton invariant mass squared) regions. However, calculating the form factors in the PQCD approach is reliable only in the large recoil regions. We obtain the $q^2$ shape of the form factors in $B_c\to \eta_c\ell^-\bar{\nu}$ decays using the HQSS symmetry. After predicting the PQCD form factors at $q^2=0$, the shape of the form factors in $B_c\to \chi_{c0}\ell\nu$, $B_c \to \chi_{c1}\ell\nu$ and $B_c \to h_c\ell\nu$ decays are obtained by using the pole expansion technique which we will discuss later. Finally, using these form factors, we have predicted the $q^2$ distribution of the respective rates, the branching fractions, and the ratios $R(\eta_c)$, $R(\chi_{c0})$, $R(\chi_{c1})$ and $R(h_c)$. In addition, we have predicted a couple of angular observables in the SM. We take this opportunity to predict the branching fractions of a couple of non-leptonic decay modes of $B_c$ meson to charmonium and a light meson.
 
 The paper is structured as follows: In section \ref{section:theoretical framework} we describe the analytic expressions of the various physical observables that we intend to predict in this work and briefly discuss about the respective form factors in modified pQCD framework and light cone distribution amplitudes(LCDAs) of the participating mesons. In section \ref{section:extraction of LCDA parameters} we extract the $B_{c}$, $J/\psi$ and $\eta_{c}$ LCDA shape parameters and present our predictions of the corresponding form factors at $q^{2}=0$. In section \ref{section:Analysis Bc to P wave} we obtain information of $B_{c}\rightarrow P$ semileptonic form factors over the full physical $q^{2}$ region utilising a suitable extrapolation technique and present predictions of some physical observables. In section \ref{section:Nonleptonic} we present our predictions of branching ratios of a number of nonleptonic decays of $B_{c}$ meson into $S$ and $P$ wave charmonia. Finally, in section \ref{section:summary and conclusion} we briefly summarize our work.

\section{Theoretical background for $B_{c}\rightarrow$ charmonium semileptonic modes}
\label{section:theoretical framework}
In this section, we will focus primarily on the theoretical aspects of our work. We present the theoretical expressions of the different observables related to the semileptonic decays in the SM, which we will predict in this work.
 
\subsection{Physical Observables}
\label{subsection:Physical observables}

In the SM, the effective Hamiltonian for $b\rightarrow c\ell^-\bar{\nu}$ decay can be written as
\begin{equation}
	\mathcal{H}_{eff}=\frac{G_{F}}{\sqrt{2}}V_{cb}^{*}\bar{b}\gamma_{\mu}(1-\gamma_{5})c\otimes \bar{\nu_{l}}\gamma^{\mu}(1-\gamma_{5})l,
\end{equation}	 
where $G_{F}=1.16637\times 10^{-5}\hspace{1.0mm}GeV^{-2}$ is the Fermi coupling constant, and $V_{cb}$ is one of the CKM matrix elements. Using the above effective Hamiltonian, we have the following differential decay widths \cite{Tanaka:2010se,Tanaka:2012nw,Hu:2019bdf}
	\begin{equation}
	\label{eq:dtauP}
	\begin{split}
		\frac{d\Gamma(B_{c}\rightarrow Pl^-\bar{\nu}_{l})}{dq^{2}}=\frac{G_{F}^{2}|V_{cb}|^{2}}{384\pi^{3}m_{B_{c}}^{3}} \sqrt{\lambda(q^{2})} \left(1-\frac{m_{l}^{2}}{q^{2}}\right)^{2} & \Bigg[ 3m_{l}^{2} (H^s_{V,t}(q^2))^2 
		+(m_{l}^{2}+2q^{2})(H_{V,0}(q^{2}))^{2}\Bigg],
	\end{split}
	\end{equation}
for the decay to a pseudoscalar or scalar mesons $P$, like $\eta_{c}$ or $\chi_{c0}$, respectively, and
	\begin{equation}
	\label{eq:dtauV}
	\begin{split}
	\frac{d\Gamma(B_{c}\rightarrow V l^-\bar{\nu}_{l}) }{dq^{2}} =  \frac{G_{F}^{2}|V_{cb}|^{2}}{384\pi^{3}m_{B_{c}}^{3}}\sqrt{\lambda(q^{2})}\left(1-\frac{m_{l}^{2}}{q^{2}}\right)^{2} & \Bigg[(m_{l}^{2} + 2 q^2) (H_{V,+}^{2}+H_{V,-}^{2}+H_{V,0}^{2}) \\
	& + 3 m_l^2 H_{V,t}^{2} \Bigg],
	\end{split}
	\end{equation}

for a final state meson $V$, where $V$ can be an axial-vector meson, like $\chi_{c1}$ or $h_{c}$, or a vector meson $J/\psi$. In the above equations, the phase space factor is expressed as 
	\begin{equation}
	\label{eqn:phasefactor}
	\lambda(q^{2})=(m_{B_{c}}^{2}+m_{P/V}^{2}-q^{2})^{2}-4m_{B_{c}}^{2}m_{P/V}^{2},
\end{equation}	 
with $m_l$, $m_{P/V}$ are the masses of the respective lepton and the final state meson. The total decay width is obtained by integrating $d\Gamma/dq^{2}$ over the physical $q^{2}$ region, which ranges from $m_{l}^{2}$ to $(m_{B_{c}}-m_{P/V})^{2}$.

In the expressions above the rates are written as a functions of the helicity amplitudes $H_{V,0}^{s},H_{V,t}^{s},H_{V,\pm},$ $H_{V,0},H_{V,t}$, which are related to the QCD form factors as given below: 
\begin{equation}
\begin{split}
H_{V,0}^{s}(q^{2})&=\sqrt{\frac{\lambda(q^{2})}{q^{2}}}F_{+}(q^{2}),\\
H_{V,t}^{s}(q^{2})&=\frac{m_{B_{c}}^{2}-m_{P}^{2}}{\sqrt{q^{2}}}F_{0}(q^{2}),\\
H_{V,\pm}(q^{2})&=(m_{B_{c}}\pm m_{V})A_{1}(q^{2})\mp \frac{\sqrt{\lambda(q^{2})}}{m_{B_{c}}\pm m_{V}}V(q^{2}),\\
H_{V,0}(q^{2})&=-\frac{m_{B_{c}}\pm m_{V}}{2m_{V}\sqrt{q^{2}}}\left[(m_{B_{c}}^{2}-m_{V}^{2}-q^{2})A_{1}(q^{2})-\frac{\lambda(q^{2})}{(m_{B_{c}}\pm m_{V})^{2}}A_{2}(q^{2})\right],\\
H_{V,t}(q^{2})&=-\sqrt{\frac{\lambda(q^{2})}{q^{2}}}A_{0}(q^{2}),
\end{split}
\label{eqn:helicity}
\end{equation}
where $(m_{B_{c}}+m_{V})$ and $(m_{B_{c}}-m_{V})$ are for $B_{c}\rightarrow S$ and $B_{c}\rightarrow P$ wave channels respectively. The QCD form factors, which are obtained as the transition matrix elements of the charged weak quark current, are defined above. Depending on the final state charmonium mesons, the corresponding transition matrix elements can be parametrized in terms of the appropriate form factors.
In case the final state meson is a pseudoscalar or scalar meson, the transition matrix element can be parametrized in terms of two form factors $F_{+}$ and $F_{0}$,
\begin{equation}
\begin{split}
\label{eqn:etaC matrix element}
\left\langle P (p_{2})|\bar{c}\gamma^{\mu}b|B_{c}(p_{1})\right\rangle=&\left[(p_{1}+p_{2})^{\mu}-\frac{m_{B_{c}}^{2}-m_P^{2}}{q^{2}}q^{\mu}\right]F_{+}(q^{2})+\frac{m_{B_{c}}^{2}-m_P^{2}}{q^{2}}q^{\mu}F_{0}(q^{2}),
\end{split}
\end{equation}
with $q^\mu = p_1^\mu-p_2^\mu$. In case the final state meson is a vector or axial-vector meson, the transition matrix elements are parametrised in terms of four form factors $A_{0}$, $A_{1}$, $A_{2}$ and $V$,
\begin{equation}
\label{eqn:JPsi matrix element 1}
\left\langle V(p_{2})|\bar{c}\gamma^{\mu}b|B_{c}(p_{1})\right\rangle=\frac{2i V(q^{2})}{m_{B_{c}}\pm m_V}\epsilon^{\mu \nu \rho \sigma}\epsilon_{\nu}^{*} p_{2\rho} p_{1\sigma},
\end{equation}
and,
\begin{equation}
\begin{split}
\label{eqn:JPsi matrix element 2}
\left\langle V(p_{2})|\bar{c}\gamma^{\mu}\gamma_{5}b|B_{c}(p_{1})\right\rangle&=2m_V A_{0}(q^{2})\frac{\epsilon^{*}\cdot q}{q^{2}}q^{\mu}+(m_{B_{c}}\pm m_V)A_{1}(q^{2})\left[\epsilon^{*\mu}-\frac{\epsilon^{*}\cdot q}{q^{2}}q^{\mu}\right]  \\
& -A_{2}(q^{2})\frac{\epsilon^{*}\cdot q}{m_{B_{c}}\pm m_V}   \left[(p_{1}+p_{2})^{\mu}-\frac{m_{B_{c}}^{2}-m_V^{2}}{q^{2}}q^{\mu}\right].
\end{split}
\end{equation}
There are relations among these form factors at maximum recoil, i.e., $q^{2}=0$, are as
\begin{equation}
\label{eqn:constant 1}
F_{+}(0)=F_{0}(0),
\end{equation}
for form factors defined in Eqn.\eqref{eqn:etaC matrix element}, and for those defined in Eqn.\eqref{eqn:JPsi matrix element 2},
\begin{equation}
\begin{split}
\label{eqn:constraint2}
\text{For S wave:}\qquad 2rA_{0}(0)=&(1+r)A_{1}(0)-(1-r)A_{2}(0),\\
\text{For P wave:}\qquad 2rA_{0}(0)=&(1-r)A_{1}(0)-(1+r)A_{2}(0),
\end{split}
\end{equation}
hold. These form factors are the non-perturbative unknowns. To get the $q^2$ distributions of the decay rates, we need to know the $q^2$ shapes of these form factors, which we will discuss in the next subsection. 

Integrating the differential decay rates over the kinematically allowed ranges of $q^2$, we will get the total decay rates, hence, we will obtain the branching fractions by multiplying these decay rates by the lifetime of the $B_c$ meson. In addition to branching ratios, there are three additional observables for the considered $B_{c}$ semileptonic decays that find significance in probing contributions to physics beyond the standard model. These are the longitudinal polarisation of the $\tau$ lepton, $P_{\tau}$, vector and axial-vector meson longitudinal polarisation fraction, $F_{L}(V)$, and forward backward asymmetry for lepton modes, $A_{FB}(l)$.
\begin{itemize}
\item For the first of the three observables, the tau lepton polarisation's definition depends on the frame considered. We follow the framework considered by the authors in \cite{Tanaka:2010se}, in which the spatial components of the momentum transfer $q^{\mu}=p_{B_{c}}^{\mu}-p_{P/V}^{\mu}$ vanish, $p_{B_{c}}^{\mu}$ and $p_{M}^{\mu}$ being the four-momenta of the initial state $B_{c}$ and the final state mesons respectively. The coordinate system they have considered is such that the direction of momenta of the initial and the final state mesons are along the z- axis, and that for the $\tau$ lepton it lies in the x-z plane. We consider the definition of $P_{\tau}$ from previous works \cite{Tanaka:2010se,Tanaka:2012nw,Hu:2019bdf}, which has the form
\begin{equation}
P_{\tau}=\frac{\Gamma_{+}-\Gamma_{-}}{\Gamma_{+}+\Gamma_{-}},
\label{eqn:tau polarization}
\end{equation}
where $\Gamma_{\pm}$ denotes the decay rate of the decay $B_{c}\rightarrow P(V)\tau \nu_{\tau}$ with the $\tau$ lepton helicity $\pm 1/2$. The explicit expressions of $\Gamma_{\pm}$ has been taken from \cite{Sakaki:2013bfa}. In our present work, however, we will only be focusing on the SM contributions, which have the form
\begin{equation}
\label{eqn:Ptau1}
\begin{split}
\frac{d\Gamma_{+}}{dq^{2}}=&\frac{G_{F}^{2}|V_{cb}|^{2}}{192\pi^{3}m_{B_{c}}^{3}}q^{2}\sqrt{\lambda(q^{2})}\left(1-\frac{m_{\tau}^{2}}{q^{2}}\right)^{2}\frac{m_{\tau}^{2}}{2q^{2}}(H_{V,0}^{s\text{ }2}+3H_{V,t}^{s\text{ }2}),\\
\frac{d\Gamma_{-}}{dq^{2}}=&\frac{G_{F}^{2}|V_{cb}|^{2}}{192\pi^{3}m_{B_{c}}^{3}}q^{2}\sqrt{\lambda(q^{2})}\left(1-\frac{m_{\tau}^{2}}{q^{2}}\right)^{2}(H_{V,0}^{s\text{ }2}),
\end{split}
\end{equation}
for $B_{c}\rightarrow P\tau \bar{\nu}_{\tau}$ decays, and
\begin{equation}
\label{eqn:Ptau2}
\begin{split}
\frac{d\Gamma_{+}}{dq^{2}}&=\frac{G_{F}^{2}|V_{cb}|^{2}}{192\pi^{3}m_{B_{c}}^{3}}q^{2}\sqrt{\lambda(q^{2})}\left(1-\frac{m_{\tau}^{2}}{q^{2}}\right)^{2}\frac{m_{\tau}^{2}}{2q^{2}}(H_{V,+}^{2}+H_{V,-}^{2}+H_{V,0}^{2}+3H_{V,t}^{2}),\\
\frac{d\Gamma_{-}}{dq^{2}}&=\frac{G_{F}^{2}|V_{cb}|^{2}}{192\pi^{3}m_{B_{c}}^{3}}q^{2}\sqrt{\lambda(q^{2})}\left(1-\frac{m_{\tau}^{2}}{q^{2}}\right)^{2}(H_{V,+}^{2}+H_{V,-}^{2}+H_{V,0}^{2}),
\end{split}
\end{equation} 
for $B_{c}\rightarrow V\tau \bar{\nu}_{\tau}$ decays.
\item For the second observable, the $V$ logitudinal polarization fraction, the definition has been taken from \cite{Hu:2019bdf}, and is defined as

\begin{equation}
F_{L}(V)=\frac{\Gamma^{0}}{\Gamma^{0}+\Gamma^{+1}+\Gamma^{-1}},
\label{eqn:long pol}
\end{equation}
with the corresponding differential rates having the following forms
\begin{equation}
\label{eqn:FL1}
\begin{split}
\frac{d\Gamma^{\pm1}}{dq^{2}}&=\frac{G_{F}^{2}|V_{cb}|^{2}}{192\pi^{3}m_{B_{c}}^{3}}q^{2}\sqrt{\lambda(q^{2})}\left(1-\frac{m_{l}^{2}}{q^{2}}\right)^{2}\left(1+\frac{m_{l}^{2}}{2q^{2}}\right)(H_{V,\pm}^{2}),\\
\frac{d\Gamma^{0}}{dq^{2}}&=\frac{G_{F}^{2}|V_{cb}|^{2}}{192\pi^{3}m_{B_{c}}^{3}}q^{2}\sqrt{\lambda(q^{2})}\left(1-\frac{m_{l}^{2}}{q^{2}}\right)^{2}\left[\left(1+\frac{m_{l}^{2}}{2q^{2}}\right)H_{V,0}^{2}+\frac{3}{2}\frac{m_{l}^{2}}{q^{2}}H_{V,t}^{2}\right],
\end{split}
\end{equation}
and $V$ signifying that the final state meson is either a vector or an axial-vector meson.
\item Finally for the lepton forward-backward asymmetry, $A_{FB}(l)$ is defined in the $l\bar{\nu_{l}}$ rest frame. The expression has been taken from \cite{Sakaki:2013bfa} and has the following form

\begin{equation}
A_{FB}=\frac{\int_{0}^{1}\frac{d\Gamma}{d\cos\theta}d\cos\theta-\int_{-1}^{0}\frac{d\Gamma}{d\cos\theta}d\cos\theta}{\int_{-1}^{1}\frac{d\Gamma}{d\cos\theta}d\cos\theta}=\frac{\int b_{\theta}(q^{2})dq^{2}}{\Gamma_{B_{c}}},
\label{eqn:forward backward asymmetry}
\end{equation}

where $\theta$ is the angle between the three momentum of the lepton and the $B_{c}$ meson in the $l\bar{\nu_{l}}$ rest frame. $b_{\theta}(q^{2})$ represents the angular coefficient, whose explicit expression has already been shown in \cite{Sakaki:2013bfa}. Here, in this work, we extract the SM contributions, which have the form
\begin{equation}
\label{eqn:AFB1}
\begin{split}
b_{\theta}^{P}(q^{2})&=\frac{G_{F}^{2}|V_{cb}|^{2}}{128\pi^{3}m_{B_{c}}^{3}}q^{2}\sqrt{\lambda(q^{2})}\left(1-\frac{m_{l}^{2}}{q^{2}}\right)^{2}\frac{m_{l}^{2}}{q^{2}}(H_{V,0}^{s}H_{V,t}^{s}),\\
b_{\theta}^{V}(q^{2})&=\frac{G_{F}^{2}|V_{cb}|^{2}}{128\pi^{3}m_{B_{c}}^{3}}q^{2}\sqrt{\lambda(q^{2})}\left(1-\frac{m_{l}^{2}}{q^{2}}\right)^{2}\left[\frac{1}{2}(H_{V,+}^{2}-H_{V,-}^{2})+\frac{m_{l}^{2}}{q^{2}}(H_{V,0}H_{V,t})\right],
\end{split}
\end{equation}
for $B_{c}\rightarrow Pl\bar{\nu_{l}}$ and $B_{c}\rightarrow Vl\bar{\nu_{l}}$ decays respectively, P representing $\eta_{c}$ and $\chi_{c0}$, and V representing $J/\psi$, $\chi_{c1}$ and $h_{c}$ as the final state mesons.
\end{itemize}

\subsection{Form Factors}
\label{subsection:Form Fcators}

\begin{figure}[htb!]
	\centering
	\begin{tikzpicture}
	\begin{feynman}
	\vertex[crossed dot](a){$$};
	\vertex[left=1.0cm of a](a1);
	\vertex[left=1.0cm of a1](a2);
	\vertex[right=1.8cm of a](a3);
	\vertex[below left=1.0cm and 1.0cm of a](b1);
	\vertex[left=1.0cm of b1](b2);
	\vertex[right=2.8cm of b1](b3);
	\vertex[crossed dot][above=0.5cm of a](c1){$$};
	\vertex[above right=1.3cm and 0.5cm of c1](c2){$l$};
	\vertex[above right=0.6cm and 1.0cm of c1](c3){$\bar{\nu_{l}}$};
	\diagram*{(a3)--[arrow size=1pt,fermion](a)--[arrow size=0pt,fermion](a1)--[arrow size=1pt,fermion](a2),(b2)--[arrow size=0pt,fermion](b1)--[arrow size=1pt,fermion](b3),(a1)--[style=red,gluon](b1),(c2)--[arrow size=1pt,fermion](c1),(c1)--[arrow size=1pt,fermion](c3),(a2)--[fill=cyan,bend right, plain,edge label'={\(B_c\)}](b2),(a2)--[fill=cyan,bend left,plain](b2),(a3)--[fill=cyan,bend left,plain,edge label={$X_{c\bar{c}}$}](b3),(a3)--[fill=cyan,bend right, plain](b3)};
	\end{feynman}
	\end{tikzpicture}
	\begin{tikzpicture}
	\begin{feynman}
	\vertex[crossed dot](a){$$};
	\vertex[left=1.7cm of a](a1);
	\vertex[right=1.0cm of a](a2);
	\vertex[right=1.0cm of a2](a3);
	\vertex[below right=1.0cm and 1.0cm of a](b1);
	\vertex[right=1.0cm of b1](b2);
	\vertex[left=2.7cm of b1](b3);
	\vertex[crossed dot][above=0.5cm of a](c1){$$};
	\vertex[above right=1.3cm and 0.5cm of c1](c2){$l$};
	\vertex[above right=0.6cm and 1.0cm of c1](c3){$\bar{\nu_{l}}$};
	\diagram*{(a3)--[arrow size=1pt,fermion](a2)--[arrow size=0pt,fermion](a)--[arrow size=1pt,fermion](a1),(b3)--[arrow size=1pt,fermion](b1)--[arrow size=0pt,fermion](b2),(a2)--[style=red,gluon](b1),(c2)--[arrow size=1pt,fermion](c1),(c1)--[arrow size=1pt,fermion](c3),(a1)--[fill=cyan,bend right, plain,edge label'={\(B_c\)}](b3),(a1)--[fill=cyan,bend left,plain](b3),(a3)--[fill=cyan,bend left,edge label={$X_{c\bar{c}}$}](b2),(a3)--[fill=cyan,bend right, plain](b2)};
	\end{feynman}
	\end{tikzpicture}
	
	\caption{Leading order Feynman diagrams for semileptonic decays of $B_{c}$ meson.}
\label{fig:Semileptonic Feynman Diagrams}
\end{figure}
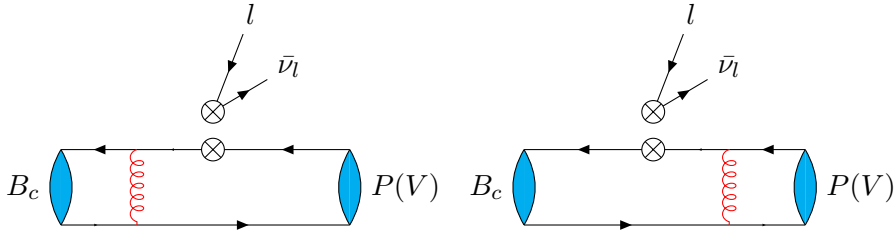

In this subsection, we focus our discussion on the form factors left to be elaborated at the end of the previous subsection. We will consider the analytical expressions of the relevant form factors calculated in the PQCD factorization framework  \cite{Li:1994iu,Li:1995jr,Li:1999hx,Kurimoto:2001zj,PhysRevD.67.034001,Wang:2012lrc,Rui:2017pre,Rui:2018kqr,PhysRevD.97.113001,Hu:2019bdf}. For the semileptonic decays of $B_c$ meson to S and P-wave charmonium, the leading order factorizable Feynman diagrams are shown in Fig.\ref{fig:Semileptonic Feynman Diagrams}. According to the PQCD factorization theorem, we can express the form factors as a convolution of initial and final state meson distribution amplitudes with a hard amplitude. The detailed analytical expressions of these form factors are given in Appendix \ref{section:Appendix pQCD form factors}. In those definitions, the distribution amplitudes absorb nonperturbative dynamics of the process and are process-independent. These distribution amplitudes are introduced in the definition of the non-local matrix elements of the longitudinally and transversely polarised axial-vector and scalar charmonium mesons. The distribution amplitudes relevant to this work are shown in the next subsection.

The hard amplitude, on the other hand, encodes all the hard sub-processes occurring, such as the exchange of hard gluons between the decaying quark and the spectator quark, and is perturbatively calculable and process-dependent. The higher-order radiative corrections to the diagrams shown in Fig. \ref{fig:Semileptonic Feynman Diagrams} generate large logarithms, which can be absorbed
into the meson wave functions. Also, due to the overlapping of the soft and collinear divergences, one will encounter double logarithmic divergences $\propto \alpha_s \ln^2(k_T)$, $k_{T}$ representing the transverse momenta of the quarks. These large double logarithms can be summed to all orders to give a Sudakov exponent factor \cite{Nagashima:2002ia}. The Sudakov factor thus obtained fixes the infrared divergences in $k_{T}$ space. After absorbing all the soft dynamics, the initial and final state meson wave functions can be treated as nonperturbative inputs, which are not calculable but universal. 

As has been explained in \cite{Li:1994iu,Li:1995jr,Li:1999hx,Kurimoto:2001zj} for $B\rightarrow \pi$ form factor and in \cite{Kurimoto:2002sb} for $B\rightarrow D^{(*)}$ form factors, radiative corrections to the meson wave functions and hard amplitudes with the processes having kinematics as shown in appendix \ref{section:Appendix kinematics} also generate double logarithms $\alpha_{s}\ln^{2}(x)$, where $x$ is fraction of the spectator momentum fraction. This term will be divergent at end point regions of $x\to 0$. This double logarithm can be organised into a jet function $S_{t}(x)$ as a consequence of threshold resummation \cite{Li:2001ay}. The jet function is expressed as
\begin{equation}
\label{eqn:jet function}
S_{t}(x)=\frac{2^{1+2c}\Gamma(\frac{3}{2}+c)}{\sqrt{\pi}\Gamma(1+c)}[x(1-x)]^{c},
\end{equation}
with c=0.3. The factorization formulae thus obtained in $k_{T}$ space upon Fourier transform gets translated to the impact parameter space. Details of the derivation for form factors has been shown in \cite{Li:1994iu,Li:1995jr,Li:1999hx,Kurimoto:2001zj}. Following the same procedure, authors of \cite{Rui:2016opu,Rui:2017pre} have derived the form factors for $B_{c}\rightarrow P,V$ channels, which we have adopted in this work. 

Analysis of $B_{c}$ meson, being a heavy-heavy system, involves multiple scales and may be studied in the formalism for heavy quarkonium decays. Resummation of such systems is much more complicated than for B meson decays. However, taking the limit $m_{b}\rightarrow \infty$ but keeping $m_{c}$ finite, the $B_{c}$ meson can be treated as a heavy-light system and analysis of the decays can be carried out in conventional PQCD approach for B meson decays \cite{Kurimoto:2002sb}. However, following the details of the formalism mentioned in \cite{PhysRevD.97.113001} there will be a modification which we have incorporated in the Sudakov factor arising from $k_{T}$ resummation. Details of the formalism have been shown in \cite{PhysRevD.97.113001}. The Sudakov exponent thus derived taking the charm quark mass effect in the impact parameter space has been derived to have the form
\begin{equation}
\begin{split}
s_{c}(Q,b) = & s(Q,b)-s(m_{c},b), \\
=&\int_{m_{c}}^{Q}\frac{d\mu}{\mu}\left[\int_{1/b}^{\mu}\frac{d\bar{\mu}}{\bar{\mu}}A(\alpha_{s}(\bar{\mu}))+B(\alpha_{s}(\bar{\mu}))\right],
\end{split}
\end{equation}
where the expressions for $s(Q,b)$ representing the Sudakov exponent obtained by $k_{T}$ resummation of an energetic light quark, $A(\alpha_{s}(\bar{\mu}))$ and $B(\alpha_{s}(\bar{\mu}))$ has been taken from \cite{Li:1999kna}. Accordingly, we will obtain the expressions for the total Sudakov exponential factors for $B_c$ and other charmonium meson distribution amplitudes \cite{PhysRevD.97.113001}.

\subsection{Light Cone Distribution Amplitudes}
\label{subsection:LCDAs}

In the last subsection, the form factors were expressed as functions of light cone distribution amplitudes (LCDAs) of the initial and final mesons. The dependence is better presented in the analytic expressions of form factors in Appendix \ref{section:Appendix pQCD form factors}. In this subsection we carry forward our discussion with a description of the various LCDAs that has been considered in this work. 

\paragraph{\underline{$\boldsymbol{B_c}$ meson LCDA:}}
The form of the $B_{c}$ meson distribution amplitude we would be considering here is an approximate Gaussian form that has been taken from \cite{PhysRevD.97.113001}
\begin{equation}
\begin{split}
	\phi_{B_{c}}(x,b)=N_{B_{c}} x (1-x) \exp\left[ -\frac{(1-x)m_{c}^{2}+xm_{b}^{2}}{8\omega_{B_{c}}^{2}x(1-x)}\right]\exp[-2\hspace{0.05cm}\omega_{B_{c}}^{2}x(1-x)b^{2}],
	\end{split}
\end{equation}
The normalization constant $N_{B_{c}}$ is fixed by the relation
\begin{equation}
	\int_{0}^{1}\phi_{B_{c}}(x,b=0)dx=	\int_{0}^{1}\phi_{B_{c}}(x) dx=1,
\end{equation}
and the parameter $b$ being the impact parameter, or transverse separation between the quarks, and is infact Fourier conjugate to the transverse momentum $k_{T}$, $\omega_{B_{c}}$ being the shape parameter of the $B_{c}$ meson distribution amplitude which is an model dependent parameter and can be treated as an unknown parameter.

\paragraph{\underline{LCDA of S-wave charmoniums:}}
For the LCDA of the $J/\psi$ and $\eta_{c}$ mesons, we consider a potential model, which effectively performs the action of binding the valence quarks, namely the charm quark and the charm anti-quark into a single bound state. But, before bringing forth the potential model, we would consider probing into the structure of the charmonium meson, be it the vector meson or the pseudoscalar meson, in a little detail. In the case of a charmonium meson, the $c\bar{c}$ system can be considered analogous to an atom's nucleus, both systems having their own spectroscopy. Now, the most realistic potential that can describe the nucleons inside the nucleus is the Wood-Saxon Potential, which despite being realistic, turns out to pose a difficulty when attempts are made to solve the Schr$\ddot{o}$dinger equation analytically. Alternatively, the system can be treated numerically. Further simplification to the computation is achieved when the energy levels and other properties are achieved by approximating the potential model with a three-dimensional harmonic oscillator potential \cite{Sun:2008ew}. In the ground state, i.e., the 1S state, the radial wavefunction of the corresponding Schr$\ddot{o}$dinger state is expressed as
\begin{equation}
	\psi_{n_{r}L}(r)=\psi_{1S}(r)=R_{1S}(0) \exp\left(-\frac{\alpha^{2}r^{2}}{2}\right),
	\label{eqn:radial wavefunction 1S}
\end{equation}
where $R_{1S}(0)$, serving as the normalization constant, represents the wave function at $r=0$, $\alpha^{2}=\frac{m_{c}\omega}{2}$, and $\omega$ represents the frequency of oscillations. The quantum numbers $n_{r}L$ for the $J/\psi$ and $\eta_{c}$ mesons represent the radial quantum number and the orbital angular momentum quantum number, respectively.

Next, we apply the Fourier transform on the above radial wavefunction to obtain the wavefunction in the momentum space,
\begin{equation}
	\psi_{1S}(\vec{k})\approx \int d^{3}\vec{r}\hspace{1mm}exp(-i\vec{r}\cdot \vec{k})\psi_{1S}(r)\propto \exp\left(-\frac{\vec{k}^{2}}{2\alpha^{2}}\right),
\end{equation}
Next, we apply the substitutions proposed by the authors in their work \cite{Brodsky:1982nx}
\begin{equation}
	\begin{split}
		\vec{k}_{T}&\rightarrow \vec{k}_{T},\\
		k_{z}&\rightarrow (x-\bar{x})\frac{m_{0}}{2},\\
		m_{0}^{2}&=\frac{m_{c}^{2}+\vec{k}_{T}^{2}}{x\bar{x}},
	\end{split}
\end{equation}
where $\bar{x}=1-x$, and x are the longitudinal momentum fractions carried away by the valence quarks of the meson. Upon performing the above-mentioned substitutions, we obtain the wavefunction as 
\begin{equation}
	\psi_{1S}(\vec{k})\rightarrow \psi_{1S}(x,\vec{k}_{T})\propto \exp\left(-\frac{(x-\bar{x})^{2}m_{c}^{2}+\vec{k}_{T}^{2}}{8\alpha^{2}x\bar{x}}\right).
\end{equation}
In the next step, we again perform Fourier Transform on the above wavefunction to obtain the wavefunction in terms of the impact parameter b, which is the Fourier conjugate to the transverse momentum $\vec{k_{T}}$. The 1S oscillator wavefunction form comes out to be of the form
\begin{equation}
	\begin{split}
		\psi_{1S}(x,b)&\approx \int d^{2}\vec{k}_{T}\hspace{1mm} \exp\left(-i\vec{b}\cdot\vec{k}_{T}\right)\psi_{1S}(x,\vec{k}_{T})\\
		&\propto x\bar{x}\hspace{1mm} \exp\left[-\frac{m_{c}}{\omega}x\bar{x}\left\lbrace\left(\frac{x-\bar{x}}{2x\bar{x}}\right)^{2}+\omega^{2}b^{2}\right\rbrace\right],
	\end{split}
\end{equation}
with the modified wavefunction can be written as
\begin{small}
\begin{equation}
	\psi_{X_{c\bar{c}}(1S)}\left(x,b\right)\propto \Phi^{asy}(x)\hspace{1mm}\exp\left[-\frac{m_{c}}{\omega}x\bar{x}\left\lbrace\left(\frac{x-\bar{x}}{2x\bar{x}}^{2}+\omega^{2}b^{2}\right)\right\rbrace\right],
\end{equation}
\end{small}
with $\Phi^{asy}\left(x\right)$ represents the wavefunction of the corresponding twist for the light meson, when set to the asymptotic limit. Thus, the LCDAs of the $J/\psi$ and the $\eta_{c}$ meson can be expressed as
\begin{equation}\label{eq:wfJetac}
\begin{split}
		&\psi^{L,T}_{J/\psi}(x,b)=\frac{f_{J/\psi}}{2\sqrt{2N_{c}}}N^{L}_{J/\psi}\hspace{1mm}x\bar{x}\times f(x),\\
		\vspace{0.5cm}		
		&\psi^{t}_{J/\psi}(x,b)=\frac{f_{J/\psi}}{2\sqrt{2N_{c}}}N^{t}_{J/\psi}\hspace{1mm}(x-\bar{x})^{2}\hspace{1mm}\times f(x),\\
		&\psi^{v}_{\eta_{c}}(x,b)=\frac{f_{\eta_{c}}}{2\sqrt{2N_{c}}}N^{v}_{\eta_{c}}\hspace{1mm}x\bar{x}\hspace{1mm}\times f(x),\\
		&\psi^{s}_{\eta_{c}}(x,b)=\frac{f_{\eta_{c}}}{2\sqrt{2N_{c}}}N^{s}_{\eta_{c}}\hspace{1mm}\times f(x),
		\end{split}
\end{equation}
where 
\begin{equation}
f(x)=\exp\left[-\frac{m_{c}}{\omega_{J/\psi(\eta_{c})}}x\bar{x}\left\lbrace\left(\frac{x-\bar{x}}{2x\bar{x}}\right)^{2}+\omega_{J/\psi(\eta_{c})}^{2}b^{2}\right\rbrace\right].
\end{equation}
In the above equations, $f_{J/\psi}$, $\omega_{J/\psi}$, $f_{\eta_{c}}$ and $\omega_{\eta_{c}}$ are the decay constants and meson distribution amplitude shape parameters of $J/\psi$ and $\eta_{c}$ mesons respectively. The normalization constants 
$N^{L,t}_{J/\psi}$ and $N^{v,s}_{\eta_{c}}$ can be fixed by the relations
\begin{equation}
	\begin{split}
		&\int_{0}^{1}dx\hspace{1mm}\psi^{L,t}_{J/\psi}(x,0)=\frac{f_{J/\psi}}{2\sqrt{2N_{c}}},\\
		&\int_{0}^{1}dx\hspace{1mm}\psi^{v,s}_{\eta_{c}}(x,0)=\frac{f_{\eta_{c}}}{2\sqrt{2N_{c}}},
	\end{split}
\end{equation}
 where $N_{c}$ represents the color number. Using the above normalization conditions we will obtain $N^{L,t}_{J/\psi}$ and $N^{v,s}_{\eta_{c}}$ as functions of the shape parameters $\omega_{J/\psi}$ and $\omega_{\eta_{c}}$, respectively. Therefore, to get information on the meson LCDA, we need inputs on the decay constants and the shape parameters. 

\paragraph{\underline{LCDA of P-wave charmoniums:}}

To extract the LCDA defined above, we will need information on the decay constants and the wave function shape parameters. For the P-wave charmonia, relatively less information is available. Minimal inputs are available to extract them simultaneously. Therefore, we will take a different approach to define the LCDA related to P-wave charmonium compared to S-wave charmonium.   

For P wave charmonia we consider the LCDAs as taken in ref. \cite{Rui:2017pre}, where the authors have considered the valence quarks bound by Coulombic potential. For scalar charmonium, $\chi_{c0}$, the leading and next-to-leading twist LCDAs are considered to have the same form as pseudoscalar mesons \cite{Cheng:2005nb}

\begin{equation}
\begin{split}
	\psi_{S}^{s}(x)=&\frac{f_{S}}{2\sqrt{2N_{c}}}N_{S}\mathcal{T}(x),\\
	\psi_{S}^{v}(x)=&\frac{f_{S}}{2\sqrt{2N_{c}}}N_{T}x(1-x)(2x-1)\mathcal{T}(x),
\end{split}
\label{eqn:LCDA chic0}
\end{equation}
with the normalization conditions
\begin{equation}
	\begin{split}
		\int_{0}^{1}\psi_{S}^{s}(x)dx&=\frac{f_{S}}{2\sqrt{2N_{c}}},\\
		\int_{0}^{1}(2x-1)\psi_{S}^{v}(x)dx&=\frac{f_{S}}{2\sqrt{2N_{c}}}, 
\end{split}
\end{equation}
fixing the constants $N_{S}$ and $N_{T}$. For axial-vector mesons, the leading and next-to-leading twist LCDAs have the form
\begin{equation}
\begin{split}
\psi^{L}(x)&=\frac{f_{A}}{2\sqrt{2N_{c}}}N_{L}x(1-x)\mathcal{T}(x),\\
\psi^{T}(x)&=\frac{f_{A}^{\perp}}{2\sqrt{2N_{c}}}N_{T}x(1-x)(2x-1)\mathcal{T}(x),\\
\psi^{t}(x)&=\frac{f_{A}^{\perp}}{2\sqrt{2N_{c}}}\frac{N_{T}}{6}(2x-1)(1-6x+6x^{2})\mathcal{T}(x),\\
\psi^{V}(x)&=\frac{f_{A}}{2\sqrt{2N_{c}}}\frac{N_{L}}{8}[1+(1-2x)^{2}]\mathcal{T}(x),
\end{split}
\label{eqn:LCDA chic1}
\end{equation}
for $\chi_{c1}$ meson, and
\begin{equation}
\begin{split}
\psi^{L}(x)&=\frac{f_{A}}{2\sqrt{2N_{c}}}N_{T}x(1-x)(2x-1)\mathcal{T}(x),\\
\psi^{T}(x)&=\frac{f_{A}^{\perp}}{2\sqrt{2N_{c}}}N_{L}x(1-x)\mathcal{T}(x),\\
\psi^{t}(x)&=\frac{f_{A}^{\perp}}{2\sqrt{2N_{c}}}\frac{N_{L}}{2}(1-2x)^{2}\mathcal{T}(x),\\
\psi^{V}(x)&=\frac{f_{A}}{2\sqrt{2N_{c}}}\frac{N_{T}}{12}(2x-1)^{3}\mathcal{T}(x),
\end{split}
\label{eqn:LCDA hc}
\end{equation}
for $h_{c}$ meson. The constants $N_{L}$ and $N_{T}$ are fixed by normalization conditions
\begin{equation}
\begin{split}
&\int_{0}^{1}N_{L}x(1-x)\mathcal{T}(x)dx=1,\\
&\int_{0}^{1}N_{T}x(1-x)(2x-1)^{2}\mathcal{T}(x)dx=1,
\end{split}
\end{equation}
and $f_{S}$ and $f_{A}$ representing the corresponding decay constants and $\mathcal{T}(x)$ has the form \cite{Rui:2017pre}
\begin{equation}
\mathcal{T}(x)=\left[\frac{\sqrt{x(1-x)(1-4x(1-x))^{3}}}{\lbrace1-4x(1-x)(1-\frac{v^{2}}{4})\rbrace^{2}}\right]^{1-v^{2}},
\end{equation}
with $v^{2}=0.3$ representing the square of the relative velocity between the quark pair. 
We do not have any input from the lattice on the decay constants; only a few measured branching fractions are available, which we will discuss in the coming sections. In this form, the P-wave charmonium DAs are known in the asymptotic limit. Apart from the decay constants, there are no free parameters to be extracted as such. 

\section{Extraction of LCDA parameters for $B_{c}$ and $S$ wave charmonia}
\label{section:extraction of LCDA parameters}

\begin{table}[t!]		
	\centering
	\renewcommand{\arraystretch}{0.9}
	\setlength{\tabcolsep}{3pt}
	\begin{tabular}{|l|lll|}
		\hline
		\textbf{Mass (GeV)} & $m_{B_{c}}=6.277$ &&  \\ 
		& $m_{J/\psi}=3.097$ & $m_{\eta_{c}}=2.981$ &\\
		& $m_{\chi_{c0}}=3.415$ & $m_{\chi_{c1}}=3.511$ & $m_{h_{c}}=3.525$ \\ 
		& $m_{e}=0.511\times 10^{-3}$&$m_{\nu}=0.105$&$m_{\tau}=1.776$\\
		&&&\\
		\hline
		\textbf{CKM matrix}&$V_{cb}=0.0411(12)$&&\\
		\textbf{elements}&$V_{ud}=0.97370(14)$&$V_{us}=0.2245(8)$&\\
		&&&\\
		\hline
		\textbf{Decay}&$f_{B_{c}}=0.427(6)$ \cite{McNeile:2012qf}&&\\
		\textbf{Constants}&$f_{\eta_{c}}=0.3947(24)$ \cite{Becirevic:2013bsa}&$f_{J/\psi}=0.405(6)$		 \cite{Donald:2012ga}&\\
		\textbf{(GeV)}&$f_{\pi}=0.130(38)$&$f_{K}=0.160(25)$&\\
		&&&\\
		\hline
	\end{tabular}
	\caption{Values of input parameters used in this work.}
	\label{table:input parameters}
	\end{table}

Having discussed the framework and the analytic expressions for the various observables, we next move onto the first step in our analysis, which involves the extraction of the shape parameters of the LCDAs of the participating mesons. We do so by the method of chi-square minimization. The chi-square function is generally defined as
\begin{equation}
	\chi^{2}=\sum_{i}(\mathcal{O}_{i}^{data}-\mathcal{O}_{i}^{theory})^{T}V_{ij}^{-1}(\mathcal{O}_{j}^{data}-\mathcal{O}_{j}^{theory})+\chi^{2}_{nuis},
	\label{eqn:chi square function}
\end{equation}

with $\mathcal{O}_{i}^{data}$ representing the synthetic data values of the corresponding inputs. For this section the inputs are going to be estimates of form factors at $q^{2}=0$. Now, as for the form factors of $B_{c}\rightarrow J/\psi$ channel, HPQCD \cite{Harrison:2020gvo} has supplied with the BCL parameters which let us extrapolate them from high $q^{2}$ region to $q^{2}=0$. As for the form factor of $B_{c}\rightarrow \eta_{c}$ channel, the analysis from the lattice come with an incomplete error treatment rendering it unusable in our current analysis \cite{Colquhoun:2016osw}. Thus in order to obtain information on $B_{c}\rightarrow \eta_{c}$ form factors, an indirect approach needs to be utilised which would connect the available $B_{c}\rightarrow J/\psi$ form factors to the $B_{c}\rightarrow \eta_{c}$ form factors. One such approach is to utilise the Heavy Quark Spin Symmetry (HQSS) \cite{Jenkins:1992nb}  that exists between the two states through a universal Isgur Wise function. The approach and steps involved to obtain the form factor for the same has been discussed in detail in \cite{Biswas:2023bqz,Cohen:2019zev,Colangelo:2022lpy}. We have redone the steps and independently have arrived at our estimate of the corresponding form factor. These inputs are shown in Table \ref{table:form factor q2=0}. $V_{ij}$ represents the covariance matrix between the inputs.

\begin{table}[htb!]
	\centering
	\begin{tabular}{|c|cc|cccc|}
		\hline
		\textbf{Decay}&\textbf{Form} & \textbf{Values} &&\textbf{Correlation}&& \\
		\cline{4-7}
		\textbf{Channel}& \textbf{Factors}&\textbf{at $\boldsymbol{q^{2}=0}$}&$A_{0}(0)$&$A_{1}(0)$&$A_{2}(0)$&$V(0)$\\
		\hline
		$B_{c}\rightarrow \eta_{c}\hspace{0.01cm}l\nu_{l}$&$F_{+}(0)$ & 0.521(197)&1.0&0.0&0.0& 0.0\\ 
		&$A_{0}(0)$ & 0.477(43) &&1.0&0.466&0.018\\ 
		$B_{c}\rightarrow J/\psi\hspace{0.01cm}l\nu_{l}$&$A_{1}(0)$ & 0.457(28) &&&1.0&0.029\\ 
		&$V(0)$ & 0.725(67) &&&&1.0\\ 
		\hline
	\end{tabular} 
	\caption{Form Factor inputs at $q^{2}=0$ for all the channels along-with their correlation.}
	\label{table:form factor q2=0}
\end{table}

As for $\mathcal{O}^{theory}$, the analytic expressions for the respective form factors in pQCD at $q^{2}=0$ are taken. These expressions along-with appropriate references has been shown in appendix \ref{section:Appendix pQCD form factors}. In addition to these, there is $\chi^{2}_{nuis}$, which is the chi-square function formed by the relevant nuisance parameters. In this section, these nuisance parameters involve the decay constants of the participating mesons, estimates of which we have taken from Table \ref{table:input parameters}. In addition to decay constants, we have masses of charm and bottom quarks which has been taken to be average of the masses in pole mass, $\overline{MS}$ and kinetic schemes along-with a 10\% and 25\% error for $m_{b}$ and $m_{c}$ respectively. This has been done to account for an inclusive and scheme independent approach to the choice of the relevant quark masses. The quark masses has been presented in Table \ref{table:scheme dependent masses}.

\begin{table}[htb!]
	\centering
	\begin{tabular}{|c|cc|}
		\hline
		\textbf{Scheme}&$\boldsymbol{m_{b}}$ (GeV)&$\boldsymbol{m_{c}}$ (GeV)\\
		\hline
		Pole mass&4.78&1.67\\
		$\overline{MS}$&4.18&1.273\\
		Kinetic&4.56&1.091\\
		\hline
		Average&4.506(451)&1.345(336)\\
		\hline
	\end{tabular}
	\caption{Values of $m_{b}$ and $m_{c}$ (in GeV) in three different schemes and their average value \cite{ParticleDataGroup:2022pth}.}
	\label{table:scheme dependent masses}
\end{table}

Along-with $m_{b}$ and $m_{c}$, the shape parameters $\omega_{J/\psi}$ and $\omega_{\eta_{c}}$ of $J/\psi$ and $\eta_{c}$ LCDAs are also taken into the chi-square function as nuisance parameters. They are fixed by solving the radial wave-functions of the respective charmonium states at the origin with their corresponding numerical estimates. The radial wave function has already been discussed in \cref{eqn:radial wavefunction 1S} in section \ref{subsection:LCDAs}. An analytic expression for $R_{1S}(0)$ is extracted by simply following the normalization condition of the wavefunction, giving us
\begin{equation}
	R_{1S}(0)=\left [\frac{2}{\pi}(m_{c}\omega)^{3}\right ]^{1/4},
	\label{eqn:radial wavefunction 1S at r=0}
\end{equation}
following which we extract our preliminary estimates for $\omega_{J/\psi}$ and $\omega_{\eta_{c}}$ by simply equating \cref{eqn:radial wavefunction 1S at r=0} to the numerical values of $R_{1S}(0)$ already extracted in \cite{Biswas:2023bqz} and solving for $\omega_{J/\psi}$ and $\omega_{\eta_{c}}$. The values thus obtained are as
\begin{table}
	\centering
	\begin{tabular}{|c|ccc|}
		\hline
		&$\omega_{J/\psi}$&$\omega_{\eta_{c}}$&$m_{c}$\\
		\hline
		$\omega_{J/\psi}$&1.0&0.938&-0.960\\
		$\omega_{\eta_{c}}$&&1.0&-0.977\\
		$m_{c}$&&&1.0\\
		\hline
	\end{tabular}
	\caption{Correlation matrix between $\omega_{J/\psi}$, $\omega_{\eta_{c}}$ and $m_{c}$.}
	\label{table:correlation nuisance}
\end{table}

\begin{equation}
	\begin{split}
		\omega_{J/\psi}=0.681(184)~GeV,\qquad
		\omega_{\eta_{c}}=0.898(227)~GeV,
	\end{split}
	\label{eqn:nuisance shape parameters of charmonium}
\end{equation}

with correlation between them and $m_{c}$ as shown in Table \ref{table:correlation nuisance}.

These extracted values of $\omega_{J/\psi}$ and $\omega_{\eta_{c}}$ along with the correlation in Table \ref{table:correlation nuisance} are fed into $\chi^{2}_{nuis}$. The chi-square function thus constructed is then minimised to extract the required shape parameter of the $B_{c}$ meson LCDA along-with the other nuisance parameters. The results of the chi-square minimisation has been presented in Table \ref{table:LCDA1} and the corresponding correlation matrix has been presented in Table \ref{table:Appendix correlation LCDA}.

\begin{table}[htb!]
	\centering
	\begin{tabular}{|cc|cc|}
		\hline
		\multicolumn{2}{|c|}{\textbf{Free Parameters}}  &  \multicolumn{2}{c|}{\textbf{Nuisance Parameters}} \\
		\hline
		\textbf{Parameters}& \textbf{Fit Results (in GeV)} & \textbf{Parameters}& \textbf{Fit Results (in GeV)} \\
		\hline
		$\omega_{B_{c}}$&0.998(34)&$\omega_{J/\psi}$&0.667(80)\\
		&&$\omega_{\eta_{c}}$&0.783(82)\\
		& & $f_{B_{c}}$ & 0.429(4) \\
		& & $f_{J/\psi}$ & 0.405(4)\\
		& & $f_{\eta_{c}}$ & 0.3947(17)\\
		&  & $\Lambda_{QCD}$ & 0.2797(4)\\
		&&$m_{c}$&1.343(111)\\
		&&$m_{b}$&4.506(6)\\
		\hline
		\textbf{D.O.F}&\multicolumn{3}{|c|}{3}\\
		\hline
		\textbf{p-Value}&\multicolumn{3}{|c|}{8.72\%}\\
		\hline
	\end{tabular}
	\caption{Extracted values of LCDA parameters obtained by fitting pQCD form factors of $B_{c}\rightarrow J/\psi(\eta_{c})$ transition with corresponding lattice input at $q^{2}=0$.}
	\label{table:LCDA1}
\end{table}

We now discuss the results of Table \ref{table:LCDA1}, separating it into two parts. First, we discuss our estimates of the nuisance parameters. As for the error estimates of $\omega_{J/\psi}$, $\omega_{\eta_{c}}$, $m_{c}$ and $m_{b}$, we have obtained a significant reduction in error compared to what we had initially fed into $\chi^{2}_{nuis}$. This is mainly due to relatively small errors in the form factors used as inputs as compared to the other inputs used as nuisance parameters. These inputs on the form factors play an essential role in constraining the parameters. Also, it would be interesting to look at the correlation matrices in Tables \ref{table:correlation nuisance} and \ref{table:Appendix correlation LCDA}. For $m_{b}$, a strong correlation is observed in Table \ref{table:Appendix correlation LCDA} with most of the other parameters, inferring its error estimate propagating into the error estimates of the other parameters. Moreover, the pQCD expressions of form factors are highly sensitive on $m_{b}$, because of which it gets tightly constrained during the chi-square optimisation. For $m_{c}$, the reduction in error has a similar reason, that the form factors have a high sensitivity on it, and hence a chi-square optimisation tightly constrains the $1\sigma$ region of $m_{c}$. This reduction in error of both $m_{b}$ and $m_{c}$ in turn reduces the error estimate of $\omega_{J/\psi}$ and $\omega_{\eta_{c}}$ due to the correlation between them.

\begin{figure*}[t]	
	\centering
	\includegraphics[width=9.0cm]{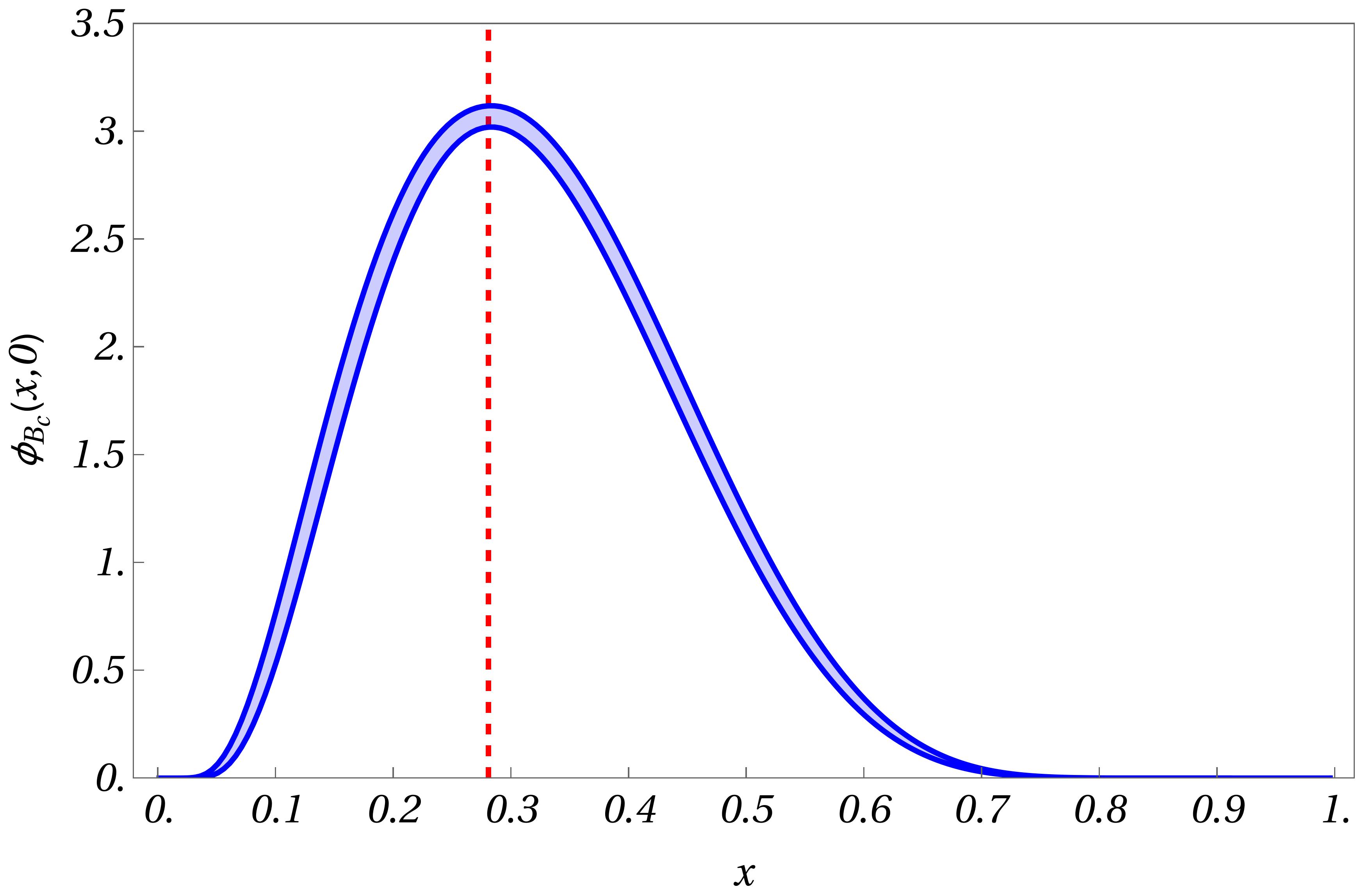}
	\caption{Shape of $B_{c}$ meson distribution amplitude with extracted value of $\omega_{B_{c}}$. The blue curve denotes $\phi_{B_{c}}(x,0)$ obtained using our extracted value of $\omega_{B_{c}}$ and the red dashed line denotes the point $x\sim m_{c}/m_{b}\sim 0.298(19)$.}
	\label{fig:phiBc DA}
\end{figure*}

Second, for the free parameter $\omega_{B_{c}}$, our estimate can be verified by plotting the distribution amplitude of the $B_{c}$ meson for our constrained value of $\omega_{B_{c}}$. Authors of \cite{PhysRevD.97.113001} in their work had explained about the kinematic constraints on the shape of the $B_{c}$ meson distribution amplitude and that it would attain a peak at around $x\sim m_{c}/m_{b}$, which with numerical values of $m_{c}$ and $m_{b}$, should be at $x\sim 0.298(19)$. For the shape of $\phi_{B_{c}}(x,0)$ that we have obtained with our extracted value of $\omega_{B_{c}}$, as can be seen in Fig \ref{fig:phiBc DA}, the peak is at $x\sim 0.281$, being consistent with the ratio $m_{c}/m_{b}$ within 1$\sigma$ error range. Thus we can safely accept the extracted value of $\omega_{B_{c}}$ as an acceptable one.

With the LCDA parameters extracted, we now use them as inputs into the analytic expressions of the form factors showcased in appendix \ref{section:Appendix pQCD form factors} and obtain our predictions for the form factors at $q^{2}=0$. In calculating the form factors, we set the cut-off of the impact parameter, $b_{c}$ in the form factor expressions at 90\% of $1/\Lambda_{QCD}$. This is done to keep our predictions in a region well within maximum value of $b$ upto which pQCD is valid, i.e., $1/\Lambda_{QCD}$. These predictions are presented in Table \ref{table:LCDA 2} which are very much consistent with the lattice inputs used in the analysis. 

\begin{table}[t]	
	\centering	
	\begin{tabular}{|c|cc|}
		\hline
		\textbf{Form Factors}& \textbf{This work}& \textbf{Lattice Input at $\boldsymbol{q^{2}=0}$ \cite{Harrison:2020gvo}}  \\
		\hline
		$F_{+}^{B_{c}\rightarrow \eta_{c}}(0)$& 0.527(118) &-\\
		
		$A_{0}^{B_{c}\rightarrow J/\psi}(0)$& 0.452(39) & 0.477(43)\\
		
		$A_{1}^{B_{c}\rightarrow J/\psi}(0)$& 0.439(32) & 0.457(28)\\
		
		$A_{2}^{B_{c}\rightarrow J/\psi}(0)$& 0.411(68) & 0.417(87) \\
		
		$V^{B_{c}\rightarrow J/\psi}(0)$& 0.746(57) & 0.725(67) \\
		\hline
	\end{tabular}
	\caption{Prediction of form factors of $B_{c}\rightarrow J/\psi(\eta_{c})$ transition at $q^{2}=0$ and comparison with lattice values.}	
	\label{table:LCDA 2}
\end{table}

Thus to conclude in this section, we have extracted the shape parameters of $B_{c}$, $J/\psi$ and $\eta_{c}$ meson distribution amplitudes to be used during the predictions of form factors concerning $B_{c}\rightarrow \text{ }P$ wave charmonia in later sections.

\section{Numerical analysis of some $B_{c}\rightarrow P$ wave semileptonic channels}
\label{section:Analysis Bc to P wave}

Now that the analysis involving the decay of $B_{c}\rightarrow \text{ }S$ wave charmonium states is accomplished, in this section we move on to analysing the $B_{c}\rightarrow P$ wave semileptonic decay channels. For these channels, the information on the form factors are not available from lattice. Hence, it would be interesting to obtain the shapes of the form factors for further phenomenological studies. This void in the present phenomenological arena motivates us to explore these decay channels and present predictions on some fundamental quantities, such as the form factors and some physical observables like semileptonic branching ratios, which can be verified in the near future when sufficient inputs become available. In this section, our focus is primarily going to be on the scalar $\chi_{c0}$ and the axial-vector $\chi_{c1}$ and $h_{c}$ states. We are going to perform the analysis in three steps, separated into three subsections. In the first step in subsection \ref{subsection: P wave decay constants}, we will be extracting the decay constants of the charmonium states. In the second step in subsection \ref{subsection:P wave form factors}, we will be predicting the relevant form factors first at $q^{2}=0$ and then extrapolating them to the full physical $q^{2}$ region utilising a suitable extrapolation technique. And in the final step in subsection \ref{subsection: P wave observables} we will be calculating and predicting a number of relevant physical observables utilising the form factor information obtained in the second step. The form factors for the same channels along-with predictions of various relevant physical observables, has also been calculated in \cite{Colangelo:2022awx} using HQSS.

\subsection{Extraction of decay constants of charmonium states}
\label{subsection: P wave decay constants}

 In this subsection, we extract the decay constants of $\chi_{c0}$, $\chi_{c1}$ and $h_{c}$ states, through a rather data-driven approach, steering ourselves away from taking the currently available model dependent estimates as inputs. The primary motivation for doing so can be explained by revisiting \cref{eqn:LCDA chic0,eqn:LCDA chic1,eqn:LCDA hc}. We can clearly see that, unlike the LCDAs of $S$ wave charmonia, these do not have any shape parameters to introduce any degree of flexibility to their shapes. Even if they had such a parameter, extracting them would be difficult owing to the unavailability of sufficient data at present. Therefore, we consider the decay constants as a free parameter in the LCDAs and extract them following the approach discussed in the following text. Revisiting \cref{eqn:LCDA chic1,eqn:LCDA hc} we make an assumption $f_{A}\approx f_{A}^{\perp}$ due to the lack of enough inputs available at present to extract them separately.
\begin{enumerate}
	\item \textbf{\underline{Extracting $\boldsymbol{f_{\chi_{c0}}}$}:}  As for the theoretical expressions we consider the results presented by Li and Vary in their work \cite{Li:2021ejv}, where they have presented the two photon decay width of $\chi_{c0}$ in Basis Light-Front Quantization (BLFQ) approach. In their work, they have expressed the transition amplitude for the process by parametrizing it in terms of two transition form factors (TFF)
	\begin{equation}
		\begin{split}
			\mathcal{M}^{\mu\nu}=&\frac{4\pi\alpha}{M_{\chi_{c0}}^{2}}\Biggl\lbrace M_{\chi_{c0}}^{2}\left[(q_{1}\cdot q_{2})g^{\mu\nu}-q_{2}^{\mu}q_{1}^{\nu}\right]F_{1}^{S}(q_{1}^{2},q_{2}^{2})+
			\\&\left[q_{1}^{2}q_{2}^{2}g^{\mu\nu}+(q_{1}\cdot q_{2})q_{1}^{\mu}q_{2}^{\nu}-q_{1}^{2}q_{2}^{\mu}q_{2}^{\nu}-q_{2}^{2}q_{1}^{\mu}q_{1}^{\nu}\right]F_{2}(q_{1}^{2},q_{2}^{2})\Biggr\rbrace,
		\end{split}
	\end{equation}
	with $\alpha=1/137$. The subsequent two photon decay width can then be expressed as
	\begin{equation}
		\Gamma_{S\rightarrow \gamma\gamma}=\frac{\pi \alpha^{2}}{4}M_{\chi_{c0}}^{3}|F_{1}^{S}(0,0)|^{2},
	\end{equation}
	considering both the photons are on-shell, and $F_{1}^{s}(0,0)$ represents the TFF in that case. In case if one of the photons is considered to be off-shell, the TFF would be a function of the momentum transferred, $Q^{2}$, which, as has been expressed in \cite{Li:2021ejv}, will have the form
	\begin{equation}
		F_{s\gamma}(Q^{2})=e_{c}^{2}f_{\chi_{c0}}\int_{0}^{1}\frac{(1-2x)\phi_{s}(x,\mu)}{x(1-x)Q^{2}+m_{f}^{2}}dx,
	\end{equation} 
	$f_{\chi_{c0}}$ being the decay constant and $\phi_{s}(x,\mu)$ the leading twist LCDA of the $\chi_{c0}$ meson respectively, the form for which, in our work, has been considered to be the same as taken while calculating the for $B_{c}\rightarrow \chi_{c0}$ form factors. The branching ratio is expressed as
	\begin{equation}
		\mathcal{B}(\chi_{c0}\rightarrow \gamma\gamma)=\frac{\Gamma(\chi_{c0}\rightarrow \gamma\gamma)}{\Gamma_{\chi_{c0}}},
	\end{equation}
	where $\Gamma_{\chi_{c0}}=10.8(6)$ MeV is the total decay width of $\chi_{c0}$ meson. Experimentally the numerical value of the branching ratio of the channel $\chi_{c0}\rightarrow \gamma\gamma$ has been determined to be 2.04(9)$\times 10^{-4}$ \cite{Belle:2006mzv}. With the expressions for the branching ratio and the corresponding experimental input at our hand we extract $f_{\chi_{c0}}$ and finally get
	
	\begin{equation}
		f_{\chi_{c0}}=0.147(20)~GeV.
	\end{equation}

	\item \textbf{\underline{Extracting $\boldsymbol{f_{\chi_{c1}}}$}:} Extracting $f_{\chi_{c1}}$ involves a bit more discussion since the process will involve a chi-square minimization taking three radiative decay channels of $\chi_{c1}$ as inputs. First we discuss the three channels used in this analysis in three separate bullet points and then
	present our final results.
	\begin{itemize}
		\item Channel 1: $\chi_{c1}\rightarrow J/\psi \gamma$: This transition is an electric dipole (E1) transition, with the P wave charmonium state $\chi_{c1}$ decaying into an S wave charmonium state with the emission of a photon. Being an E1 transition, it involves a change in orbital angular momentum quantum number (L) by 1 while the spin quantum number (S) remains unchanged. The process follows E1 selection rules, allowing a change of $\Delta L=\pm 1$ and $\Delta J=0,\pm 1$ other than $J=0$ to $J=1$. The branching ratio for this process has been calculated using the potential NRQCD (pNRQCD) approach, where the transition amplitude depends on the overlap between the initial $\chi_{c1}$ and final $J/\psi$ state. In this framework, the branching ratio is expressed as \cite{MartinezNeira:2017prz}
		
		\begin{equation}
			\mathcal{B}(\chi_{c1}\rightarrow J/\psi\gamma)=\frac{1}{\Gamma_{\chi_{c1}}}\cdot\frac{4}{9}e_{c}^{2}\alpha\omega_{\chi_{c1}}^{3}|I_{SP}|^{2}\left(1+\delta^{\prime}\right),
			\label{eqn:branching ratio ChiC1 JPsi gamma}
		\end{equation}
		with $e_{c}$ representing the charge of the charm quark, $\omega_{\chi_{c1}}$ representing the energy of the emitted photon, $\Gamma_{\chi_{c1}}=0.84(4)$ MeV and $I_{SP}$ representing the overlap integral of the radial states of  the participating mesons expressed as
		\begin{equation}
			I_{SP}=\int_{0}^{\infty}r^{3}R_{S}(r)R_{P}(r)dr,
			\label{eqn:overlap integral}
		\end{equation}
		where $R_{S}(r)$ and $R_{P}(r)$ represents the normalized radial solution of Schr$\ddot{o}$dinger equation for $J/\psi$ and $\chi_{c1}$ mesons respectively which in this work has been calculated by considering the potential binding the quark and anti-quark to be a Coulombic potential. The normalized radial solutions have the form
		\begin{equation}
			\begin{split}
				R_{S}(r)&=\sqrt{\frac{q_{B_{S}}^{3}}{2}}\exp\left(-\frac{q_{B_{S}}r}{2}\right),\\
				R_{P}(r)&=\sqrt{\frac{q_{B_{P}}^{5}}{24}}r\exp\left(-\frac{q_{B_{P}}r}{2}\right),
			\end{split}
			\label{eqn:radial solution of charmonia}
		\end{equation}
		with $q_{B_{S}}$ and $q_{B_{P}}$ representing the Bohr momenta for S and P wave states, respectively. To introduce decay constants into Eqn \eqref{eqn:branching ratio ChiC1 JPsi gamma} we are going to express $q_{B_{S}}$ and $q_{B_{P}}$ in terms of respective decay constants. The decay constants of $J/\psi$ and $\chi_{c1}$ can be expressed in terms of $R_{S}(0)$ and $R_{P}^{\prime}(0)$ as
		\begin{itemize}
			\item For $J/\psi$ meson: The decay constant $f_{J/\psi}$ can be expressed in terms of necessary NLO and relativistic corrections by taking the expression for the decay width of $J/\psi\rightarrow e^{+}e^{-}$ channel in \cite{Bodwin:2002cfe} and expressing the decay width as proportional to the square of the decay constant 
			\begin{equation}
				\scriptsize
				f_{J/\psi}=\sqrt{\frac{3}{2m_{c}\pi}}|R_{J/\psi}(0)|\left[1-\frac{8}{3}\frac{\alpha_{s}}{\pi}-\frac{1}{6}\langle v^{2}\rangle_{J/\psi}+\frac{\alpha_{s}}{3\pi}\left\lbrace\frac{8}{9}+\frac{8}{3}\ln\left(\frac{\mu^{2}}{m_{c}^{2}}\right)\right\rbrace\langle v^{2}\rangle_{J/\psi}+\frac{29}{18}\langle v^{4}\rangle_{J/\psi}\right],
				\label{eqn:decay constant of JPsi}
			\end{equation}
			where $\alpha_{s}$ representing the running coupling constant has been calculated at scale $\mu=2m_{c}$. $\langle v^{2}\rangle_{J/\psi}$ represents the average of the square of relative velocity between the quark pair in the charmonium, its value being taken to be $\langle v^{2}\rangle_{J/\psi}=0.267$ \cite{Zhu:2017lqu}.
			\item and for $\chi_{c1}$ meson: The expression for the decay constant of $\chi_{c1}$ in terms of $|R^{\prime}_{\chi_{c1}}|$ has been taken from \cite{Chung:2021efj} and is expressed as
			\begin{equation}
				f_{\chi_{c1}}=\sqrt{\frac{9N_{c}}{\pi M_{\chi_{c1}}}}\frac{|R_{\chi_{c1}}^{\prime}(0)|}{m_{c}}\left[1+\alpha_{s}c_{a}^{(1)}+\alpha_{s}^{2}c_{a}^{(2)}+\delta_{C}+\alpha_{s}c_{a}^{(1)}\delta_{C}+\delta_{NC}\right],
				\label{eqn:decay constant of ChiC1}
			\end{equation}
			where $c_{a}^{(1)}$ and $c_{a}^{(2)}$ represent the short distance coefficinets at order $\alpha_{s}$ and $\alpha_{s}^{2}$ respectively and their explicit expressions have been taken from \cite{Chung:2021efj}. $\delta_{C}$ and $\delta_{NC}$ represent Coulombic and Non-Coulombic corrections to the binding potential, respectively, and $\alpha_{s}$ has been calculated at the same scale as has been done for $J/\psi$. The expression for $f_{\chi_{c1}}$ in \cite{Chung:2021efj} has two additional correction terms $\delta_{RS^{\prime}}$ and $\delta_{m_{RS^{\prime}}}$ which in this work we have neglected, since our analysis does not consider the charm quark mass to be in any particular renormalization scheme, but an average of the three schemes. Therefore, any correction term introduced that is renormalization scheme dependent has been neglected.   
		\end{itemize}
		Setting r=0 in Eqn \eqref{eqn:radial solution of charmonia} and putting them in Eqns \eqref{eqn:decay constant of JPsi} and \eqref{eqn:decay constant of ChiC1}, we can finally express the respective $q_{B}$ in terms of $f_{J/\psi}$ and $f_{\chi_{c1}}$. We get
		\begin{equation}
			\scriptsize
			q_{B_{J/\psi}}=\left[\frac{4m_{c}\pi}{3}f_{J/\psi}^{2}\left\lbrace 1-\frac{8}{3}\frac{\alpha_{s}}{\pi}-\frac{1}{6}\langle v^{2}\rangle_{J/\psi}+\frac{\alpha_{s}}{3\pi}\left(\frac{8}{9}+\frac{8}{3}\ln\left(\frac{\mu^{2}}{m_{c}^{2}}\right)\right)\langle v^{2}\rangle_{J/\psi}+\frac{29}{18}\langle v^{4}\rangle_{J/\psi}\right\rbrace^{-2}\right]^{1/3},
			\label{eqn:Bohr momentum Jpsi}
		\end{equation}
		and 
		\begin{equation}
			q_{B_{\chi_{c1}}}=\left[\frac{24\pi M_{\chi_{c1}} m_{c}^{2}}{9N_{c}}f_{\chi_{c1}}^{2}\frac{1}{|1+\alpha_{s}c_{a}^{(1)}+\alpha_{s}^{2}c_{a}^{(2)}+\delta_{C}+\alpha_{s}c_{a}^{(1)}\delta_{C}+\delta_{NC}|^{2}}\right]^{1/5}.
			\label{eqn:Bohr momentum chic1}
		\end{equation}
		Putting these equations in eqn.\eqref{eqn:radial solution of charmonia} we can express the radial solutions in terms of decay constants, using which we can, in turn, express the overlap integral in eqn.\eqref{eqn:overlap integral} and hence the branching ratio in terms of the relevant decay constants. Going back to eqn.\eqref{eqn:branching ratio ChiC1 JPsi gamma}, the $\delta^{\prime}$ term, which has been added to represent the relativistic corrections to leading order expression of the branching ratio. This correction term comprises of 
		\begin{itemize}
			\item Correction arising from higher order operators in pNRQCD Lagrangian.
			\item Corrections arising due to the interference between the higher order terms to the initial and final quarkonium states.
		\end{itemize}
		Details of the sources of these correction terms have been discussed in \cite{MartinezNeira:2017prz}.
		\item Channel 2: $\chi_{c1}\rightarrow V \gamma$: The $\chi_{c1}\rightarrow V\gamma$ transition, with V representing a light vector meson $\rho$ and $\phi$, is a radiative decay of charmonium involving the emission of a photon and the production of a light meson. These decays are different from E1 transition, the process involving a quark annihilation process, followed by hadronisation of gluons into light mesons. For the expression of decay width of the respective decay channel, we consider the analysis done by N. Kivel and M. Vanderhaegen in \cite{Kivel:2017nrw} using the QCD factorization approach. The decay width is expressed as
		\begin{equation}
			\mathcal{B}(\chi_{c1}\rightarrow V\gamma)=\frac{1}{\Gamma_{\chi_{c1}}}\frac{1}{12\pi}\frac{\omega_{\chi_{c1}}^{5}}{M_{\chi_{c1}}^{4}}\left\lbrace|\mathcal{A}_{1V}^{\parallel}|^{2}+|\mathcal{A}_{1V}^{\perp}|^{2}\right\rbrace,
		\end{equation}
		where $\mathcal{A}_{1V}^{\parallel}$ is the decay amplitude into longitudinal light meson and has the form as
		\begin{equation}
			\mathcal{A}_{1V}^{\parallel}=-i \langle\mathcal{O}(\chi_{c1})\rangle\frac{f_{V}M_{\chi_{c1}}^{2}}{m_{c}^{6}}Q_{V}\sqrt{4\pi\alpha}\alpha_{s}^{2}\frac{N_{c}^{2}-1}{2N_{c}^{2}}\int_{0}^{1}dx \phi_{V}^{\parallel}(x)T_{1}(x),
			\label{eqn:amplitude parallel}
		\end{equation}
		with the hard kernel 
		\begin{equation}
			T_{1}(x)=\text{Re }T_{1}(x)+\text{Im }T_{1}(x),
		\end{equation}
		with the explicit expressions of $\text{Re }T_{1}(x)$ and $\text{Im }T_{1}(x)$ has been taken from \cite{Kivel:2017nrw} and the twist 2 distribution amplitude of the light vector meson $\phi_{V}^{\parallel}(x)$ has the form
		\begin{equation}
			\phi_{V}(x,\mu)=6x(1-x)\left\lbrace 1+a_{2}^{V}(\mu)C_{2}^{3/2}(2x-1)\right\rbrace,
		\end{equation}
		with the coefficient $a_{2}^{V}(\mu)$ defined at scale $\mu$. In addition, $A_{1V}^{\perp}$ representing the decay amplitude into transverse light meson is expressed as 
		\begin{equation}
			\begin{split}
				\mathcal{A}_{1V}^{\perp}=i \langle\mathcal{O}(\chi_{c1})\rangle& \frac{f_{V}m_{V}}{m_{c}^{5}}\frac{M_{\chi_{c1}}^{2}}{m_{c}^{2}}\sqrt{4\pi\alpha}\frac{\pi\alpha_{s}}{N_{c}}\int D\alpha_{i}\biggl\lbrace \delta_{I0}9e_{Q}\frac{G(\alpha_{i})}{\alpha_{1}\alpha_{2}\alpha_{3}^{2}}\\&-\frac{Q_{V}}{4}\left(\frac{\alpha_{1}-\alpha_{2}}{\alpha_{1}\alpha_{2}\alpha_{3}^{2}}V(\alpha_{i})+\frac{1-\alpha_{3}}{\alpha_{1}\alpha_{2}\alpha_{3}^{2}}A(\alpha_{i})\right)\biggr\rbrace,
			\end{split}
			\label{eqn:amplitude perpendicular}
		\end{equation}
		with
		\begin{equation}
			\int D\alpha_{i}f(\alpha_{i})=\int_{0}^{1}d\alpha_{1}\int_{0}^{1}d\alpha_{2}\int_{0}^{1}d\alpha_{3}\delta(1-\alpha_{1}-\alpha_{2}-\alpha_{3})f(\alpha_{1},\alpha_{2},\alpha_{3}),
		\end{equation}
		and $A(\alpha_{i})$, $V(\alpha_{i})$ and $G(\alpha_{i})$ are twist 3 distribution amplitudes and have the form as \cite{Kivel:2017nrw}
		\begin{equation}
			\begin{split}
				A(\alpha_{i})&=360\zeta_{3}\alpha_{1}\alpha_{2}\alpha_{3}^{2}\left(1+\omega_{3}^{A}\frac{1}{2}(7\alpha_{3}-3)\right),\\
				V(\alpha_{i})&=540\zeta_{3}\omega_{3}^{V}\alpha_{1}\alpha_{2}\alpha_{3}^{2}(\alpha_{2}-\alpha_{1}),\\
				G(\alpha_{i})&=5040\zeta_{3}\omega_{3}^{G}\alpha_{1}^{2}\alpha_{2}^{2}\alpha_{3}^{2},
			\end{split}
		\end{equation}
		with $\zeta_{3}$, $\omega_{3}^{A,V,G}$ being the non-perturbative parameters of the twist 3 DAs with the scale $\mu$ set at $\mu=2m_{c}$ and the evolution of the parameters from $\mu=1.0$ GeV to $2m_{c}$ has been taken from \cite{Kivel:2017nrw}. In eqns.\eqref{eqn:amplitude parallel} and \eqref{eqn:amplitude perpendicular} $f_{V}$ represent the decay constant of the light meson having the values
		\begin{equation}
			f_{\rho}=0.221\text{ }GeV, \qquad f_{\phi}=0.161\text{ }GeV,
		\end{equation}
		\begin{table}[t!]
			\centering
			\begin{tabular}{|c|c|}
				\hline
				\textbf{Decay mode}&\textbf{ Measured branching fractions}\\
				\hline
				$\chi_{c1}\rightarrow J/\psi \gamma$&34.3(1.3)\%\\
				$\chi_{c1}\rightarrow \rho^{0} \gamma$&2.16(17)$\times 10^{-4}$\\
				$\chi_{c1}\rightarrow \phi \gamma$&2.4(5)$\times 10^{-5}$\\
				\hline
			\end{tabular}
			\caption{Inputs used to extract $f_{\chi_{c1}}$.}
			\label{table:inputs chic1 decay constant}
		\end{table}
		and other parameters include $e_{Q}=2/3$, the charge of the heavy quark, $Q_{V}$ represents an appropriate combination of the quark charges
		\begin{equation}
			Q_{\rho^{0}}=\frac{1}{2}(e_{u}-e_{d})=\frac{1}{2},\qquad Q_{\phi}=e_{s}=-\frac{1}{3},
		\end{equation}
		$m_{V}$ represents the mass of the light vector mesons, having values
		\begin{equation}
			m_{\rho}=0.775\text{ }GeV, \qquad m_{\phi}=1.019\text{ }GeV,
		\end{equation}
		and the NRQCD matrix element $\langle\mathcal{O}(\chi_{c1})\rangle$ is related to $R^{\prime} (0)$ as
		\begin{equation}
			\langle\mathcal{O}(\chi_{c1})\rangle=\sqrt{2N_{c}}\sqrt{2M_{\chi_{c1}}}\sqrt{\frac{3}{4\pi}}R_{\chi_{c1}}^{\prime}(0),
		\end{equation} 
		where we again substitute $R^{\prime}(0)$ using eqn.\eqref{eqn:decay constant of ChiC1} to introduce $f_{\chi_{c1}}$ into the expression.
	\end{itemize}
	With the above expressions for the branching ratios and corresponding inputs from Table \ref{table:inputs chic1 decay constant} we construct a chi-square function. As for the nuisance parameters, we take the prior estimates of $f_{J/\psi}$, $m_{c}$ and $\Lambda$ from Table \ref{table:LCDA1}, that of Coulombic and Non-Coulombic corrections terms from \cite{Chung:2021efj} with a 10\% error as
	\begin{equation}
		\delta_{C}=0.266(27), \qquad \delta_{NC}=0.493(49),
	\end{equation}
	and that of the coefficints of twist-2 and twist-3 DAs at $\mu=1.0$ GeV as \cite{Kivel:2017nrw}
	\begin{align*}
		&a_{2\rho}=0.15(7), &a_{2\phi}=0.18(8),\\
		&\zeta_{\rho}=0.03(1),  &\zeta_{\phi}=0.024(8),\\
		&\omega_{\rho}^{A}=-3.0(1.4), &\omega_{\phi}^{A}=-2.6(1.3),\\
		&\omega_{\rho}^{V}=5.0(2.4), &\omega_{\phi}^{V}=5.3(3.0).
	\end{align*}
	With these inputs, once the chi-square function is constructed, we optimize it to finally extract $f_{\chi_{c1}}$ which we present in Table \ref{table:results for chic1 decay constant}, along-with the corresponding correlation matrix between the extracted parameters in Table \ref{table:correlation chic1 decay constant}.

	\begin{table}[htb!]
		\centering
		\begin{tabular}{|cc|cc|}
			\hline
			\multicolumn{2}{|c|}{\textbf{Free Parameters}}  &  \multicolumn{2}{c|}{\textbf{Nuisance Parameters}} \\
			\hline
			\textbf{Parameters}& \textbf{Fit Results} & \textbf{Parameters}& \textbf{Fit Results} \\
			\hline
			$f_{\chi_{c1}}$&0.169(28) GeV&$f_{J/\psi}$&0.405(3) GeV\\
			$\delta^{\prime}$&-0.624(22)&$m_{c}$&1.342(31) GeV\\
			& & $\Lambda$ & 0.2789(8) GeV\\
			& & $\Gamma_{\chi_{c1}}$ & 0.840(28) GeV\\
			& & $\delta_{C}$ & 0.266(18)\\
			& & $\delta_{NC}$ & 0.493(35)\\
			& & $a_{2\rho}$ & 0.154(49)\\
			& & $a_{2\phi}$ & 0.172(56)\\
			& & $\zeta_{\rho}$ & 0.033(6)\\
			& & $\zeta_{\phi}$ & 0.023(6)\\
			& & $\omega_{\rho}^{A}$ & -3.85(89)\\
			& & $\omega_{\rho}^{V}$ & 2.11(88)\\
			& & $\omega_{\phi}^{A}$ & -2.60(99)\\
			& & $\omega_{\phi}^{V}$ & 5.30(2.12)\\
			\hline
			\textbf{D.O.F}&\multicolumn{3}{|c|}{1}\\
			\hline
			\textbf{p-Value}&\multicolumn{3}{|c|}{15.36\%}\\
			\hline
		\end{tabular}
		\caption{Extracted value of decay constant of $\chi_{c1}$ meson, along-with estimates of other relevant parameters. We extract $a_{2\rho}$, $a_{2\phi}$, $\zeta_{\rho}$, $\zeta_{\phi}$, $\omega_{\rho}^{A}$, $\omega_{\phi}^{A}$, $\omega_{\rho}^{V}$ and $\omega_{\phi}^{V}$ at $\mu=1.0~GeV$.}
		\label{table:results for chic1 decay constant}
	\end{table}
	
	\item \textbf{\underline{Extracting $\boldsymbol{f_{h_{c}}}$}:} Extracting $f_{h_{c}}$ is comparatively straightforward due to the unavailability of enough decay channels to be considered as inputs. We take a single decay channel, namely $h_{c}\rightarrow \eta_{c}\gamma$ radiative channel for this purpose. Being an E1 transition, the branching ratio for this channel can also be calculated in pNRQCD framework, the expression being the same as we had previously shown for $\chi_{c1}\rightarrow J/\psi \gamma$. The expression for the branching ratio is as

	\begin{equation}
		\mathcal{B}(h_{c}\rightarrow \eta_{c}\gamma)=\frac{1}{\Gamma_{h_{c}}}\cdot\frac{4}{9}e_{c}^{2}\alpha\omega_{h_{c}}^{3}|I_{SP}|^{2}\left(1+\delta^{\prime}\right),
		\label{eqn:branching ratio of hc etac gamma}
	\end{equation}
	with $\Gamma_{h_{c}}=0.78(28)\text{ }MeV$ being the total width of $h_{c}$ meson, $e_{c}$, $\alpha$ and $\omega_{h_{c}}$ having the same meaning as has been discussed before. $I_{SP}$ represents the overlap integral between the initial $h_{c}$ state and the final $\eta_{c}$ state wavefunctions. The general expression for the overlap integral, and its expression in terms of the decay constant, is the same as we had calculated for $\chi_{c1}\rightarrow J/\psi \gamma$. $f_{h_{c}}$ can be incorporated into the expression for the overlap integral same as before, utilising eqn. \eqref{eqn:decay constant of ChiC1} with $\chi_{c1}$ being replaced by $h_{c}$ in this case and the estimates of Coulombic and Non-Coulombic correction terms taken from Table \ref{table:results for chic1 decay constant}. As for $\eta_{c}$ meson, the decay constant can be connected to $R_{\eta_{c}}(0)$, the radial wave function at origin as
	\begin{equation}
		\begin{split}
			f_{\eta_{c}}=\sqrt{\frac{3}{2m_{c}\pi}}R_{\eta_{c}}(0)\biggl[&1+\frac{\alpha_{s}(\mu)}{\pi}\frac{\pi^{2}-20}{3}+\langle v^{2}\rangle_{\eta_{c}}\biggl\lbrace-\frac{4}{3}+\frac{\alpha_{s}(\mu)}{\pi}\frac{1}{27}(48\ln(\frac{\mu^{2}}{m_{c}^{2}})-96\ln 2\\&-15\pi^{2}+196)\biggr\rbrace+\frac{68}{45}\langle v^{4}\rangle_{\eta_{c}}\biggr]^{1/2},
		\end{split}
	\end{equation} 
	by expressing the decay width of $\eta_{c}\rightarrow\gamma\gamma$ radiative channel derived in \cite{Bodwin:2002cfe,Guo:2011tz} as proportional to the square of $f_{\eta_{c}}$. Following the same procedure as previously shown in eqns. \eqref{eqn:radial solution of charmonia} and \eqref{eqn:Bohr momentum Jpsi}, we arrive at $q_{B_{\eta_{c}}}$ in terms of $f_{\eta_{c}}$ as
	\begin{equation}
		\begin{split}
			q_{B_{\eta_{c}}}=\biggl[\frac{4m_{c}\pi}{3}f_{\eta_{c}}^{2}\biggl[&1+\frac{\alpha_{s}(\mu)}{\pi}\frac{\pi^{2}-20}{3}+\langle v^{2}\rangle_{\eta_{c}}\biggl\lbrace-\frac{4}{3}+\frac{\alpha_{s}(\mu)}{\pi}\frac{1}{27}(48\ln(\frac{\mu^{2}}{m_{c}^{2}})-96\ln 2\\&-15\pi^{2}+196)\biggr\rbrace+\frac{68}{45}\langle v^{4}\rangle_{\eta_{c}}\biggr]^{-1}\biggr]^{1/3},
		\end{split}
	\end{equation}
	with $\alpha_{s}$ being calculated at $\mu=2m_{c}$, $\Lambda$ taken from Table \ref{table:results for chic1 decay constant} and $\langle v^{2}\rangle_{\eta_{c}}= 0.267$. The correction term $\delta^{\prime}$ has the same meaning as for the former channel and its value has been taken from Table \ref{table:results for chic1 decay constant} as inputs into eqn.\eqref{eqn:branching ratio of hc etac gamma}. As for the value of $f_{\eta_{c}}$ we take the estimate previously extracted in Table \ref{table:LCDA1} as input. Finally we extract $f_{h_{c}}$ by solving eqn. \eqref{eqn:branching ratio of hc etac gamma} with experimentally observed value of the branching ratio, i.e., $\mathcal{B}(h_{c}\rightarrow \eta_{c})_{exp}=60(4)\%$ \cite{ParticleDataGroup:2022pth} and arrive at
	\begin{equation}
		f_{h_{c}}=0.182(47)~GeV.
	\end{equation}

\end{enumerate}

\subsection{Prediction of $B_{c}\rightarrow \text{ }P$ wave semileptonic form factors}
\label{subsection:P wave form factors}

 Following the extraction of $f_{\chi_{c0}}$, $f_{\chi_{c1}}$ and $f_{h_{c}}$, we are now in a position to calculate the $B_{c}\rightarrow P$ wave semileptonic form factors. First we calculate the relevant form factors at $q^{2}=0$ through the modified pQCD approach. Once we have the form factors at $q^{2}=0$, we then extrapolate them to $q_{max}^{2}$ through the extrapolation technique discussed in the subsequent text in this subsection. After introducing the extrapolation technique, we then actually extract the relevant extrapolation parameters, essentially using the inputs on $B_{c}\rightarrow \text{ }S$ wave form factor, motivation for which has been discussed in the text, and then propagate the thus extracted parameters to predict the $B_{c}\rightarrow \text{ }P$ wave form factors.
 
	\paragraph{\underline{Form factors at $\boldsymbol{q^{2}=0}$:}}
	
	With the estimates of the decay constants extracted in the previous subsection and that of the shape parameter of $B_{c}$ meson, $\omega_{B_{c}}$, and other relevant parameters from Table \ref{table:LCDA1}, we take them as inputs in the expressions for the form factors in modified pQCD shown in Appendix \ref{section:Appendix pQCD form factors} and calculate the numerical values of the relevant form factors at $q^{2}=0$. Similar to what we did while calculating the form factors in Table \ref{table:LCDA 2}, here too we consider a cut-off in the upper limit of impact parameter used in the pQCD expressions of form factors, $b_{c}$  at 90\% of $1/\Lambda_{QCD}$. The predicted values of the form factors at thus calculated are presented in Table \ref{table:LCDA 3}.
	
	
	\begin{table}[t]
		\centering
		\setlength{\tabcolsep}{3pt}
		\begin{tabular}{|c|c|ccccc|}
			\hline
			\textbf{Decay}&\textbf{Form}&\textbf{This}&\textbf{Previous}&\textbf{QCDSR}&\textbf{LFQM}&\textbf{NRQCD}\\
			\textbf{Channel}&\textbf{Factor}&\textbf{work}&\textbf{pQCD\textcolor{blue}{ [1]}}&\textbf{\textcolor{blue}{[2]}}&\textbf{\textcolor{blue}{[3]}}&\textbf{\textcolor{blue}{[4]}}\\
			\hline
			$B_{c}\rightarrow \chi_{c0}$&$F_{+}(0)$&0.431(70)&0.41$^{+0.09}_{-0.08}$&0.673(195)&$0.47^{+0.03}_{-0.06}$&$1.25^{+0.15}_{-0.12}$\\
			\hline
			&$A_{0}(0)$&0.166(26)&0.18$^{+0.03}_{-0.03}$&0.03(1)&$0.13^{+0.01}_{-0.01}$&$0.12^{+0.01}_{-0.01}$\\
			&$A_{1}(0)$&0.583(119)&0.86$^{+0.17}_{-0.16}$&0.30(9)&$0.85^{+0.02}_{-0.04}$&$2.34^{+0.21}_{-0.22}$\\
			$B_{c}\rightarrow \chi_{c1}$&$A_{2}(0)$&0.054(16)&0.11$^{+0.02}_{-0.01}$&0.06(2)&$0.15^{+0.01}_{-0.01}$&$0.47^{+0.07}_{-0.06}$\\
			&$V(0)$&0.290(48)&0.18$^{+0.04}_{-0.04}$&0.13(4)&$0.36^{+0.02}_{-0.04}$&$0.99^{+0.19}_{-0.15}$\\
			\hline
			&$A_{0}(0)$&0.322(61)&0.22$^{+0.05}_{-0.03}$&0.03(1)&$0.64^{+0.10}_{-0.02}$&$1.63^{+0.25}_{-0.19}$\\
			&$A_{1}(0)$&0.761(188)&0.46$^{+0.07}_{-0.07}$&0.30(9)&$0.50^{+0.05}_{-0.08}$&$0.46^{+0.07}_{-0.03}$\\
			$B_{c}\rightarrow h_{c}$&$A_{2}(0)$&-0.023(10)&-0.03$^{+0.0}_{-0.01}$&0.06(2)&$-0.32^{+0.06}_{-0.05}$&$-0.75^{+0.17}_{-0.17}$\\
			&$V(0)$&0.323(77)&0.10$^{+0.02}_{-0.01}$&0.13(4)&$0.07^{+0.00}_{-0.01}$&$0.07^{+0.00}_{-0.01}$\\
			\hline
		\end{tabular}
		\caption{Prediction of form factors of $B_{c}\rightarrow$ P wave transition at $q^{2}=0$.}
		\label{table:LCDA 3}
	\end{table}
	
	From Table \ref{table:LCDA 3} we see that the error estimates of the form factors range from a minimum of 15.66\% for $A_{0}(0)$ of $B_{c}\rightarrow \chi_{c1}$ channel to a maximum of 43.47\% for $A_{2}(0)$ of $B_{c}\rightarrow h_{c}$ channel. The reason for the error estimate being comparatively larger than $B_{c}\rightarrow J/\psi(\eta_{c})$ channels is primarily due to the large error estimate of the decay constants, being about 13.60\%, 16.56\% and 25.82\% for $f_{\chi_{c0}}$, $f_{\chi_{c1}}$ and $f_{h_{c}}$ respectively. Availability of information on branching ratios of more radiative decay channels of P wave charmonia or other inputs, such as moments of charmonium LCDAs, in future might help us better constrain the decay constants, thereby enabling us to predict the form factors to a greater degree of precision.

	\paragraph{\underline{Extrapolation of form factors over full semileptonic region:}}

\indent Having extracted the form factor information at $q^{2}=0$, the next course of action is to extrapolate the form factors to full physical $q^{2}$ region. The pCQD framework in itself is more reliable at the lower $q^{2}$ region, specifically in the range of about $0.0$ to $0.2(m_{B_{c}}-m)^{2}$ \cite{Rui:2016opu}. But as we saw in subsection \ref{subsection:Physical observables}, prediction of the physical observables require information of the form factors over the full physical $q^{2}$ region. Thus in order to make our predictions of form factors reliable over the full $q^{2}$ region, we need to extrapolate them to the high $q^{2}$ region. There are a number of extrapolation techniques available in literature that has been used previously. We use the two parameter pole expansion parametrization considered in \cite{Melikhov:2000yu} for our work, which has the form as
\begin{equation}
	\label{eqn:pole1}
	f_{i}(q^{2})=\frac{f_{i}(0)}{P_{i}(q^{2})\left(1-\alpha_{i} \frac{q^{2}}{M_{i}^{2}}+\beta\frac{q^{4}}{M_{i}^{4}}\right)},
\end{equation}
where $\alpha_{i}$ and $\beta$ represent the parameters that we intend to extract and $M_{i}$ represents the masses of the low-lying $B_{c}$ resonances.
Their values have been taken from \cite{Leljak:2019eyw} and has been shown in Table \ref{table:pole masses}. The form factors $f_{i}(q^{2})$ have an additional weight factor $P_{i}(q^{2})$ having the form 
\begin{equation}
	P_{i}(q^{2})=1-\frac{q^{2}}{M_{i}^{2}},
\end{equation}
which accounts for the contributions due to low-lying resonances present below the threshold production of $B_{c}-V$ pairs at $q^{2}=(m_{B_{c}}+m_{V})^{2}$, V representing the final charmonium state. The inclusion of $P(q^{2})$ into the parametrization equations for all the form factors is justified as the resonance points for each of the masses presented in Table \ref{table:pole masses} lies well below the pair production threshold.

Note that the slope of the shapes of the form factors will be highly dependent on the respective masses of the low-lying resonances. The parameter $\alpha_{i}$, being the leading order coefficient in $q^{2}$ is taken to be different for each form factor. In contrast, for simplicity, the parameter $\beta$ being subleading in $q^{2}$ and only exhibiting control over the form factors at the very end of the semileptonic region is considered the same for all the form factors. 

\begin{table}[t!]
	\centering
	\begin{tabular}{|c|cccccc|}
		\hline
		Form Factor& $F_{+}$& $F_{0}$& $A_{0}$ & $A_{1}$ & $A_{2}$ & V\\
		\hline
		$M_{i}$ in GeV&6.34&6.71&6.28&6.75&6.75&6.34\\
		\hline
	\end{tabular}	
	\captionof{table}{Masses of the low lying $B_{c}$ resonances.}
	\label{table:pole masses}
\end{table}


 The primary objective of this section is to extract the pole expansion parameters introduced in \cref{eqn:pole1} utilising the $B_{c}\rightarrow J/\psi$ and $B_{c}\rightarrow \eta_{c}$ form factors as inputs, and then use the same extracted parameters to obtain information on $B_{c}\rightarrow \text{ }P$ wave form factors. 
 To do this, a chi-square function is constructed with the form factor inputs at certain discreet $q^{2}$ points shown in Table \ref{table:Appendix Pole Data 1} in Appendix \ref{section:Appendix form factor data} and then minimized to extract the required pole expansion parameters. The parameters thus extracted are shown in Table \ref{table:pole2}, and the correlation between the thus extracted parameters has been shown in Table \ref{table: correlation pole}.

\begin{table}[t]
	
	\centering
	
	\begin{tabular}{|cc|}
		\hline
		\textbf{Parameters}& \textbf{Fit Results}  \\
		\hline
		$\alpha_{A_{0}}$ & 1.600(138) \\
		
		$\alpha_{A_{1}}$ & 1.553(121) \\
		
		$\alpha_{A_{2}}$& 1.589(255)  \\
		
		$\alpha_{V}$& 1.647(124)  \\
		
		$\alpha_{F_{0}}$& 0.916(287)\\
		
		$\alpha_{F_{+}}$& 1.465(197)\\
		
		$\beta$& 1.063(218)\\
		
		\hline
		\textbf{D.O.F}&\multicolumn{1}{|c|}{9}\\
		\hline
		\textbf{p-Value}&\multicolumn{1}{|c|}{95.14\%}\\
		\hline
		
	\end{tabular} 
	\captionof{table}{Results of extraction of Pole expansion parameters of $B_{c}\rightarrow J/\psi(\eta_{c})$ form factors.}
	
	\label{table:pole2}
\end{table}

The $q^{2}$ distribution of the form factors can be easily determined with the pole expansion parameters extracted. To check the predictivity of our fit, we have reproduced the $q^2$ distributions of the $B_c\to \eta_c$ and $B_c\to J/\psi$ form factors, which we have shown in the plots of Fig. \ref{fig:Pole Bc_JPsi} (appendix), respectively. The predicted $q^2$ shapes comfortably explain all the inputs used in the fits.

\begin{figure*}[t]
	
	\centering
	\subfloat[\centering]{\includegraphics[width=7.1cm]{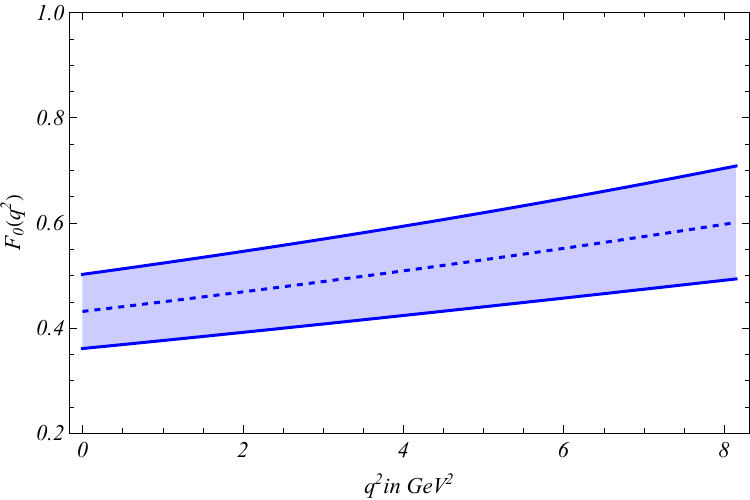}}
	\qquad
	\subfloat[\centering]{\includegraphics[width=7.1cm]{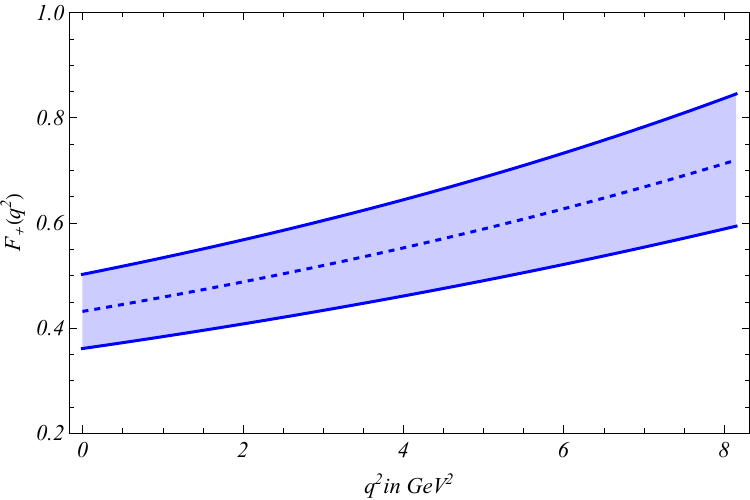}}
	\caption{Plots showing the $q^{2}$ dependence of $B_{c}\rightarrow \chi_{c0}$ semileptonic form factors.}
	\label{fig:Pole Bc_chic0}
\end{figure*}

\begin{figure*}[htb!]
	
	\centering
	\subfloat[\centering]{\includegraphics[width=7.1cm]{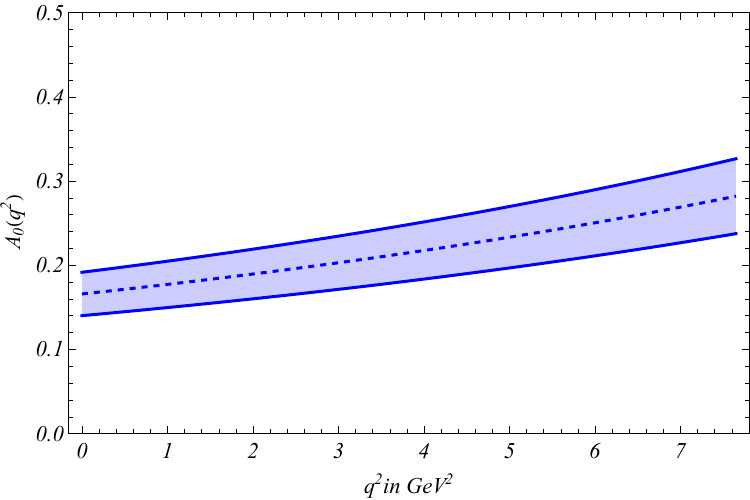}}
	\qquad
	\subfloat[\centering]{\includegraphics[width=7.1cm]{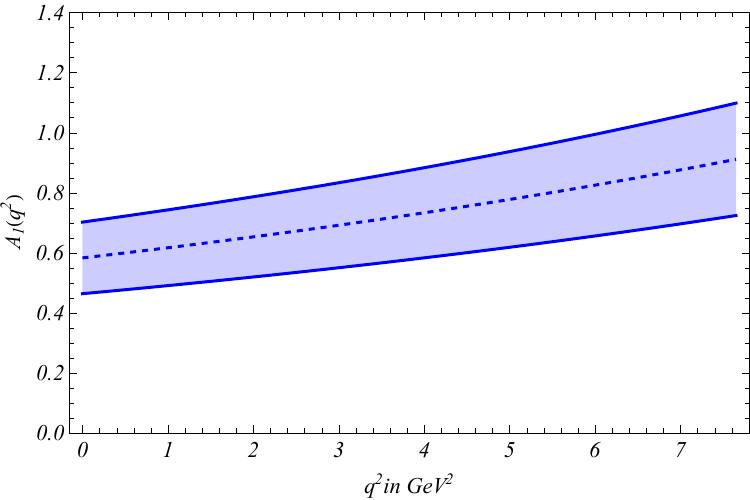}}
	\qquad
	\subfloat[\centering]{\includegraphics[width=7.1cm]{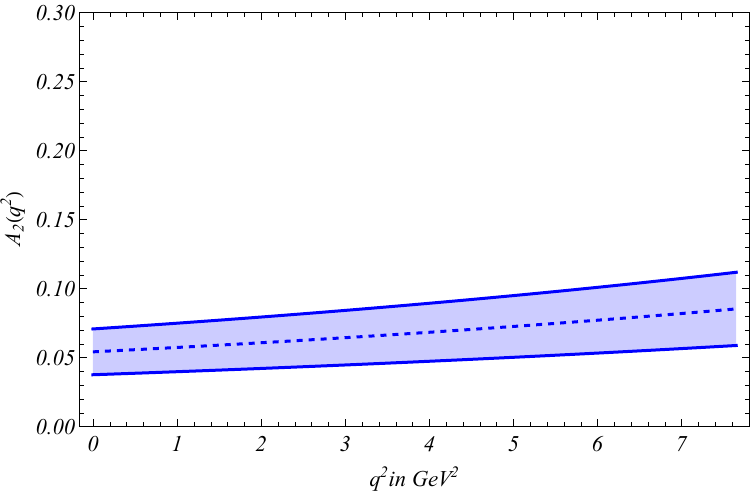}}
	\qquad
	\subfloat[\centering]{\includegraphics[width=7.1cm]{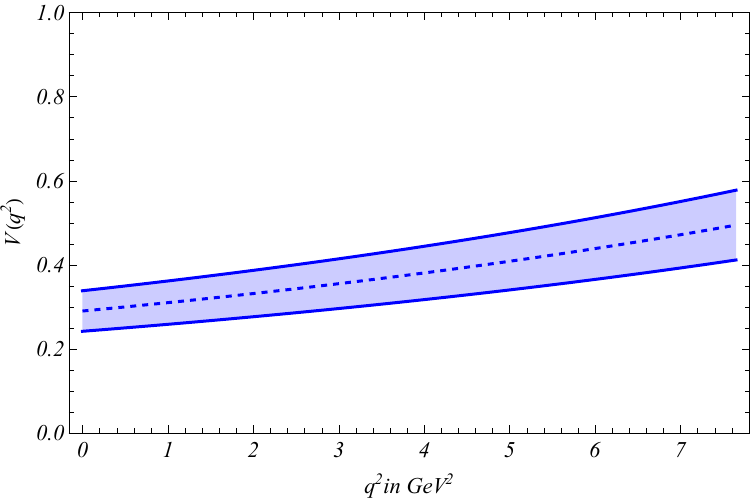}}
	\caption{Plots showing the $q^{2}$ dependence of $B_{c}\rightarrow \chi_{c1}$ semileptonic form factors.}
	\label{fig:Pole Bc_chic1}
\end{figure*}

\begin{figure*}[htb!]
	
	\centering
	\subfloat[\centering]{\includegraphics[width=7.1cm]{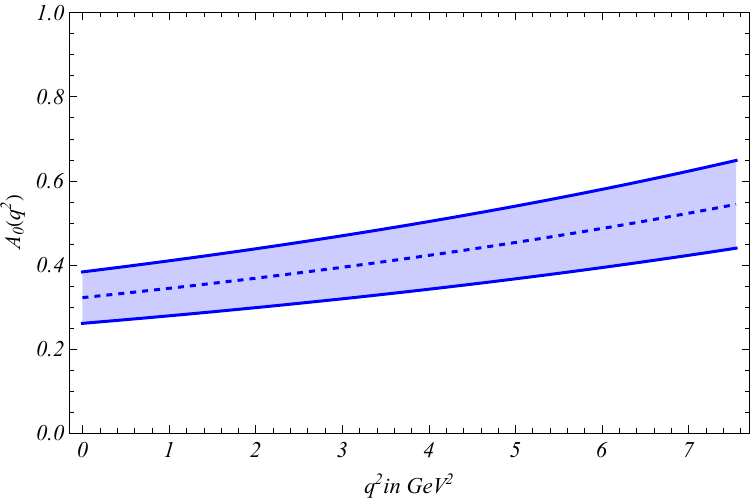}}
	\qquad
	\subfloat[\centering]{\includegraphics[width=7.1cm]{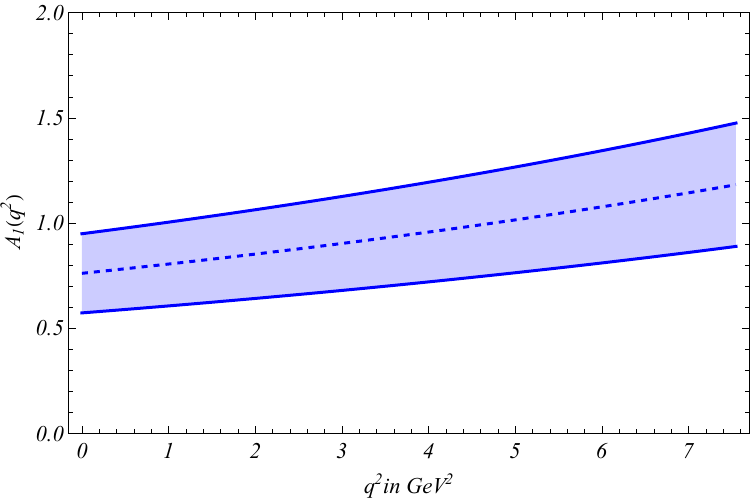}}
	\qquad
	\subfloat[\centering]{\includegraphics[width=7.1cm]{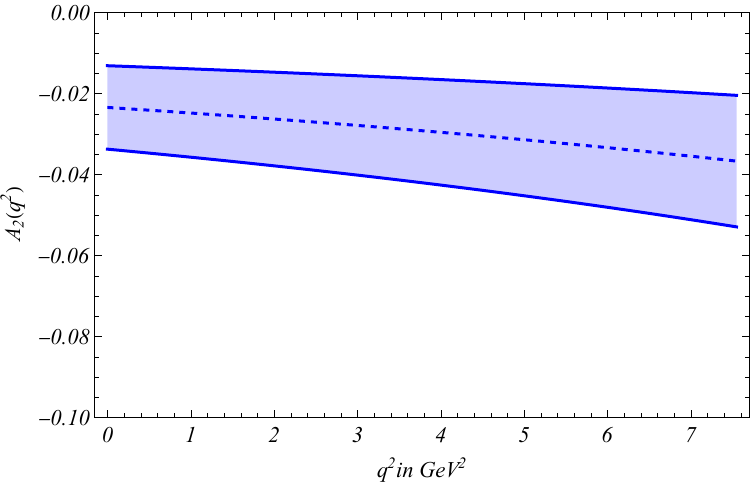}}
	\qquad
	\subfloat[\centering]{\includegraphics[width=7.1cm]{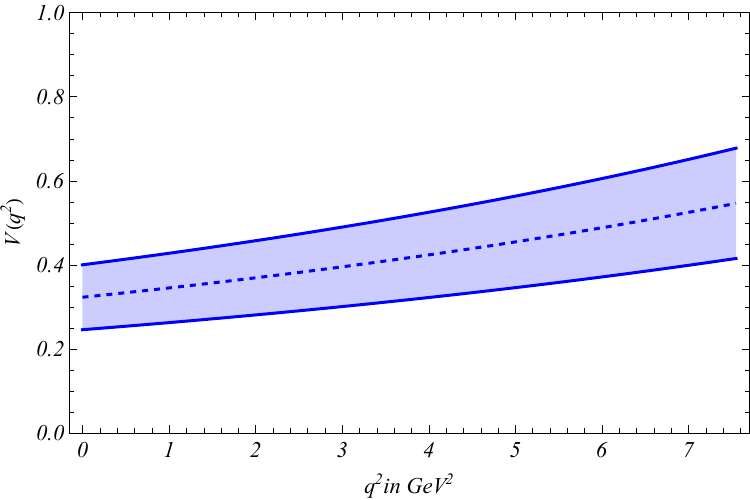}}
	\caption{Plots showing the $q^{2}$ dependence of $B_{c}\rightarrow h_{c}$ semileptonic form factors.}
	\label{fig:Pole Bc_hc}
\end{figure*}

As for the $q^{2}$ distribution of $B_{c}\rightarrow \text{ }P$ wave semileptonic form factors, we are going to utilise the same pole expansion parametrization already discussed in \cref{eqn:pole1}. We have mentioned earlier that the slope of the $q^2$ shapes of the form factors is highly dependent on the low-lying resonances. Therefore, as an approximate approach, we have utilised the connection between the total angular momentum of the final meson states, enabling us to connect the slope of the form factors of $P$ wave scalar state, $\chi_{c0}$ with $S$ wave pseudoscalar state, $\eta_{c}$, both having total angular momentum $J=0$, and also the form factors of $P$ wave axial-vector meson states, $\chi_{c1}$ and $h_{c}$ with S wave vector meson state, $J/\psi$, both having total angular momentum $J=1$ \cite{Hernandez:2006gt,Ivanov:2005fd}\footnote{Following this approach, we have extracted the $q^2$ shapes of $B_c\to \chi_{c0}$, $B_c\to \chi_{c1}$ and $B_c\to h_c$. The mesons $J/\psi$, $\chi_{c1}$ and $h_{c}$ have total quantum number $J = 1$ and $\eta_c$ and $\chi_{c0}$ have $J=0$. Our approach may be helpful to get the $q^2$ shapes of $B_c\to \chi_{c0,1},h_c$ form factors using the available information on the shapes of the $B_c\to J/\psi,\eta_c$ form factors. A similar approach may not be directly applicable to obtain the $B_{c}\rightarrow \chi_{c2}$ form factors since $\chi_{c2}$ is a tensor meson with $J=2$. The relevant $B_c\to \chi_2$ form factors are $h(q^2)$, $k(q^2)$, $b_+(q^2)$ and $b_-(q^2)$, respectively, for a detail see the ref. \cite{Wang:2009mi}. We have obtained the DA for $\phi_{B_c}$ and we can use the available data on $\Gamma(\chi_{c2}\to \gamma\gamma)$ to obtain the shape of $\chi_{c2}$ wave function. In principle, we can use both this information to obtain the PQCD prediction for $B_c\to \chi_{c2}$ form factors at $q^2=0$. However, for the predictions of the semileptonic or non-leptonic rates, we need the respective $q^2$ shapes of the form factors or values at $q^2\ne 0$. Currently, we do not have any additional inputs that help us get the $q^2$ shapes of the relevant $B_c \to \chi_{c2}$ form factors. This is why, in this study, we have refrained from including channels involving $\chi_{c2}$ meson.}. For simplicity, the parameter $\beta$ stays the same for all the form factors, while we take $\alpha_{F_{+}}$ and $\alpha_{F_{0}}$ given in table \ref{table:pole2} as inputs to extrapolate the $B_{c}\rightarrow \chi_{c0}$ form factors. On the other hand, we have used $\alpha_{A_{0,1,2}}$ and $\alpha_{V}$ given in Table \ref{table:pole2} as inputs to extrapolate the $B_{c}\rightarrow \chi_{c1}(h_{c})$ form factors. The $q^{2}$ distribution of the form factors thus obtained through extrapolation has been shown in Figs. \ref{fig:Pole Bc_chic0}, \ref{fig:Pole Bc_chic1} and \ref{fig:Pole Bc_hc} which can be tested once we have results from lattice. 


\subsection{Prediction of some physical observables}
\label{subsection: P wave observables}

With information of form factors over the entire physical $q^{2}$ obtained, we are now in a position to perform predictions on some of the relevant physical observables. These include the branching ratios of some of the semileptonic transitions involving the emission of both light as well as heavy lepton, ratios of the respective branching ratios, and an angular observables, the forward backward asymmetry. The explicit expressions for the $q^{2}$ distribution of these observables have already been shown in subsection \ref{subsection:Physical observables}.
\begin{itemize}
	\item In Fig \ref{fig:differential decay width} we showcase $q^{2}$ distribution of $B_{c}\rightarrow\chi_{c0},\chi_{c1}$ and $h_{c}$ semileptonic decay widths. In Table \ref{table:branching3} we present our predictions of the branching ratios, obtained by integrating the differential decay width over the physical $q^{2}$ region, along-with comparison with predictions from other approaches. In the second column the actual error estimates obtained in \cite{PhysRevD.98.033007} are added up in quadrature and shown here.


	\begin{table*}[t]
		
		\centering
		\begin{tabular}{|c|cccc|}
			
			\hline
			\textbf{Decay Channels}&\textbf{This work}&\textbf{Previous pQCD\textcolor{blue}{[1]}}&\textbf{QCDSR\textcolor{blue}{[2]}}&\textbf{LFQM\textcolor{blue}{[3]}}\\
			\hline
			
			$\mathcal{B}(B_{c}\rightarrow \chi_{c0}l\nu_{l})$&2.00(65)&2.22$^{+1.18}_{-0.83}$&1.82(51)&$2.1^{+0.2}_{-0.4}$\\
			$\mathcal{B}(B_{c}\rightarrow \chi_{c0}\tau\nu_{\tau})$&0.339(114)&0.48$^{+0.25}_{-0.18}$&0.49(16)&$0.24^{+0.01}_{-0.03}$\\
			$\mathcal{B}(B_{c}\rightarrow \chi_{c1}l\nu_{l})$&1.39(51)&1.53$^{+0.69}_{-0.54}$&1.46(42)&$1.4^{+0.0}_{-0.1}$\\
			$\mathcal{B}(B_{c}\rightarrow \chi_{c1}\tau\nu_{\tau})$&0.170(65)&0.20$^{+0.09}_{-0.06}$&0.147(44)&$0.15^{+0.01}_{-0.02}$\\
			$\mathcal{B}(B_{c}\rightarrow h_{c}l\nu_{l})$&2.61(1.16)&1.06$^{+0.31}_{-0.32}$&1.42(40)&$3.1^{+0.5}_{-0.8}$\\
			$\mathcal{B}(B_{c}\rightarrow h_{c}\tau\nu_{\tau})$&0.259(115)&0.13$^{+0.03}_{-0.04}$&0.137(38)&$0.22^{+0.02}_{-0.04}$\\
			\hline
		\end{tabular}
		\caption{Branching ratios ($\times 10^{-3}$) of some $B_{c}\rightarrow P$ wave semileptonic channels predicted in this work along-with comparison with other predictions in existing literature.}
		\label{table:branching3}
	\end{table*} 	

	Checking Table \ref{table:branching3} we can see that our predictions have attained values that agree well to the existing predictions within the error bars. However, a comparison between our predictions to the previous pQCD predictions \cite{PhysRevD.98.033007} shows an improvement of 49.9\%, 47.9\%, 35.5\% and 30.5\% for the first four rows, respectively. But there is an increase in the error estimate in the last two rows. Additionally, the error estimates for the last two channels are higher compared to those of the other channels. The reason can be traced back to Table \ref{table:LCDA 3} where the error estimates of our $B_{c}\rightarrow h_{c}$ form factors are larger than the corresponding previous pQCD predictions and also to our $B_{c}\rightarrow \chi_{c1}$ form factor predictions. In all the predictions, the electron and muon modes have not been differentiated due to both of them having small masses and not coming up with any significant difference in the values of the observables. However, comparing the light lepton modes to the heavy tau lepton mode, we can see a significant drop in the values for the latter. This is mainly due to a suppression coming from the phase space for the heavy lepton channels. Next, we calculate ratios of the branching fraction $R(X)$, where $X$ is the final state charmonium. The general expression for $R(X)$ is as
	\begin{equation}
		R(X)=\frac{\mathcal{B}(B_{c}\rightarrow X\tau\nu_{\tau})}{\mathcal{B}(B_{c}\rightarrow X l\nu_{l})}.
	\end{equation}	
These ratios are much cleaner observables compared to the branching ratios due to the reduction of theoretical uncertainties coming from the form factors. Our predictions along with comparison with predictions from the previous pQCD approach, are shown in Table \ref{table:branching4}.
	\begin{table}[htb!]
		\centering
		\begin{tabular}{|c|cc|}
			\hline
			\textbf{Ratios}&\textbf{This work}&\textbf{Previous pQCD\textcolor{blue}{[1]}}\\
			\hline
			$R(\chi_{c0})$&0.169(11)&0.22$^{+0.0}_{-0.01}$\\
			$R(\chi_{c1})$&0.126(2)&0.13$^{+0.01}_{-0.0}$\\
			$R(h_{c})$&0.113(3)&0.12$^{+0.01}_{-0.0}$\\
			\hline
		\end{tabular}
		\caption{Predictions for $R(\chi_{c0,1})$ and $R(h_{c})$ and comparison with existing predictions.}
		\label{table:branching4}
	\end{table}
	
We can see a significant reduction in the error estimate in all three ratios compared to the previous pQCD predictions. The central values typically lie between 0.113 and 0.169, which is significantly smaller than SM predictions of $R(\eta_{c})$ and $R(J/\psi)$. It would be interesting however to see if future measurements could indicate towards any possible anomaly in its values, just like $R(J/\psi)$. If so, it could in future hint towards possible NP effects, opening up a new arena to explore.
	
	\item  Along-with the predictions of branching fractions, we also present predictions of tau lepton polarization $\langle P_{\tau}\rangle$, $\chi_{c1}(h_{c})$ longitudinal polarization fraction $\langle F_{L}\rangle$ and forward backward asymmetry $\langle A_{FB}(l(\tau))\rangle$, explicit expressions for which has already been discussed in section \ref{subsection:Physical observables}, in Table \ref{table:PredictionsPtauFLAFBPwave} for both light lepton and heavy lepton cases.
	
		\begin{figure*}[htb!]
		
		\centering
		\subfloat[\centering]{\includegraphics[width=7.1cm]{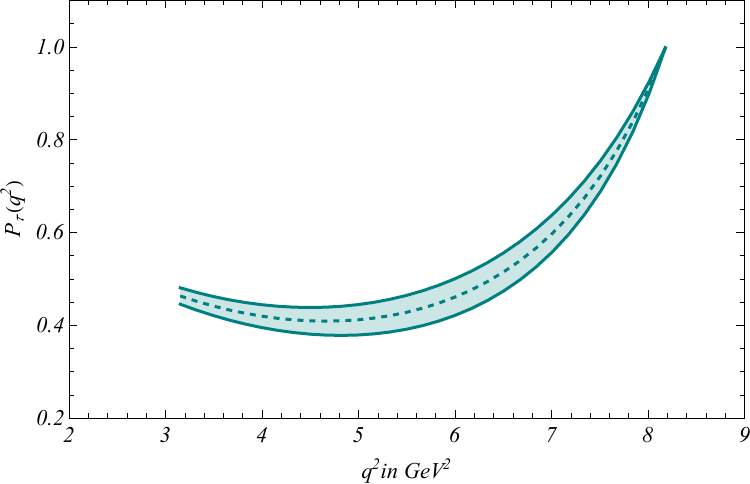}}
		\qquad
		\subfloat[\centering]{\includegraphics[width=7.1cm]{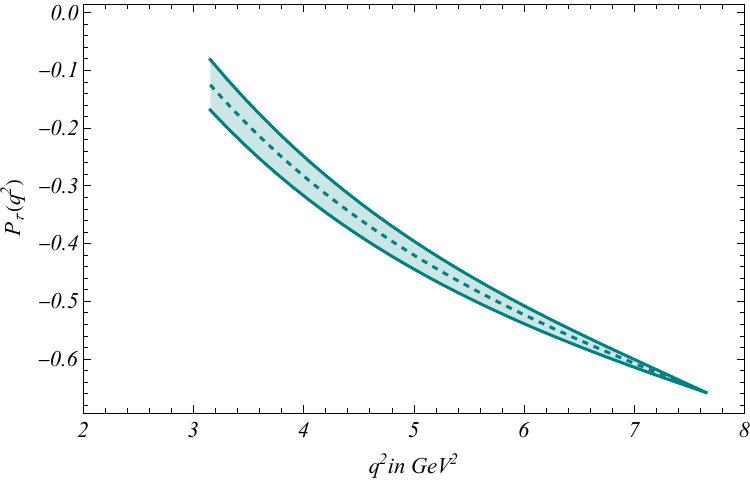}}
		\qquad
		\subfloat[\centering]{\includegraphics[width=7.1cm]{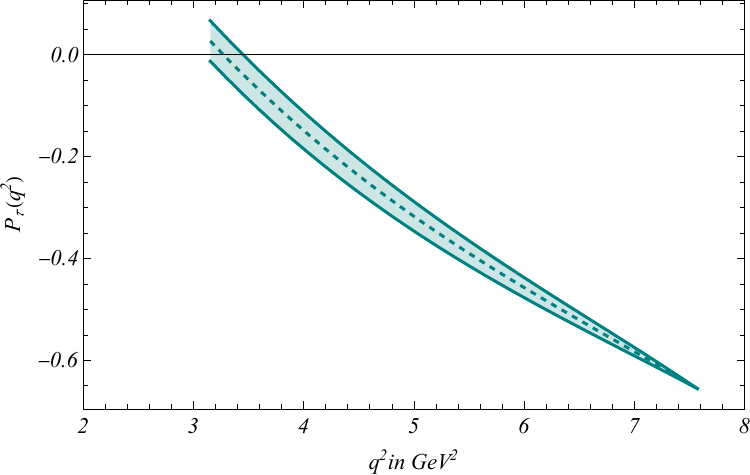}}
		\caption{$q^{2}$ distribution of $P_{\tau}(q^{2})$ for (a)$B_{c}\rightarrow \chi_{c0}$, (b) $B_{c}\rightarrow \chi_{c1}$ and (c) $B_{c}\rightarrow h_{c}$ semitauonic channels.}
		\label{fig:tau polarization}
	\end{figure*}

	\begin{figure*}[htb!]
	
	\centering
	\subfloat[\centering]{\includegraphics[width=7.1cm]{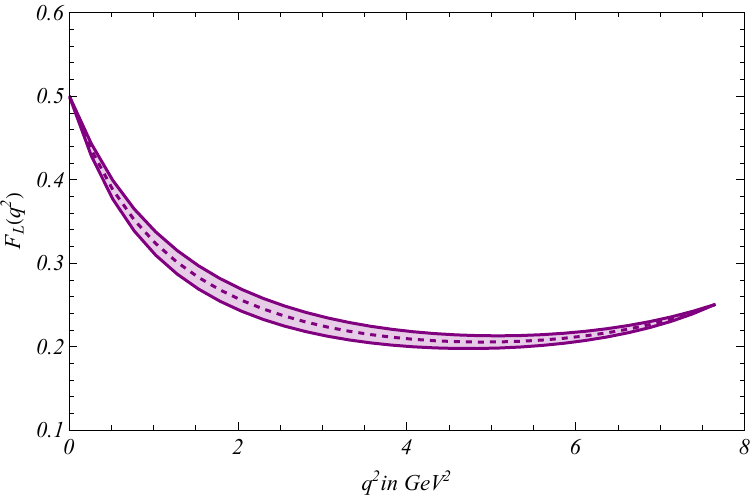}}
	\qquad
	\subfloat[\centering]{\includegraphics[width=7.1cm]{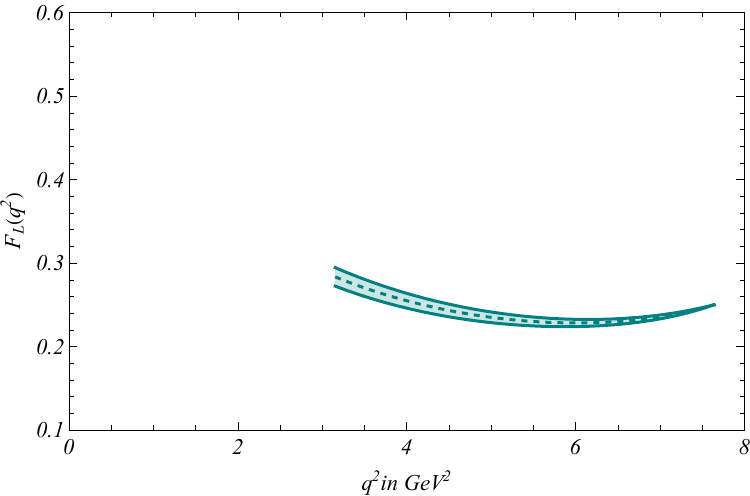}}
		\qquad
		\subfloat[\centering]{\includegraphics[width=7.1cm]{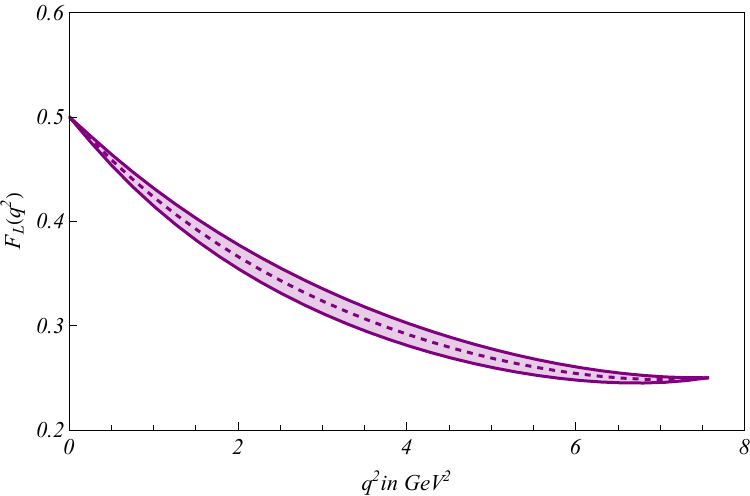}}
		\qquad
		\subfloat[\centering]{\includegraphics[width=7.1cm]{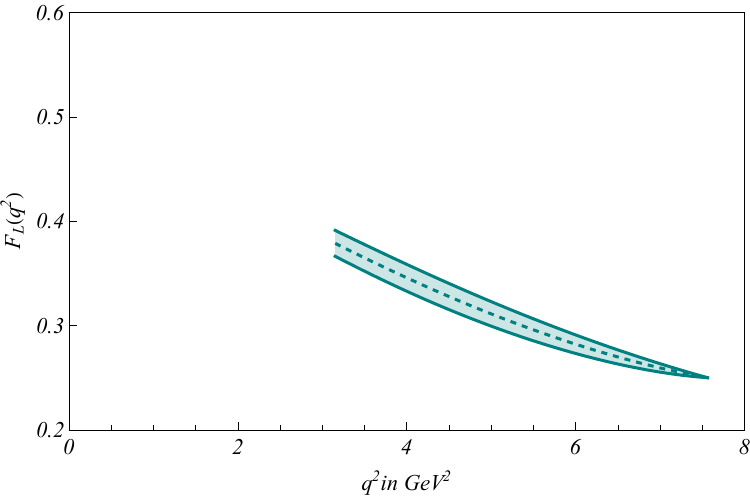}}
		\caption{$q^{2}$ distribution of $F_{L}(q^{2})$. Plots (a) and (b) are for $B_{c}\rightarrow\chi_{c1}$ and (c) and (d) are for $B_{c}\rightarrow h_{c}$ semileptonic channels. The violet and green plots denote the light lepton and heavy lepton cases respectively.}
		\label{fig:longitudinal polarizaiton fraction}
\end{figure*}

	\begin{figure*}[htb!]
	\centering
	\subfloat[\centering]{\includegraphics[width=7.1cm]{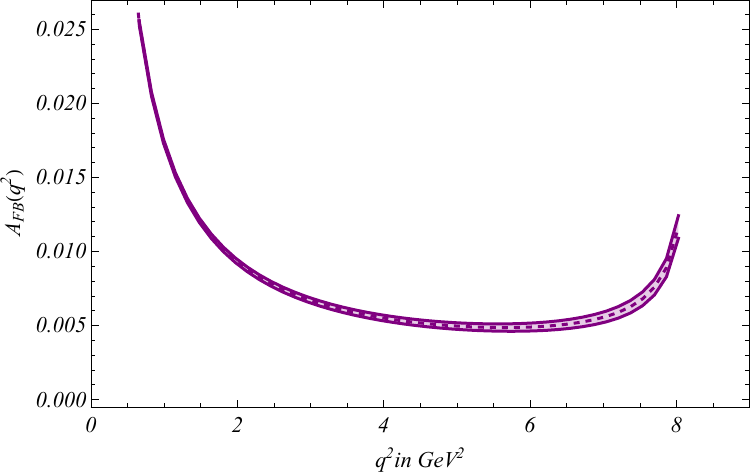}}
	\qquad
	\subfloat[\centering]{\includegraphics[width=7.1cm]{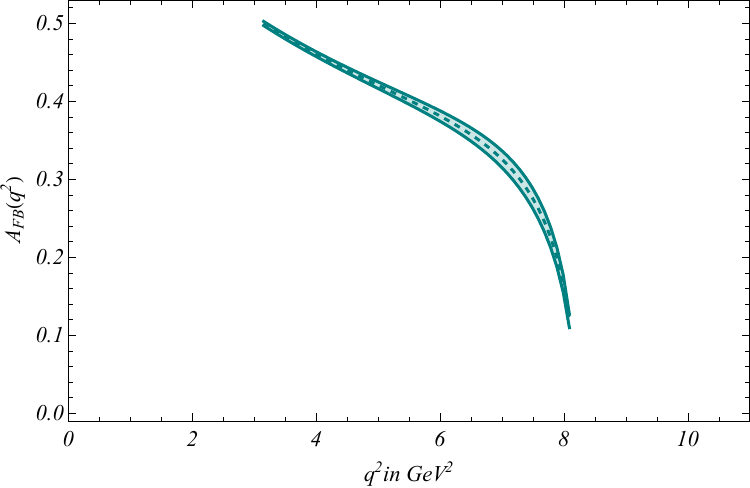}}
	\qquad
	\subfloat[\centering]{\includegraphics[width=7.1cm]{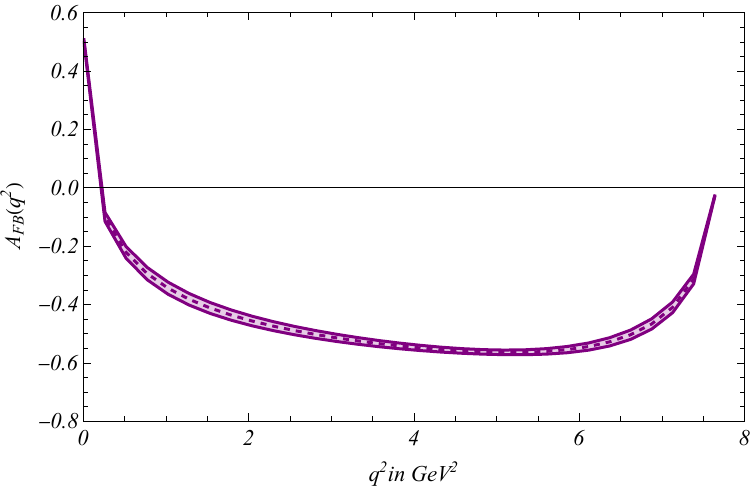}}
	\qquad
	\subfloat[\centering]{\includegraphics[width=7.1cm]{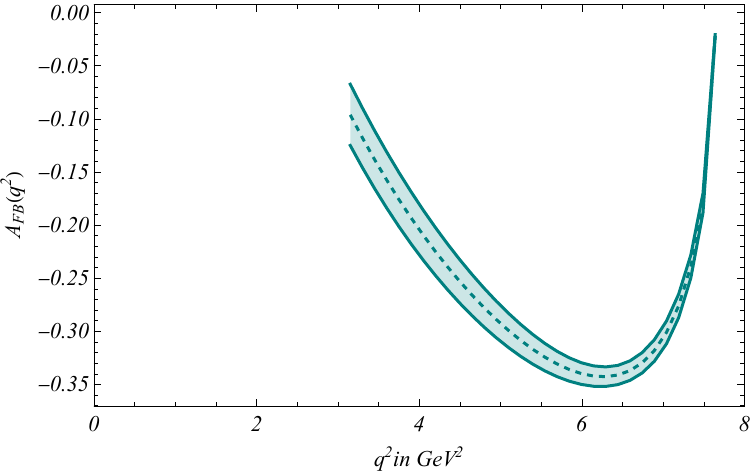}}
		\qquad
		\subfloat[\centering]{\includegraphics[width=7.1cm]{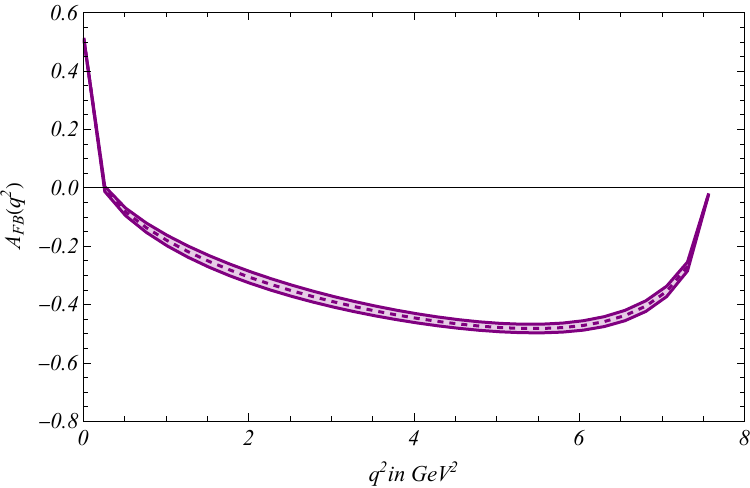}}
		\qquad
		\subfloat[\centering]{\includegraphics[width=7.1cm]{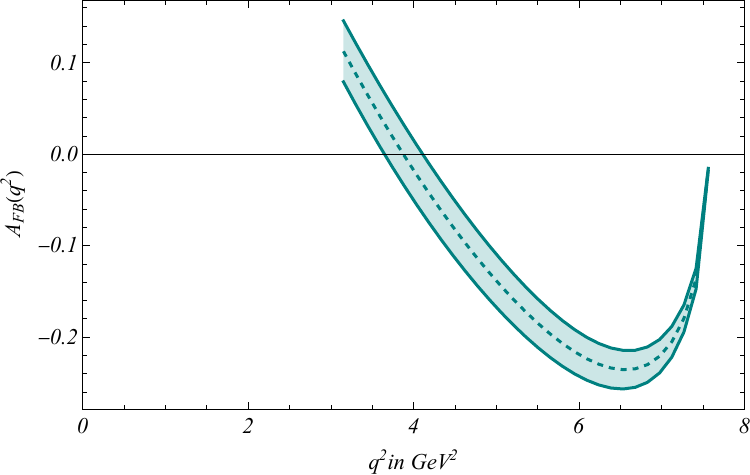}}
		\caption{$q^{2}$ distribution of $A_{FB}(q^{2})$. Plots (a) and (b) are for $B_{c}\rightarrow\chi_{c0}$, (c) and (d) are for $B_{c}\rightarrow\chi_{c1}$ and (e) and (f) are for $B_{c}\rightarrow h_{c}$ semileptonic channels. The violet and green plots denote the light lepton and heavy lepton cases respectively.}
		\label{fig:forward backward asymmetry}
	\end{figure*}
	
	
	\begin{table}[htb!]
		\centering
		\begin{tabular}{|c|c|c|c|c|c|}
			\hline
			\textbf{Channel}&$\boldsymbol{\langle P_{\tau}\rangle}$&$\boldsymbol{\langle F_{L}(l)\rangle}$&$\boldsymbol{\langle F_{L}(\tau)\rangle}$&$\boldsymbol{\langle A_{FB}(l)\rangle}$&$\boldsymbol{\langle A_{FB}(\tau)\rangle}$\\
			\hline
			$B_{c}\rightarrow \chi_{c0}$&0.507(37)&-&-&0.0185(5)&0.374(8)\\
			$B_{c}\rightarrow \chi_{c1}$&-0.498(19)&0.315(16)&0.307(8)&-0.490(11)&-0.300(11)\\
			$B_{c}\rightarrow h_{c}$&-0.415(23)&0.490(24)&0.417(23)&-0.362(18)&-0.172(25)\\
			\hline
		\end{tabular}
		\caption{Predictions on $\langle P_{\tau}\rangle$, $\langle F_{L}\rangle$ and $\langle A_{FB}\rangle$ in SM framework for $B_{c}\rightarrow P$ wave channels.}
		\label{table:PredictionsPtauFLAFBPwave}
	\end{table}
	Checking Table \ref{table:PredictionsPtauFLAFBPwave}, in the second column we can see that $\langle P_{\tau}\rangle$ for $\chi_{c0}$ is positive while that for $\chi_{c1}$ and $h_{c}$ are negative. This is because of the dominance of decay width with tau lepton helicity +1/2 over that with helicity -1/2 for the former case, while for the later case, the production of tau lepton with helicity -1/2 is favoured over one with helicity +1/2. In the third and fourth columns comparing the $\langle F_{L}\rangle$, the value for $h_{c}$ is greater than that for $\chi_{c1}$. The reason can be traced back to the helicity amplitudes, mainly $H_{V0}$ which contributes to $\Gamma_{0}$ in the numerator of \cref{eqn:long pol}. $H_{V0}$ inturn depends on $A_{1}$ and $A_{2}$ which for $B_{c}\rightarrow\chi_{c1}$ carry the same sign, hence resulting in a destructive interference between the two, while for $B_{c}\rightarrow h_{c}$ the two form factors carry opposite sign resulting in a constructive interference, thus resulting in a higher value. Also the differences between values for tau and light-lepton channels are small when checked for both $\chi_{c1}$ and $h_{c}$ modes, suggesting that longitudinal polarisation fraction still favours lepton flavor universality to some extent. In the fifth and sixth columns $\langle A_{FB}\rangle$ is positive for $\chi_{c0}$ while it is negative for $\chi_{c1}$ and $h_{c}$. This signifies that the lepton-neutrino pair is more preferebly emitted in forward direction relative to the $B_{c}$ meson for $B_{c}\rightarrow\chi_{c0}$ channel, and more preferebly in the backward direction for $B_{c}\rightarrow\chi_{c1}(h_{c})$ channels. As for the error estimtates, similar to Table \ref{table:branching5}, here too we get a significant reduction in error, ranging from a minimum of about 2\% to a maximum of about 14\%, the reason primarily being the cancellation of errors coming from all the relevant form factors.
	
	Coming to the plots, Fig \ref{fig:tau polarization} showcases $q^{2}$ distribution of $P_{\tau}$ where in every plot we can see its magnitude increasing as we move from low $q^{2}$ to high $q^{2}$. This happens due to (a) the additional $(m_{l}^{2}/2q^{2})$ term in $d\Gamma_{+}/dq^{2}$ which suppresses it at high $q^{2}$, and (b) the helicity term $3(H_{V,t}^{s})^{2}$ which falls faster with increasing $q^{2}$ compared to $(H_{V,0}^{s})^{2}$, inturn making the denominator in \cref{eqn:tau polarization} fall faster with increasing $q^{2}$ compared to the numerator, thereby increasing its value with increasing $q^{2}$.
	
	 Next in Fig \ref{fig:longitudinal polarizaiton fraction} we showcase the $q^{2}$ distribution of $F_{L}$ where the curve initially falls as $q^{2}$ increases and then rises slightly near $q_{max}^{2}$. This can be explained from \cref{eqn:long pol} where as $q^{2}$ initially increases, $\Gamma^{+1}+\Gamma^{-1}$, or rather the transverse polarization component increases at a pace faster than $\Gamma^{0}$, the longitudinal polarization component, thereby resulting in an initial negative slope. The picture however changes as $q^{2}$ approaches $q_{max}^{2}$, where the transverse polarisation component starts falling faster than the longitudinal polarisation component, thereby resulting in a slight positive slope towards the end of the plot. Additionally, comparing plots (a) with (b) and (c) with (d), we can see that the plots effectively overlap over the common kinematic region, thereby reinforcing the idea that this observable effectively follows lepton flavor universality.
	 
	 Finally in Fig \ref{fig:forward backward asymmetry} we showcase the $q^{2}$ distribution of $A_{FB}$. Starting with $B_{c}\rightarrow\chi_{c0}$ in figs (a) and (b) a significant difference in the value can be seen, with the plot for heavy lepton having larger value than that for light lepton, the reason for which can be traced back to \cref{eqn:forward backward asymmetry}, where the $m_{l}^{2}/q^{2}$ term contributes significantly for the heavy lepton case, thereby raising its value. As to why the nature of the plots have opposite curvature, for light lepton case $b_{\theta}(q^{2})$ in the numerator first falls faster compared to $d\Gamma/dq^{2}$ in the denominator at low $q^{2}$, but towards high $q^{2}$ value $d\Gamma/dq^{2}$ falls faster than $b_{\theta}(q^{2})$, thereby causing a rise in the plot. For the heavy lepton case $b_{\theta}(q^{2})$ first rises starting from $q^{2}=m_{l}^{2}$, attains a maximum value at around $q^{2}=5.5\text{ }GeV^{2}$ and then falls steeply until $q_{max}^{2}$. This when combined with the distribution of $d\Gamma/dq^{2}$ explains the nature of the plot. As for the plots (c) and (e) involving the light lepton modes for $B_{c}\rightarrow\chi_{c1}(h_{c})$ decays, at low $q^{2}$,  $b_{\theta}(q^{2})$ has a negative slope while $d\Gamma/dq^{2}$ has a positive slope, making $A_{FB}$ fall initially. The negative slope gradually reduces, until at around $q^{2}=4.5\text{ }GeV^{2}$, the slopes of both $b_{\theta}(q^{2})$ and $d\Gamma/dq^{2}$ becomes very small, thereby saturating the plot. After that the slopes change sign, with $b_{\theta}(q^{2})$ rising while $d\Gamma/dq^{2}$ falls, making $A_{FB}$ rise at large $q^{2}$. The same reason works for plots (d) and (f) except that the intial fall in $b_{\theta}(q^{2})$ is steadier, resulting in a more curved plot.
	 
	  These predictions will be verified once experimental measurements start coming up in future.
\end{itemize}

\section{Study of some non-leptonic channels}
\label{section:Nonleptonic}

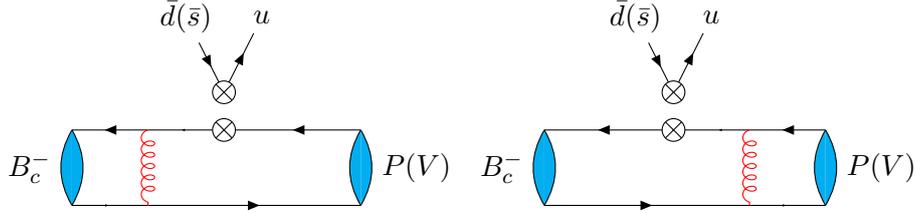
\begin{figure}[htb!]
	\centering
	\begin{tikzpicture}
		\begin{feynman}
			\vertex[crossed dot](a){$$};
			\vertex[left=1.0cm of a](a1);
			\vertex[left=1.0cm of a1](a2);
			\vertex[right=1.8cm of a](a3);
			\vertex[below left=1.0cm and 1.0cm of a](b1);
			\vertex[left=1.0cm of b1](b2);
			\vertex[right=2.8cm of b1](b3);
			\vertex[crossed dot][above=0.5cm of a](c1){$$};
			\vertex[above left=1.0cm and 0.5cm of c1](c2){$\bar{d}(\bar{s})$};
			\vertex[above right=1.0cm and 0.5cm of c1](c3){$u$};
			\diagram*{(a3)--[arrow size=1pt,fermion](a)--[arrow size=0pt,fermion](a1)--[arrow size=1pt,fermion](a2),(b2)--[arrow size=0pt,fermion](b1)--[arrow size=1pt,fermion](b3),(a1)--[style=red,gluon](b1),(c2)--[arrow size=1pt,fermion](c1),(c1)--[arrow size=1pt,fermion](c3),(a2)--[fill=cyan,bend right, plain,edge label'={\(B_c^+\)}](b2),(a2)--[fill=cyan,bend left,plain](b2),(a3)--[fill=cyan,bend left,plain,edge label={$P/V$}](b3),(a3)--[fill=cyan,bend right, plain](b3)};
		\end{feynman}
	\end{tikzpicture}
	\begin{tikzpicture}
		\begin{feynman}
			\vertex[crossed dot](a){$$};
			\vertex[left=1.7cm of a](a1);
			\vertex[right=1.0cm of a](a2);
			\vertex[right=1.0cm of a2](a3);
			\vertex[below right=1.0cm and 1.0cm of a](b1);
			\vertex[right=1.0cm of b1](b2);
			\vertex[left=2.7cm of b1](b3);
			\vertex[crossed dot][above=0.5cm of a](c1){$$};
			\vertex[above left=1.0cm and 0.5cm of c1](c2){$\bar{d}(\bar{s})$};
			\vertex[above right=1.0cm and 0.5cm of c1](c3){$u$};
			\diagram*{(a3)--[arrow size=1pt,fermion](a2)--[arrow size=0pt,fermion](a)--[arrow size=1pt,fermion](a1),(b3)--[arrow size=1pt,fermion](b1)--[arrow size=0pt,fermion](b2),(a2)--[style=red,gluon](b1),(c2)--[arrow size=1pt,fermion](c1),(c1)--[arrow size=1pt,fermion](c3),(a1)--[fill=cyan,bend right, plain,edge label'={\(B_c^+\)}](b3),(a1)--[fill=cyan,bend left,plain](b3),(a3)--[fill=cyan,bend left,edge label={$P/V$}](b2),(a3)--[fill=cyan,bend right, plain](b2)};
		\end{feynman}
	\end{tikzpicture}
	
	\caption{Leading order factorizable diagrams for $B_{c}^{+}\rightarrow X_{c\bar{c}} \hspace{0.8mm}\pi^{+}(K^{+})$.}
	\label{fig:Nonleptonic Feynman Diagrams Fac}
\end{figure}
\begin{figure}[htb!]
	\centering
	\begin{tikzpicture}
		\begin{feynman}
			\vertex[crossed dot](a){$$};
			\vertex[left=2.0cm of a](a1);
			\vertex[above left=0.6cm and 0.5cm of a](a2);
			\vertex[above left=0.6cm and 0.5cm of a2](a3){$\bar{d}(\bar{s})$};
			\vertex[crossed dot][right=0.5cm of a](c1){$$};
			\vertex[right=1.5cm of c1](c3);
			\vertex[above right=1.5cm and 0.75cm of c1](c2){u};
			\vertex[below left=1.0cm and 2.0cm of a](b1);
			\vertex[right=0.8cm of b1](b3);
			\vertex[right=3.2cm of b3](b2);
			\diagram*{(a3)--[arrow size=1pt,fermion](a2)--[arrow size=0pt,fermion](a)--[arrow size=1pt,fermion](a1),(c3)--[arrow size=1pt,fermion](c1)--[arrow size=1pt,fermion](c2),(b1)--[arrow size=0pt,fermion](b3)--[arrow size=1pt,fermion](b2),(a2)--[style=red,gluon](b3),(a1)--[fill=cyan,bend right, plain,edge label'={\(B_c^+\)}](b1),(a1)--[fill=cyan,bend left,plain](b1),(c3)--[fill=cyan,bend left,edge label={$P/V$}](b2),(c3)--[fill=cyan,bend right, plain](b2)};
		\end{feynman}
	\end{tikzpicture}
	\begin{tikzpicture}
		\begin{feynman}
			\vertex[crossed dot](a){$$};
			\vertex[left=1.5cm of a](a1);
			\vertex[above left=1.4cm and 0.75cm of a](a2){$\bar{d}(\bar{s})$};
			\vertex[crossed dot][right=0.5cm of a](c1){$$};
			\vertex[right=2.0cm of c1](c3);
			\vertex[above right=0.6cm and 0.5cm of c1](c2);
			\vertex[above right=0.6cm and 0.5cm of c2](c4){u};
			\vertex[below left=1.0cm and 1.5cm of a](b1);
			\vertex[right=3.2cm of b1](b3);
			\vertex[right=0.8cm of b3](b2);
			\diagram*{(a2)--[arrow size=1pt,fermion](a)--[arrow size=1pt,fermion](a1),(c3)--[arrow size=1pt,fermion](c1)--[arrow size=0pt,fermion](c2)--[arrow size=1pt,fermion](c4),(b1)--[arrow size=1pt,fermion](b3)--[arrow size=0pt,fermion](b2),(c2)--[style=red,gluon](b3),(a1)--[fill=cyan,bend right, plain,edge label'={\(B_c^+\)}](b1),(a1)--[fill=cyan,bend left,plain](b1),(c3)--[fill=cyan,bend left,edge label={$P/V$}](b2),(c3)--[fill=cyan,bend right, plain](b2)};
		\end{feynman}
	\end{tikzpicture}
	\caption{Leading order non-factorizable diagrams for $B_{c}^{+}\rightarrow X_{c\bar{c}} \hspace{0.8mm}\pi^{+}(K^{+})$.}
	\label{fig:Nonleptonic Feynman Diagrams Nonfac}
\end{figure}
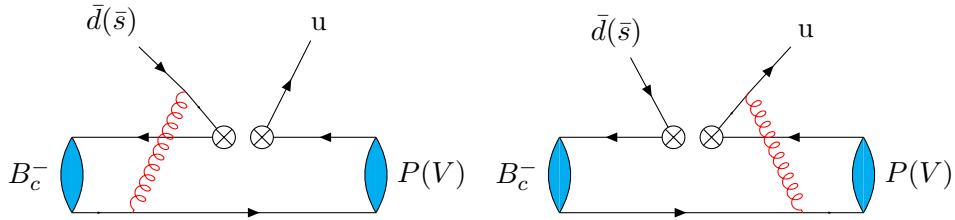

In addition to semileptonic decays of $B_{c}$ meson, we have also studied some non-leptonic decays of the $B_{c}$ meson. Non-leptonic decays of heavy mesons are particularly interesting as they present an oppurtunity to study the nature of Quantum Chromodynamics. We study the decay of $B_{c}$ meson into S wave and P wave charmonium states and a light pseudo-scalar meson. We perform our analysis in the same modified PQCD framework, whose uniqueness lies in the fact that, contrary to other approaches in the literature, where only the factorizable diagrams were calculable, in PQCD approach the non-factorizable diagrams are also calculable. The relevant leading order Feynman diagrams are shown in Figs \ref{fig:Nonleptonic Feynman Diagrams Fac} and \ref{fig:Nonleptonic Feynman Diagrams Nonfac}. The branching ratios of a few of the decay channels have been measured in this work. \\

The effective Hamiltonian for the $b\rightarrow cu\bar{d}$ transition can be expressed as \cite{Buchalla:1995vs}
\begin{equation}
	\mathcal{H}_{eff}=\frac{G_{F}}{\sqrt{2}}V_{cb}V_{ud}^{*}[C_{1}(\mu)\mathcal{O}_{1}(\mu)+C_{2}(\mu)\mathcal{O}_{2}(\mu)],
\end{equation}
with $C_{1,2}(\mu)$ representing the Wilson coefficients which encode all the short-distance contributions. These are calculated perturbatively at first $m_{W}$ scale, and then evolved down to the renormalization scale using renormalization group equations. The local four quark operators $\mathcal{O}_{1,2}$ are expressed as
\begin{equation}
	\begin{split}
		\mathcal{O}_{1}&=\bar{d}_{\alpha}\gamma_{\mu}(1-\gamma_{5})u_{\beta}\bar{c}_{\beta}\gamma^{\mu}(1-\gamma_{5})b_{\alpha},\\
		\mathcal{O}_{2}&=\bar{d}_{\alpha}\gamma_{\mu}(1-\gamma_{5})u_{\alpha}\bar{c}_{\beta}\gamma^{\mu}(1-\gamma_{5})b_{\beta},
	\end{split}
\end{equation}
with $\alpha$ and $\beta$ being the color indices. Considering the decay kinematics as in Appendix \ref{section:Appendix kinematics} at $q^{2}=0$ and the same PQCD factorization formalism as has been done in subsection \ref{subsection:Form Fcators}, the decay amplitude can be expressed as \cite{PhysRevD.90.114030}
\begin{equation}
	\mathcal{A}\propto C(t)\otimes H(x,t)\otimes \Phi(x)\otimes exp\left[-s_{c}(P,b)\right],
\end{equation}
where $C(t)$ represents the Wilson coefficients, $H(x,t)$ the hard kernel, $\Phi(x)$ the meson distribution amplitudes and $s_{c}(P,b)$ the Sudakov exponent in modified PQCD \cite{PhysRevD.97.113001}. The decay amplitude will have contributions from all the factorizable as well as the non-factorizable diagrams, and can be factorized into two parts, one having contributions from factorizable diagrams shown in fig \ref{fig:Nonleptonic Feynman Diagrams Fac} and the second having contribution from non-factorizable diagrams shown in fig \ref{fig:Nonleptonic Feynman Diagrams Nonfac}, and analytically can be expressed as
\begin{equation}
	\mathcal{A}=V_{cb}^{*}V_{ud(s)}\left[f_{\pi(K)}\left(C_{2}(t)+\frac{C_{1}(t)}{3}\right)\mathcal{F}+C_{1}(t)\mathcal{M}\right],
	\label{eqn:NLAmplitude}
\end{equation}
$\mathcal{F}$ and $\mathcal{M}$ representing contributions from factorizable and non-factorizable diagrams respectively, $f_{\pi(K)}$ the decay constants of $\pi(K)$ meson and $C_{1}(t)$ and $C_{2}(t)$ the Wilson coefficients represented as function of the hard scale $t$ \cite{Lu:2000em} shown in Appendix \ref{section:hard functions}. Analytic expressions for $\mathcal{F}$ and $\mathcal{M}$ has been shown in Appendices \ref{section:Appendix pQCD form factors}. For the LCDAs of the $B_{c}$ and charmonium mesons, we use the same forms as we had shown in subsection \ref{subsection:LCDAs} and for the $\pi$ and $K$ mesons, we consider the parametric form for the leading twist LCDAs from \cite{PhysRevD.81.014022}
\begin{equation}
	\footnotesize
	\begin{split}
		\phi_{\pi(K)}^{A}(x)=\frac{f_{\pi(K)}}{2\sqrt{2N_{c}}}6x(1-x)\bigl[1+a_{1}^{\pi(K)}C_{1}^{3/2}(2x-1)+a_{2}^{\pi(K)}C_{2}^{3/2}(2x-1)+a_{4}^{\pi(K)}C_{4}^{3/2}(2x-1)\bigr],
	\end{split}
\end{equation}
with the Gegenbauer moments being $a_{1}^{\pi}=0$, $a_{2}^{\pi}=0.115\pm 0.115$,  $a_{4}^{\pi}=-0.015$, $a_{1}^{K}=0.17\pm 0.17$, $a_{2}^{K}=0.115$ and $a_{4}^{K}=-0.015$. $C_{2,4}^{1/2,3/2}(t)$ are the corresponding Gegenbauer polynomials, and are generically expressed as
\begin{align}
	C_{1}^{3/2}(t)=3t, \qquad C_{2}^{3/2}(t)=\frac{3}{2}(5t^{2}-1), \qquad C_{4}^{3/2}(t)=\frac{15}{8}(1-14t^{2}+21t^{4}).
\end{align}

However, it is not the decay amplitudes, but the branching ratios that are the actual physical observables that we will be predicting, and are expressed as \cite{PhysRevD.97.113001}
\begin{equation}
	\mathcal{B}(B_{c}\rightarrow XP)=\frac{G_{F}^{2}\tau_{B_{c}}}{32\pi m_{B_{c}}}(1-r^{2})|\mathcal{A}|^{2}.
	\label{eqn:NL Branching}
\end{equation}

Taking these expressions and parameters previously extracted in Table \ref{table:LCDA1} as inputs, we can now calculate the branching ratios of some of the non-leptonic decay channels of $B_{c}$ meson. In this work we present our predictions for the branching ratios of $B_{c}$ decaying into S or P wave charmonium states along with a pseudoscalar meson. The predictions for the branching ratios and some ratios of the branching ratios between different modes so obtained are presented in Tables \ref{table:branching5} and \ref{table:branching6}.

\begin{table}[htb!]
	\centering
	\begin{tabular}{|c|c|ccc|}
		
		\hline
		\textbf{Charmonium}&\textbf{Decay}&\textbf{This}&\textbf{Previous}&\textbf{PDG}\\
		\textbf{state}&\textbf{Channel}&\textbf{work}&\textbf{PQCD\textcolor{blue}{[1,2]}} &\textbf{results}\\
		\hline

		&$\mathcal{B}(B_{c}^{+}\rightarrow \eta_{c} \pi^{+})$&1.448(173)&2.98$^{+0.84+0.75+0.52}_{-0.79-0.67-0.14}$&-\\
		S wave&$\mathcal{B}(B_{c}^{+}\rightarrow \eta_{c} K^{+})$&0.125(23)&0.24$^{+0.04+0.07+0.02}_{-0.05-0.06-0.01}$&-\\
		$(\times 10^{-3})$&$\mathcal{B}(B_{c}^{+}\rightarrow J/\psi \pi^{+})$&0.726(150)&2.33$^{+0.63+0.16+0.48}_{-0.58-0.16-0.12}$&-\\
		&$\mathcal{B}(B_{c}^{+}\rightarrow J/\psi K^{+})$&0.057(8)&0.19$^{+0.04+0.02+0.02}_{-0.04-0.02-0.01}$&-\\
		
		\hline
		&$\mathcal{B}(B_{c}^{+}\rightarrow \chi_{c0} \pi^{+})$&0.267(110)&16.0$^{+2.0+3.0+0.0}_{-2.0-3.0-1.0}$&0.24$^{+0.9}_{-0.8}$\\
		&$\mathcal{B}(B_{c}^{+}\rightarrow \chi_{c0} K^{+})$&0.020(4)&1.2$^{+0.2+0.3+0.0}_{-0.2-0.2-0.1}$&-\\
		P wave &$\mathcal{B}(B_{c}^{+}\rightarrow \chi_{c1} \pi^{+})$&0.121(37)&5.10$^{+0.3+1.1+0.0}_{-0.4-1.1-0.2}$&-\\
		$(\times 10^{-4})$&$\mathcal{B}(B_{c}^{+}\rightarrow \chi_{c1} K^{+})$&0.919(210)$\times 10^{-2}$&0.38$^{+0.03+0.09+0.01}_{-0.03-0.08-0.01}$&-\\
		&$\mathcal{B}(B_{c}^{+}\rightarrow h_{c} \pi^{+})$&0.149(55)&5.4$^{+0.4+1.0+0.4}_{-0.3-1.0-0.3}$&-\\
		&$\mathcal{B}(B_{c}^{+}\rightarrow h_{c} K^{+})$&0.011(4)&0.43$^{+0.03+0.07+0.03}_{-0.02-0.08-0.02}$&-\\
		\hline
		
	\end{tabular}
	
	\caption{Predictions on branching ratios of some $B_{c}\rightarrow $ S and P wave non-leptonic decay channels.}
	
	\label{table:branching5}
\end{table} 

In predictions of Table \ref{table:branching5}, the error analysis has been done by taking the errors of the $B_{c}$ meson distribution amplitude $\omega_{B_{c}}$, $\Lambda_{QCD}$, the respective decay constants of the participating mesons and quark masses, and has been added in quadrature.

A point worth mentioning here is that if we look back at the first and second columns of Table \ref{table:LCDA 3} and \ref{table:branching3}, and then at Table \ref{table:branching5}, a significant difference is observed between our estimates of nonleptonic branching ratios and earlier pQCD predictions given in \cite{Rui:2017pre}, while our results are consistent in case of semileptonic form factors and branching ratios to those given in \cite{Rui:2018kqr}. The primary reason for this difference is due to the choice of $B_{c}$ meson LCDA. In this work we are using an improved version of $B_{c}$ meson LCDA, which depends on both momentum fraction $x$ as well as transverse separation between the quarks $b$, and models the $B_{c}$ meson LCDA better. In ref. \cite{Rui:2018kqr}, the authors have taken the same form of LCDA as our work. However, the authors in \cite{Rui:2017pre} and \cite{PhysRevD.90.114030} have taken a $b$ independent form of the LCDA. This difference in choice of LCDA alone contributes to the disparity between the nonleptonic results.

Comparing the branching ratios in the first column, there are some hierarchical relations between the branching ratios that show up. First, we see that the branching ratios of processes involving pions in final states are relatively large compared to those involving kaons in the final state. This is predominantly due to CKM suppression factor $|V_{us}|^{2}/|V_{ud}|^{2}$. Further, the branching ratio of decays involving (pseudo-) scalar charmonium states also seems to be larger than their (axial-) vector counterparts. The reason can easily be checked if we compare the contributions of the twist-2 and twist-3 distribution amplitudes to the branching ratios. For the former the dominant contribution to the branching ratio comes from the twist 3 contribution of the second diagrams in Figs \ref{fig:Nonleptonic Feynman Diagrams Fac} and \ref{fig:Nonleptonic Feynman Diagrams Nonfac}, while the twist 2 contributions are suppressed, while for the later, the dominant contribution still comes from the twist-2 part which already has a small value due to the suppression caused by $r_{c}\pm r^2$. This causes the branching ratios of channels involving (pseudo-)scalar mesons to have a larger value than their (axial-)vector counterpart. This explanation has already been presented by the authors of \cite{Rui:2017pre} in their work, and has been checked to hold in this work too.

\begin{table}[htb!]
	\setlength{\tabcolsep}{3pt}
	\centering
	\begin{tabular}{|c|cc|c|cc|}
		\hline
		\multicolumn{3}{|c|}{\textbf{S wave channels}}  &  \multicolumn{3}{|c|}{\textbf{P wave channels}} \\
		\hline
		
		\textbf{Observables}&\textbf{This work}&\textbf{PDG results}&\textbf{Observables}&\textbf{This work}&\textbf{PDG results}\\
		\hline
		
		&&&&&\\
		$\frac{\mathcal{B}(B_{c}^{+}\rightarrow J/\psi \hspace{0.1cm}\pi^{+})}{\mathcal{B}(B_{c}^{+}\rightarrow J/\psi\hspace{0.1cm} l^{+}\nu_{l})}$&0.047(10)&0.0469(28)&$\frac{\mathcal{B}(B_{c}^{+}\rightarrow \chi_{c0} \hspace{0.1cm}K^{+})}{\mathcal{B}(B_{c}^{+}\rightarrow \chi_{c0}\hspace{0.1cm} \pi^{+})}$&0.075(34)&-\\
		&&&&&\\
		$\frac{\mathcal{B}(B_{c}^{+}\rightarrow \eta_{c} K^{+})}{\mathcal{B}(B_{c}^{+}\rightarrow \eta_{c} \pi^{+})}$&0.086(19)&-&$\frac{\mathcal{B}(B_{c}^{+}\rightarrow \chi_{c1} \hspace{0.1cm}K^{+})}{\mathcal{B}(B_{c}^{+}\rightarrow \chi_{c1}\hspace{0.1cm} \pi^{+})}$&0.076(29)&-\\
		&&&&&\\
		$\frac{\mathcal{B}(B_{c}^{+}\rightarrow J/\psi K^{+})}{\mathcal{B}(B_{c}^{+}\rightarrow J/\psi \pi^{+})}$&0.078(19)&0.079(7)&$\frac{\mathcal{B}(B_{c}^{+}\rightarrow h_{c} \hspace{0.1cm}K^{+})}{\mathcal{B}(B_{c}^{+}\rightarrow h_{c}\hspace{0.1cm} \pi^{+})}$&0.073(38)&-\\
		&&&&&\\
		\hline
	\end{tabular}
	
	\captionof{table}{Some ratios among the branching fractions of the $B_{c}$ decays.}
	\label{table:branching6}
\end{table}

\section{Summary and Conclusions} 
\label{section:summary and conclusion}
In this study, our goal is to analyse the semileptonic and non-leptonic decays of the $B_c$ meson with charmonium in their final state. In this analysis, we have utilised the form factors derived from the modified perturbative QCD approach. We derive the shape of the $B_c$ meson wave function using the inputs (lattice and others) on the $B_c\to J/\psi, \eta_c$ form factors. In the due process, we have extracted the decay constants of P wave charmonium states, $\chi_{c0}$, $\chi_{c1}$ and $h_{c}$ from their radiative decay modes, yielding a data-driven alternative to existing model dependent values, enabling us to use them as inputs to predict the $B_{c}\rightarrow\text{ }P$(wave) form factors at $q^{2}=0$ within the modified perturbative QCD framework. Subsequently, utilising the shapes of the $B_{c}\rightarrow\eta_{c}$ and $J/\psi$ form factors, we have obtained $q^{2}$ distribution of the $B_{c}\rightarrow\chi_{c0},\chi_{c1}$ and $h_{c}$ form factors using pole expansion parametrization, after which we obtained predictions of LFUV observables $R(\chi_{c0})=0.169(11)$, $R(\chi_{c1})=0.126(2)$ and $R(h_{c})=0.113(3)$.  

Finally, using the results of the form factors for $B_c$ to $P$ and $S$ wave charmonium decays, we have studied a few two-body non-leptonic decays of $B_c$ meson with one or two $S$ and $P$ wave charmonia in the final states.  

\section{Acknowledgements}
We would like to thank Hsiang-nan Li for helpful discussion in the initial stage of the project.

\appendix

\section{Kinematics:}
\label{section:Appendix kinematics}

In this appendix we discuss about the kinematics of the decay channels that we have considered in this work. All the notations relating to the kinematics, that find relevance in our work has been mentioned here. In addition we also, very briefly, presented a discussion on the kinematic constraints on the shape of distribution amplitudes of the participating mesons.\\

We consider that the $B_{c}$ meson is initially at rest. The initial and the final momenta are expressed in light-cone coordinate systems. Let $P_{1}$ and $P_{2}$ be the $B_{c}$ and $J/\psi$ meson momenta, then they are expressed 
\begin{equation}
	\begin{split}
		P_{1}&=\frac{m_{B_{c}}}{\sqrt{2}}(1,1,0_{T}),\\
		P_{2}&=\frac{m_{B_{c}}}{\sqrt{2}}(r\eta^{+},r\eta^{-},0_{T}),
	\end{split}
\end{equation}
the ratio $r=\frac{m}{m_{B_{c}}}$ representing the ratio of the masses of the charmonium states and the $B_{c}$ meson, and the factors $\eta^{+}=\eta+\sqrt{\eta^{2}-1}$ and  $\eta^{-}=\eta-\sqrt{\eta^{2}-1}$,\\ and $\eta$ has the form,
\begin{equation}
	\eta=\frac{1+r^{2}}{2r}-\frac{q^{2}}{2rm_{B_{c}}^{2}},
\end{equation}
with $q=P_{1}-P_{2}$ being the momentum transfer. In case the final state meson is (axial-)vector, the associated longitudinal and transverse polarisations can be written as \cite{Rui:2016opu}
\begin{equation}
\epsilon_{L}=\frac{1}{\sqrt{2}}(\eta^{+},-\eta^{-},0_{T}), \qquad \epsilon_{T}=(0,0,1),
\end{equation}
and the momenta of the valence quarks are
\begin{equation}
	\begin{split}
		k_{1}=&\left(x_{1}\frac{m_{B_{c}}}{\sqrt{2}},x_{1}\frac{m_{B_{c}}}{\sqrt{2}},\vec{k}_{1T}\right),\\
		k_{2}=&\left(\frac{m_{B_{c}}}{\sqrt{2}}x_{2}r\eta^{+},\frac{m_{B_{c}}}{\sqrt{2}}x_{2}r\eta^{-},\vec{k}_{2T}\right),
	\end{split}
\end{equation}
$k_{1T}$ and $k_{2T}$ represent the transverse momenta, and $x_{1,2}$ represent the longitudinal momentum fraction of the spectator charm quarks in $B_{c}$ and charmonium, respectively. For nonleptonic decays, the outgoing pion will carry a momentum 
\begin{equation}
P_{3}=\frac{m_{B_{c}}}{\sqrt{2}}(0,1-r^{2},0_{T}),
\end{equation} 
with the spectator quark having momentum
\begin{equation}
k_{3}=(x_{3}P_{3}^{+},x_{3}P_{3}^{-},\vec{k}_{3T}),
\end{equation}
at maximum recoil with $x_{3}$ representing the fraction of the momentum carried by the quark.

\section{PQCD form factors for semileptonic and nonleptonic $\boldsymbol{b\rightarrow c}$ decays:}
\label{section:Appendix pQCD form factors}

\indent In this appendix we present the analytical expressions of the form factors already discussed in \ref{subsection:Form Fcators}. Their expressions have been taken from \cite{Rui:2016opu,PhysRevD.98.033007}. For calculations in PQCD it is much more convenient to express the form factors $F_{+}$ and $F_{0}$ in terms of auxiliary form factors $f_{1}$ and $f_{2}$, defined as \cite{Wang:2012lrc}
\begin{equation}
\label{eqn:etaC matrix element auxillary}
\left\langle \eta_{c}(P_{2})|\bar{c}\gamma^{\mu}b|B_{c}(P_{1})\right\rangle=f_{1}(q^{2})P_{1}^{\mu}+f_{2}(q^{2})P_{2}^{\mu},
\end{equation}
and are related to $F_{+}$ and $F_{0}$ as
\begin{equation}
\begin{split}
F_{+}(q^{2})=&\frac{1}{2}[f_{1}(q^{2})+f_{2}(q^{2})],\\
F_{0}(q^{2})=&\frac{1}{2}f_{1}(q^{2})\left[1+\frac{q^{2}}{m_{B_{c}}^{2}-m^{2}}\right]+\frac{1}{2}f_{2}(q^{2})\left[1-\frac{q^{2}}{m_{B_{c}}^{2}-m^{2}}\right].
\end{split}
\end{equation}

\begin{itemize}
\item For $B_{c}\rightarrow \eta_{c},\chi_{c0}$ semileptonic decays, the auxillary form factors $f_{1}(q^{2})$ and $f_{2}(q^{2})$ have the form
\begin{equation}
	\label{eq:BcPslf1}
		\begin{split}
			f_{1}(q^{2})=&4 \sqrt{\frac{2}{3}}\pi m_{B_{c}}^{2} f_{B_{c}} C_{f} r\int_{0}^{1}dx_{1}\hspace{1mm}dx_{2}\int_{0}^{b_{c}}b_{1}db_{1}b_{2}db_{2}\hspace{1mm}\phi_{B_{c}}(x_{1},b_{1})\\
			&\biggl[\left\lbrace\psi^{v}(x_{2},b_{2})r(x_{2}-1)-\psi^{s}(x_{2},b_{2})(r_{b}-2)\right\rbrace\hspace{1mm}E_{ab}(t_{a})h(\alpha_{e},\beta_{a},b_{1},b_{2})S_{t}(x_{2})\\
			&-\left\lbrace\psi^{v}(x_{2},b_{2})(r-2\eta x_{1})+\psi^{s}(x_{2},b_{2})2(x_{1}-(\pm r_{c}))\right\rbrace\hspace{1mm}E_{ab}(t_{b})h(\alpha_{e},\beta_{b},b_{1},b_{2})S_{t}(x_{1})\biggr],
		\end{split}
	\end{equation}
	and,
	\begin{equation}
	\label{eq:BcPslf2}
		\begin{split}
			f_{2}(q^{2})=&4 \sqrt{\frac{2}{3}}\pi m_{B_{c}}^{2} f_{B_{c}} C_{f}\int_{0}^{1}dx_{1}\hspace{1mm}dx_{2}\int_{0}^{b_{c}}b_{1}db_{1}b_{2}db_{2}\hspace{1mm}\phi_{B_{c}}(x_{1},b_{1})\\
			&\biggl[\left\lbrace\psi^{v}(x_{2},b_{2})(2r_{b}-1-2r\eta (x_{2}-1))+\psi^{s}(x_{2},b_{2})2r(x_{2}-1)\right\rbrace E_{ab}(t_{a})h(\alpha_{e},\beta_{a},b_{1},b_{2})S_{t}(x_{2})\\&-\left\lbrace\psi^{v}(x_{2},b_{2})((\pm r_{c})+x_{1})-\psi^{s}(x_{2},b_{2})2r\right\rbrace E_{ab}(t_{b})h(\alpha_{e},\beta_{b},b_{1},b_{2})S_{t}(x_{1})\biggr],
		\end{split}
	\end{equation}

	\item For $B_{c}\rightarrow J/\psi,\chi_{c1},h_{c}$ semileptonic decays the form factors $A_{0,1,2}(q^{2})$ and $V(q^{2})$ have the form
	\begin{equation}
	\label{eq:BcVslA0}
		\begin{split}
			A_{0}(q^{2}&)=-2\sqrt{\frac{2}{3}}\pi m_{B_{c}}^{2} f_{B_{c}} C_{f} \int_{0}^{1}dx_{1} dx_{2}\int_{0}^{b_{c}}b_{1}db_{1}b_{2}db_{2}\phi_{B_{c}}(x_{1},b_{1})\\
			&\biggl[\left\lbrace\psi^{L}(x_{2},b_{2})(1-2r_{b}-r(x_{2}-1)(r-2\eta))-\psi^{t}(x_{2},b_{2})r(2x_{2}-r_{b})\right\rbrace E_{ab}(t_{a})h(\alpha_{e},\beta_{a},b_{1},b_{2})S_{t}(x_{2})\\&-\left\lbrace\psi^{L}(x_{2},b_{2})((\pm r_{c})+r^{2}+x_{1}(1-2r\eta))\right\rbrace E_{ab}(t_{b})h(\alpha_{e},\beta_{b},b_{1},b_{2})S_{t}(x_{1})\biggr],
		\end{split}
	\end{equation}
	\begin{equation}
	\label{eq:BcVslA1}
		\begin{split}
			A_{1}(q^{2}&)=4\sqrt{\frac{2}{3}}\frac{r}{1\pm r}\pi m_{B_{c}}^{2} f_{B_{c}} C_{f} \int_{0}^{1}dx_{1} dx_{2}\int_{0}^{b_{c}}b_{1}db_{1}b_{2}db_{2}\phi_{B_{c}}(x_{1},b_{1})\\
			&\biggl[\left\lbrace\psi^{V}(x_{2},b_{2})(-2r_{b}+\eta r(x_{2}-1)+1)+\psi^{T}(x_{2},b_{2})(\eta r_{b}-2(\eta +r(x_{2}-1)))\right\rbrace \\&E_{ab}(t_{a})h(\alpha_{e},\beta_{a},b_{1},b_{2})S_{t}(x_{2})-\left\lbrace\psi^{V}(x_{2},b_{2})((\pm r_{c})-x_{1}+\eta r)\right\rbrace E_{ab}(t_{b})h(\alpha_{e},\beta_{b},b_{1},b_{2})S_{t}(x_{1})\biggr],
		\end{split}
	\end{equation}
	\begin{equation}
	\label{eq:BcVslA2}
		\begin{split}
			A_{2}(q^{2}&)=-A_{1}(q^{2})\frac{(1\pm r)^{2}(r-\eta)}{2r(\eta^{2}-1)}-2\sqrt{\frac{2}{3}}\pi m_{B_{c}}^{2} f_{B_{c}} C_{f}\frac{1\pm r}{\eta^{2}-1} \int_{0}^{1}dx_{1} dx_{2}\int_{0}^{b_{c}}b_{1}db_{1}b_{2}db_{2}\phi_{B_{c}}(x_{1},b_{1})\\
			&\biggl[\biggl\lbrace \psi^{t}(x_{2},b_{2})(r_{b}(1-\eta r)+2r^{2}(x_{2}-1)-2\eta r(x_{2}-2)-2)\\&-\psi^{L}(x_{2},b_{2})(2r_{b}(\eta-r)-\eta+r(\eta r(x_{2}-1)-2\eta^{2}(x_{2}-1)+x_{2}))\biggr\rbrace\\& E_{ab}(t_{a})h(\alpha_{e},\beta_{a},b_{1},b_{2})S_{t}(x_{2})+\left\lbrace \psi^{L}(x_{2},b_{2})((\pm r_{c})(r-\eta)+\eta r^{2}+r(-2\eta^{2}x_{1}+x_{1}-1)+\eta x_{1})\right\rbrace\\& E_{ab}(t_{b})h(\alpha_{e},\beta_{b},b_{1},b_{2})S_{t}(x_{1})\biggr],
		\end{split}
	\end{equation}
	\begin{equation}
	\label{eq:BcVslV}
		\begin{split}
			V(q^{2}&)=2\sqrt{\frac{2}{3}}\pi m_{B_{c}}^{2} f_{B_{c}} C_{f}(1\pm r) \int_{0}^{1}dx_{1} dx_{2}\int_{0}^{b_{c}}b_{1}db_{1}b_{2}db_{2}\phi_{B_{c}}(x_{1},b_{1})\\
			&\biggl[\left\lbrace\psi^{V}(x_{2},b_{2})r(1-x_{2})+\psi^{T}(x_{2},b_{2})(r_{b}-2)\right\rbrace E_{ab}(t_{a})h(\alpha_{e},\beta_{a},b_{1},b_{2})S_{t}(x_{2})\\&-\left\lbrace\psi^{V}(x_{2},b_{2})r\right\rbrace E_{ab}(t_{b})h(\alpha_{e},\beta_{b},b_{1},b_{2})S_{t}(x_{1})\biggr].
		\end{split}
	\end{equation}
	with the $r_{c}$ and $1+ r$ are for are for $B_{c}\rightarrow S$ wave modes and $-r_{c}$ and $1-r$ terms are for $B_{c}\rightarrow P$ wave modes.
\end{itemize}

In addition, we also present the analytical expressions for the contributions from the factorizable and non-factorizable diagrams of non-leptonic decays of $B_{c}$ meson discussed in \ref{section:Nonleptonic}. Their expressions has been taken from \cite{PhysRevD.97.113001,PhysRevD.90.114030,Rui:2017pre}.
\begin{itemize}

\item For $B_{c}\rightarrow \eta_{c}\pi^{+}$ and $B_{c}\rightarrow \eta_{c}K^{+}$ decays:
\begin{equation}
\label{eq:nlffS1}
	\begin{split}
		\mathcal{F}_{S}=&2\sqrt{\frac{2}{3}}C_{F}f_{B_{c}}f_{P}\pi m_{B_{c}}^{4}\sqrt{1-r^{2}}\int_{0}^{1}dx_{1}\hspace{0.5mm}dx_{2}\int_{0}^{b_{c}}b_{1}\hspace{0.5mm}db_{1}\hspace{0.5mm}b_{2}db_{2}\hspace{0.5mm}\phi_{B_{c}}(x_{1},b_{1})\\
		&\biggl[\left\lbrace \psi_{S}^{s}(x_{2},b_{2})(r_{b}-2x_{2})r+\psi_{V}^{v}(x_{2},b_{2})(x_{2}-2r_{b})\right\rbrace h_{a}(\alpha_{e},\beta_{a},b_{1},b_{2})E_{f}(t_{a})S_{t}(x_{2})\\&-\left\lbrace \psi_{S}^{v}(x_{2},b_{2})(r_{c}+r^{2})-\psi_{S}^{s}(x_{2},b_{2})2r \right\rbrace h_{b}(\alpha_{e},\beta_{b},b_{1},b_{2})E_{f}(t_{b})S_{t}(x_{1})\biggr],
	\end{split}
\end{equation}
and
\begin{equation}
\label{eq:nlffS2}
	\begin{split}
		\mathcal{M}_{S}=&-\frac{8}{3}C_{F}f_{B_{c}}\pi m_{B_{c}}^{4}\sqrt{1-r^{2}}\int_{0}^{1}dx_{1}\hspace{0.5mm}dx_{2}\hspace{0.5mm}dx_{3}\int_{0}^{b_{c}}b_{1}\hspace{0.5mm}db_{1}b_{3}db_{3}\hspace{0.5mm}\phi_{B_{c}}(x_{1},b_{1})\phi_{\pi(K)}^{A}(x_{3})\\
		&\biggl[\left\lbrace \psi_{S}^{v}(x_{2},b_{1})((x_{1}+2x_{2}+x_{3}-2)r^{2}+x_{1}-x_{3})-\psi_{S}^{s}(x_{2},b_{1})r(x_{1}+x_{2}-1)\right\rbrace \\	&h_{c}(\alpha_{e},\beta_{c},x_{3},b_{1},b_{3})E_{f}(t_{c})+\bigl\lbrace (x_{1}+x_{2}-1)\psi_{S}^{s}(x_{2},b_{1})r+\psi_{S}^{v}(x_{2},b_{1})((x_{3}-x_{2})r^{2}\\&-2x_{1}-x_{2}-x_{3}+2)\bigr\rbrace h_{d}(\alpha_{e},\beta_{d},x_{3},b_{1},b_{3})E_{f}(t_{d})\biggr],
	\end{split}
\end{equation}

\item For $B_{c}\rightarrow J/\psi\pi^{+}$ and $B_{c}\rightarrow J/\psi K^{+}$ decays:
 	\begin{equation}
 	\label{eq:nlffS3}
	\begin{split}
		\mathcal{F}_{A}=&2\sqrt{\frac{2}{3}}C_{F}f_{B_{c}}f_{P}\pi m_{B_{c}}^{4}\sqrt{1-r^{2}}\int_{0}^{1}dx_{1}\hspace{0.5mm}dx_{2}\int_{0}^{b_{c}}b_{1}\hspace{0.5mm}db_{1}\hspace{0.5mm}b_{2}db_{2}\hspace{0.5mm}\phi_{B_{c}}(x_{1},b_{1})\\
		&\biggl[\left\lbrace \psi_{A}^{t}(x_{2},b_{2})r(r_{b}-2x_{2})+\psi_{A}^{L}(x_{2},b_{2})(x_{2}-2r_{b})\right\rbrace h_{a}(\alpha_{e},\beta_{a},b_{1},b_{2})E_{f}(t_{a})S_{t}(x_{2})\\&-\left\lbrace \psi_{A}^{L}(x_{2},b_{2})(r^{2}+r_{c}) \right\rbrace h_{b}(\alpha_{e},\beta_{b},b_{1},b_{2})E_{f}(t_{b})S_{t}(x_{1})\biggr],
	\end{split}
\end{equation}
and
\begin{equation}
\label{eq:nlffS4}
	\begin{split}
		\mathcal{M}_{A}=&-\frac{8}{3}C_{F}f_{B_{c}}\pi m_{B_{c}}^{4}\sqrt{1-r^{2}}\int_{0}^{1}dx_{1}\hspace{0.5mm}dx_{2}\hspace{0.5mm}dx_{3}\int_{0}^{b_{c}}b_{1}\hspace{0.5mm}db_{1}b_{3}db_{3}\hspace{0.5mm}\phi_{B_{c}}(x_{1},b_{1})\phi_{\pi(K)}^{A}(x_{3})\\
		&\biggl[\left\lbrace \psi_{A}^{L}(x_{2},b_{1})(r^{2}-1)(x_{1}-x_{3})-r(x_{1}+x_{2}-1)\psi_{A}^{t}(x_{2},b_{1})\right\rbrace h_{c}(\alpha_{e},\beta_{c},x_{3},b_{1},b_{3})E_{f}(t_{c})\\	&+\left\lbrace r(x_{1}+x_{2}-1)\psi^{t}(x_{2},b_{1})-\psi_{A}^{L}(x_{2},b_{1})(r^{2}(x_{2}-x_{3})+2x_{1}+x_{2}+x_{3}-2)\right\rbrace \\&h_{d}(\alpha_{e},\beta_{d},x_{3},b_{1},b_{3})E_{f}(t_{d})\biggr],
	\end{split}
\end{equation}

\item For $B_{c}\rightarrow \chi_{c0}\pi^{+}$ and $B_{c}\rightarrow \chi_{c0} K^{+}$ decays:

\begin{equation}
\label{eq:nlffP1}
\begin{split}
\mathcal{F}_{S}=&2\sqrt{\frac{2}{3}}C_{F}f_{B_{c}}f_{P}\pi m_{B_{c}}^{4}(r^{2}-1)\int_{0}^{1}dx_{1}\hspace{0.5mm}dx_{2}\int_{0}^{b_{c}}b_{1}\hspace{0.5mm}db_{1}\hspace{0.5mm}b_{2}db_{2}\phi_{B_{c}}(x_{1},b_{1})\\
		&\biggl[\left\lbrace \psi_{S}^{s}(x_{2},b_{2})r(r_{b}-2x_{2})+\psi_{S}^{v}(x_{2},b_{2})(x_{2}-2r_{b})\right\rbrace h(\alpha_{e},\beta_{a},b_{1},b_{3})\cdot E_{f}(t_{a})\cdot S_{t}(x_{2})\\&-\left\lbrace \psi_{S}^{v}(x_{2},b_{2})(r_{c}+r^{2}(x_{1}-1))-2r\psi_{S}^{s}(x_{2,b_{2}})(r_{c}+x_{1}-1)\right\rbrace h(\alpha_{e},\beta_{b},b_{1},b_{2})E_{f}(t_{b})S_{t}(x_{1})\biggr],
\end{split}
\end{equation}
\begin{equation}
\label{eq:nlffP2}
\begin{split}
\mathcal{M}_{S}=&\frac{8}{3}C_{F}f_{B_{c}}\pi m_{B_{c}}^{4}(r^{2}-1)\int_{0}^{1}dx_{1}dx_{2}dx_{3}\int_{0}^{b_{c}}b_{1}\hspace{0.5mm}db_{1}b_{3}db_{3}\phi_{B_{c}}(x_{1},b_{1})\phi_{\pi(K)}^{A}(x_{3})\\&
\biggl[\left\lbrace \psi_{S}^{v}(x_{2},b_{1})(r^{2}(x_{1}+2x_{2}+x_{3}-2)+x_{1}-x_{3})-r\psi_{S}^{s}(x_{2},b_{1})(x_{1}+x_{2}-1)\right\rbrace \\&E_{cd}(t_{c})h(\alpha_{e},\beta_{c},b_{1},b_{3})-\bigl\lbrace \psi_{S}^{v}(x_{2},b_{1})(r^{2}(x_{2}-x_{3})+2x_{1}+x_{2}+x_{3}-2)\\&-r\psi_{S}^{s}(x_{2},b_{1})(x_{1}+x_{2}-1)\bigr\rbrace E_{cd}(t_{d})h(\alpha_{e},\beta_{d},b_{1},b_{3})\biggr],
\end{split}
\end{equation}

\item and for $B_{c}\rightarrow \chi_{c1}(h_{c})\pi^{+}$ and $B_{c}\rightarrow \chi_{c1}(h_{c}) K^{+}$ decays:

\begin{equation}
\label{eq:nlffP3}
\begin{split}
\mathcal{F}_{A}=&-2\sqrt{\frac{2}{3}}C_{F}f_{B_{c}}f_{P}\pi m_{B_{c}}^{4}(r^{2}-1)\int_{0}^{1}dx_{1}\hspace{0.5mm}dx_{2}\int_{0}^{b_{c}}b_{1}\hspace{0.5mm}db_{1}\hspace{0.5mm}b_{2}db_{2}\phi_{B_{c}}(x_{1},b_{1})\\
		&\biggl[\left\lbrace r\cdot \psi_{A}^{t}(x_{2},b_{2})(r_{b}-2x_{2})+\psi_{A}^{L}(x_{2},b_{2})(x_{2}-2r_{b})\right\rbrace h(\alpha_{e},\beta_{a},b_{1},b_{3})\cdot E_{f}(t_{a})\cdot S_{t}(x_{2})\\&-\left\lbrace \psi_{A}^{L}(x_{2},b_{2})(r_{c}+r^{2}(x_{1}-1))\right\rbrace h(\alpha_{e},\beta_{b},b_{1},b_{2})E_{f}(t_{b})S_{t}(x_{1})\biggr],
\end{split}
\end{equation}
and
\begin{equation}
\label{eq:nlffP4}
\begin{split}
\mathcal{M}_{A}=&\frac{8}{3}C_{F}f_{B_{c}}\pi m_{B_{c}}^{4}(r^{2}-1)\int_{0}^{1}dx_{1}dx_{2}dx_{3}\int_{0}^{b_{c}}b_{1}\hspace{0.5mm}db_{1}b_{3}db_{3}\phi_{B_{c}}(x_{1},b_{1})\phi_{\pi(K)}^{A}(x_{3})
\\&
\biggl[\left\lbrace \psi_{A}^{L}(x_{2},b_{1})(r^{2}-1)(x_{1}-x_{3})-r\psi_{A}^{t}(x_{2},b_{1})(x_{1}+x_{2}-1)\right\rbrace E_{cd}(t_{c})h(\alpha_{e},\beta_{c},b_{1},b_{3})
\\&
+\left\lbrace \psi_{A}^{L}(x_{2},b_{1})(r^{2}(x_{2}-x_{3})+2x_{1}+x_{2}+x_{3}-2)-r\psi_{A}^{t}(x_{2},b_{1})(x_{1}+x_{2}-1)\right\rbrace \\&E_{cd}(t_{d})h(\alpha_{e},\beta_{d},b_{1},b_{3})\biggr].
\end{split}
\end{equation}

\end{itemize}

In all these expressions $r=m/m_{B_{c}}$ and $r_{c(b)}=m_{c(b)}/m_{B_{c}}$. $h(\alpha_{e},\beta_{i},b_{1},b_{2(3)})$, $E_{f}(t_{i})$, $S_{t}(x)$ and $t_{i}$ represent the hard kernels, evolution function, jet function and the hard scales respectively. Their expressions are shown in the next appendix.

	\section{Scales and relevant functions in the hard kernel:}
	\label{section:hard functions}
	
	In this appendix, we present analytic expressions for the hard functions and scales that were introduced in the previous appendix.
	
	\noindent The hard kernel $h$ comes from the Fourier transform of virtual quark and gluon propagators \cite{PhysRevD.90.114030}
	\begin{equation}
	h(\alpha_{e},\beta_{i},b_{1},b_{2})=h_{1}(\alpha_{e},b_{1})\times h_{2}(\beta_{i},b_{1},b_{2}),\\	
	\end{equation}
	with 
	\begin{equation}
	\begin{split}
	h_{1}(\alpha_{e},b_{1})&=\begin{array}{cc}
  \Biggl\{ 
    \begin{array}{cc}
      K_{0}(\sqrt{\alpha_{e}}b_{1})& \qquad \alpha_{e}>0 \\
      K_{0}(i\sqrt{-\alpha_{e}}b_{1})& \qquad \alpha_{e}<0
      \end{array},
\end{array}\\
h_{2}(\beta_{i},b_{1},b_{2})&=\begin{array}{cc}
  \Biggl\{
    \begin{array}{cc}
      \theta(b_{1}-b_{2})I_{0}(\sqrt{\beta_{i}}b_{2})K_{0}(\sqrt{\beta_{i}}b_{1})+(b_{1}\xrightarrow{} b_{2})& \qquad \beta_{i}>0 \\
      \theta(b_{1}-b_{2})J_{0}(\sqrt{-\beta_{i}}b_{2})K_{0}(i\sqrt{-\beta_{i}}b_{1})+(b_{1}\xrightarrow{} b_{2})&  \qquad \beta_{i}<0
      \end{array},
\end{array}
\end{split}
	\end{equation}
	where $J_{0}$ is the Bessel function and $K_{0}$ and $I_{0}$ are the modified Bessel functions, and
	\begin{equation}
	\begin{split}
	\alpha_{e}&=-m_{B_{c}}^{2}[x_{1}+\eta^{+}r(x_{2}-1)][x_{1}+\eta^{-}r(x_{2}-1)],\\
	\beta_{a}&=m_{b}^{2}-m_{B_{c}}^{2}[1+\eta^{+}r(x_{2}-1)][1+\eta^{-}r(x_{2}-1)],\\
	\beta_{b}&=m_{c}^{2}-m_{B_{c}}^{2}(\eta^{+}r-x_{1})(\eta^{-}r-x_{1}),
	\end{split}
	\end{equation}
	for Eqns \eqref{eq:BcPslf1}-\eqref{eq:BcVslV}, and
	\begin{equation}
	\begin{split}
	\alpha_{e}&=[x_{1}+r^{2}(x_{2}-1)][x_{1}+x_{2}-1+r^{2}(1-x_{2})]m_{B_{c}}^{2},\\
	\beta_{a}&=[r_{b}^{2}+\lbrace 1+r^{2}(x_{2}-1)\rbrace\lbrace x_{2}-r^{2}(x_{2}-1)\rbrace]m_{B_{c}}^{2},\\
	\beta_{b}&=[r_{c}^{2}+(r^{2}-x_{1})(x_{1}-1+r^{2})]m_{B_{c}}^{2},\\
	\beta_{c}&=[x_{1}+x_{2}-1+r^{2}(1-x_{2}-x_{3})][x_{3}-x_{1}-r^{2}(x_{2}+x_{3}-1)]m_{B_{c}}^{2},\\
	\beta_{d}&=[x_{1}+x_{2}-1-r^{2}(x_{2}-x_{3})][1-x_{3}-x_{1}-r^{2}(x_{2}-x_{3}-1)]m_{B_{c}}^{2},
	\end{split}
	\end{equation}
	for Eqns \eqref{eq:nlffS1}-\eqref{eq:nlffP4}. The evolution functions $E_{ij}(t)$ are expressed as
	\begin{equation}
	E_{ab,cd}(t)=\alpha_{s}(t)S_{ab,cd}(t),
	\end{equation}
	where $S_{ab,cd}(t)$ represents the Sudakov factors in modified PQCD framework has been taken from \cite{PhysRevD.97.113001}. The hard scale $t$ is chosen to be the maximum of the virtuality of internal momentum transition in the hard amplitudes \cite{PhysRevD.90.114030},
	\begin{equation}
	\begin{split}
	t_{a}&=max(\alpha_{e},\beta_{a},1/b_{1},1/b_{2}), \qquad t_{b}=max(\alpha_{e},\beta_{b},1/b_{1},1/b_{2}),\\
	t_{c}&=max(\alpha_{e},\beta_{c},1/b_{1},1/b_{3}), \qquad t_{d}=max(\alpha_{e},\beta_{d},1/b_{1},1/b_{3}),
	\end{split}
	\end{equation}
	and the jet function $S_{t}(x)$ has the same form as Eqn \ref{eqn:jet function}.

\section{Synthetic data of Form Factors:}
\label{section:Appendix form factor data}

\begin{itemize}

	\item In Table \ref{table:Appendix Pole Data 1} we present the synthetic data of $B_{c}\rightarrow J/\psi$ and $B_{c}\rightarrow\eta_{c}$ form factors at $q^{2}=$5.0, 7.5 and 10.0 $GeV^{2}$ and at $q^{2}=$6.0 and 10.0 $GeV^{2}$ respectively used as inputs to extract the parameters in Table \ref{table:pole2}.
	\begin{landscape}
		\scriptsize
		\centering
		\renewcommand{\arraystretch}{1.2}
		\setlength{\tabcolsep}{4pt}
		
		\begin{tabular}{|c|c|cccccccccccc|}
			\hline
			\textbf{Form Factors} & \textbf{Value} &  &  &  & &&\textbf{Correlation}&& &  &  &  &  \\ 
			\cline{3-14}
			\textbf{at $q^{2}$ ($GeV^{2}$)}&\textbf{from HPQCD}&$A_{0}(5.0)$&$A_{0}(7.5)$&$A_{0}(10.0)$&$A_{1}(5.0)$&$A_{1}(7.5)$&$A_{1}(10.0)$&$A_{2}(5.0)$&$A_{2}(7.5)$&$A_{2}(10.0)$&$V(5.0)$&$V(7.5)$&$V(10.0)$\\
			\hline 
			
			$A_{0}(5.0)$ & 0.686(83) & 1.0 & 0.994 & 0.977 & 0.305 & 0.147 & 0.023 & -0.764 & -0.665&-0.494&-0.026&-0.028&-0.030 \\ 
			
			$A_{0}(7.5)$ & 0.823(89) &  & 1.0 & 0.994 & 0.315 & 0.162 & 0.042 & -0.739 &-0.650 &-0.491&-0.026&-0.028&-0.030 \\ 
			
			$A_{0}(10.0)$ & 0.988(97) &  &  & 1.0 & 0.315 & 0.166 & 0.048 & -0.711 & -0.632 &-0.487&-0.026&-0.028&-0.030\\ 
			
			$A_{1}(5.0)$ & 0.594(22) &  &  &  & 1.0 & 0.921 & 0.756 & 0.060 & 0.024&0.008&-0.025&-0.024& -0.024\\ 
			
			$A_{1}(7.5)$ & 0.685(24) &  &  &  &  & 1.0 & 0.948 & 0.050 & 0.015 &0.0003&-0.011&-0.010&-0.010\\ 
			
			$A_{1}(10.0)$ & 0.796(27) &  &  &  &  &  & 1.0 & 0.035 & 0.006 &-0.006&0.005&0.005&0.005\\ 
			
			$A_{2}(5.0)$ & 0.551(115) &  &  &  &  &  &  & 1.0 & 0.961 &0.837&0.003&0.004&0.005\\ 
			
			$A_{2}(7.5)$ & 0.636(117) &  &  &  &  &  &  &  & 1.0 &0.955&0.002&0.003&0.005\\ 
			
			$A_{2}(10.0)$ & 0.735(131) &  &  &  &  &  &  &  & & 1.0 &0.003&0.003& 0.003\\
			
			$V(5.0)$ & 1.058(76) &  &  &  &  &  &  &  & &&1.0&0.977& 0.902\\
			
			$V(7.5)$ & 1.278(76) &  &  &  &  &  &  &  & &&&1.0& 0.973\\
			
			$V(10.0)$ & 1.547(81) &  &  &  &  &  &  &  & &&&&1.0 \\
			\hline
			\textbf{Form Factors} & \textbf{Value} &&&\textbf{Correlation}& &&&&&&&&\\ 
			\cline{3-14}
			\textbf{at $q^{2}$ ($GeV^{2}$)}& \textbf{from BCL}&$F_{+}(6.0)$&$F_{+}(10.0)$&$F_{0}(6.0)$&$F_{0}(10.0)$&&&&&&&&\\
			
			\hline
			$F_{+}(6.0)$ & 0.767(208) &1.0&0.998&0.995&0.991&&&&&&&& \\ 
			$F_{+}(10.0)$ & 1.010(213) &&1.0&0.991&0.986&&&&&&&&  \\ 
			
			$F_{0}(6.0)$ & 0.665(181) &&&1.0&0.998&&&&&&&& \\ 
			$F_{0}(10.0)$ & 0.793(167) &&&&1.0&&&&&&&&  \\ 
			\hline
			
		\end{tabular} 
		
		\captionof{table}{HPQCD data for $B_{c}\rightarrow J/\psi$ and $\eta_{c}$ semileptonic form factors, along-with their correlation.}
		\label{table:Appendix Pole Data 1}
	\end{landscape}

\end{itemize}

\section{Correlation matrices:}
\label{section:Appendix correlation matrices}

In this appendix, we present the correlation matrices representing the correlation between the parameters that we have extracted in this work.

\begin{itemize}

	\item In Table \ref{table:Appendix correlation LCDA}, the correlation matrix between the $B_{c}$, $J/\psi$ and $\eta_{c}$ LCDA shape parameters obtained after minimizing the chi-square function constructed in section \ref{section:extraction of LCDA parameters} is shown.

			\begin{table}[htb!]
		
		\centering
		\footnotesize
		\begin{tabular}{|c|ccccccccc|}
			\hline
			&$\omega_{B_{c}}$&$\omega_{J/\psi}$&$\omega_{\eta_{c}}$&$f_{B_{c}}$&$f_{J/\psi}$&$f_{\eta_{c}}$&$\Lambda_{QCD}$&$m_{c}$&$m_{b}$\\
			\hline 
			
			$\omega_{B_{c}}$ & 1.0 & -0.617 & -0.557 & 0.156& 0.147 & 0.007 & -0.261& 0.607 & 0.601\\ 
			
			$\omega_{J/\psi}$ &  & 1.0 & 0.810 & -0.087 & -0.069 & -0.009 & -0.473& -0.886 & -0.829 \\ 
			
			$\omega_{\eta_{c}}$ &  &  & 1.0 & -0.101 & -0.085 & -0.009 &-0.524 & -0.913& -0.863\\ 
			
			$f_{B_{c}}$ &  &  &  & 1.0 & -0.002 & 0.0007 & 0.022 &0.111&0.097\\ 
			
			$f_{J/\psi}$ &  &  &  &  & 1.0 & 0.001 & 0.012&0.093& 0.079\\ 
			
			$f_{\eta_{c}}$ &  &  &  &  &  & 1.0 &0.005 &0.010& 0.009\\ 
			
			$\Lambda_{QCD}$ &  &  &  &  &  &  & 1.0&0.575& 0.518\\ 
			
			$m_{c}$ &  &  &  &  &  &  &   & 1.0 & 0.944\\
			
			$m_{b}$ &  &  &  &  &  &  &  &  & 1.0\\
			\hline
		\end{tabular} 
		
		\captionof{table}{Correlation Matrix between extracted LCDA parameters.}
		\label{table:Appendix correlation LCDA}
	\end{table}
	
	\item  In Table \ref{table: correlation pole}, the correlation matrix between the pole expansion parameters obtained in Table 	\ref{table:pole2} is shown.
	\begin{table}[htb!]
		\fontsize{8}{8.2}
		\centering
		
		\begin{tabular}{|c|ccccccccccc|}
			\hline
			&$\beta$&$\alpha_{A_{0}}$&$\alpha_{A_{1}}$&$\alpha_{A_{2}}$&$\alpha_{V}$&$\alpha_{F_{0}}$&$\alpha_{F_{+}}$&$A_{0}(0)$&$A_{1}(0)$&$V(0)$&$F_{+}(0)$\\
			\hline 
			
			$\beta$ & 1.0 & 0.921 & 0.813 & 0.374 & 0.844 & 0.217 & 0.334 & -0.374 & -0.370 & -0.303  & -0.042\\
			
			$\alpha_{A_{0}}$ &  & 1.0 & 0.831 & 0.173 & 0.776 & 0.200 & 0.308 & -0.560 & -0.532 & -0.277  & -0.039\\ 
			
			$\alpha_{A_{1}}$ &  &  & 1.0 & 0.019 & 0.695 & 0.177 & 0.272 & -0.557 & -0.696 & -0.238  & -0.034 \\ 
			
			$\alpha_{A_{2}}$ &  &  &  & 1.0 & 0.331 & 0.081 & 0.125 & 0.293 & 0.418 & -0.091  & -0.016\\ 
			
			$\alpha_{V}$ &  &  &  &  & 1.0 & 0.184 & 0.282 & -0.317 & -0.323 & -0.554 & -0.036\\ 
			
			$\alpha_{F_{0}}$ &  &  &  &  &  & 1.0 & 0.946 & -0.081 & -0.080 & -0.066  & -0.960\\ 
			
			$\alpha_{F_{+}}$ &  &  &  &  &  &  & 1.0 & -0.125 & -0.124 & -0.101 & -0.928\\

			$A_{0}(0)$ &  &  &  &  &  &  &  & 1.0 & 0.687 & 0.114  & 0.016\\
			
			$A_{1}(0)$&&&&&&&&& 1.0 & 0.111 & 0.016\\
			
			$V(0)$&&&&&&&&&& 1.0  & 0.013\\
			
			$F_{+}(0)$&&&&&&&&&&& 1.0 \\
			\hline
			
		\end{tabular} 
		
		\captionof{table}{Correlation Matrix between extracted pole expansion parameters.}
		\label{table: correlation pole}
	\end{table}
	
	\item In Table \ref{table:correlation chic1 decay constant}, correlation matrix between the parameters extracted in Table \ref{table:results for chic1 decay constant} in subsection \ref{subsection: P wave decay constants} is presented.

	\begin{landscape}
		\scriptsize
		\centering
		
		\begin{tabular}{|c|cccccccccccccccc|}
			\hline
			&$f_{\chi_{c1}}$&$\delta^{\prime}$&$f_{J/\psi}$&$m_{c}$&$\Lambda$&$\Gamma_{\chi_{c1}}$&$\delta_{C}$&$\delta_{NC}$&$a_{2\rho}(1.0)$&$a_{2\phi}(1.0)$&$\zeta_{\rho}(1.0)$&$\zeta_{\phi}(1.0)$&$\omega_{\rho A}(1.0)$&$\omega_{\rho V}(1.0)$&$\omega_{\phi A}(1.0)$&$\omega_{\phi V}(1.0)$\\
			\hline 
			
			$f_{\chi_{c1}}$ & 1.0 & -0.698 & -0.009 & 0.863 & 0.013 & 0.109 & 0.083 & 0.177 & 0.006 & -0.183  & -0.043 &-0.136&0.034&0.103&0.0&0.0\\ 
			
			$\delta^{\prime}$ &  & 1.0 & 0.214 & -0.634 & -0.215 & 0.433 & -0.0008 & 0.0001 & -0.006 & 0.171  & 0.043&0.125&-0.033&-0.104&0.0&0.0 \\ 
			
			$f_{J/\psi}$ &  &  & 1.0 & -0.013 & -0.989 & 0.0 & 0.018 & 0.049 & -0.004 & 0.019  &0.006 &0.015&-0.005&-0.012&0.0&0.0\\ 
			
			$m_{c}$ &  &  &  & 1.0 & 0.017 & 0.007 & 0.006 & 0.012 & 0.003 & -0.017 & 0.012&-0.019&-0.008&-0.039&0.0&0.0\\ 
			
			$\Lambda$ &  &  &  &  & 1.0 & 0.0 & -0.019 & -0.049 & 0.004 & -0.019  &-0.006 &-0.015&0.005&0.012&0.0&0.0\\ 
			
			$\Gamma_{\chi_{c1}}$ &  &  &  &  &  & 1.0 & 0.0007 & 0.0016 & 0.0 & -0.002 & -0.0003&-0.001&0.0003&0.0009&0.0&0.0\\ 
			
			$\delta_{C}$ &  &  &  &  &  &  & 1.0 & 0.0021 & 0.0 & -0.001 & -0.0002&-0.001&0.0001&0.0004&0.0&0.0\\

			$\delta_{NC}$ &  &  &  &  &  &  &  & 1.0 & -0.001 & -0.002  &-0.0003 &-0.001&0.0002&0.0008&0.0&0.0\\
			
			$a_{2\rho}(1.0)$&&&&&&&&& 1.0 & -0.003 & -0.023&-0.002&0.017&0.059&0.0&0.0\\
			
			$a_{2\phi}(1.0)$&&&&&&&&&& 1.0  &0.043 &0.006&-0.032&-0.108&0.0&0.0\\
			
			$\zeta_{\rho}(1.0)$&&&&&&&&&&& 1.0&0.030&0.100&0.341&0.0&0.0 \\
			
			$\zeta_{\phi}(1.0)$&&&&&&&&&&& &1.0&-0.023&-0.077&0.0&0.0 \\
			
			$\omega_{\rho A}(1.0)$&&&&&&&&&&& &&1.0&-0.779&0.0&0.0 \\
			
			$\omega_{\rho V}(1.0)$&&&&&&&&&&& &&&1.0&0.0&0.0 \\
			
			$\omega_{\phi A}(1.0)$&&&&&&&&&&& &&&&1.0&0.0 \\
			
			$\omega_{\phi V}(1.0)$&&&&&&&&&&& &&&&&1.0 \\
			\hline
			
		\end{tabular} 
		
		\captionof{table}{Correlation Matrix between extracted parameters in Table \ref{table:results for chic1 decay constant}.}
		\label{table:correlation chic1 decay constant}
	\end{landscape}
	
\end{itemize}

\section{$q^{2}$ distribution of $B_{c}\rightarrow S$ wave semileptonic form factors:}
\label{section:q2 distribution of Bc to S form factors}

\begin{figure*}[htb!]
	
	\centering
	\subfloat[\centering]{\includegraphics[width=6.1cm]{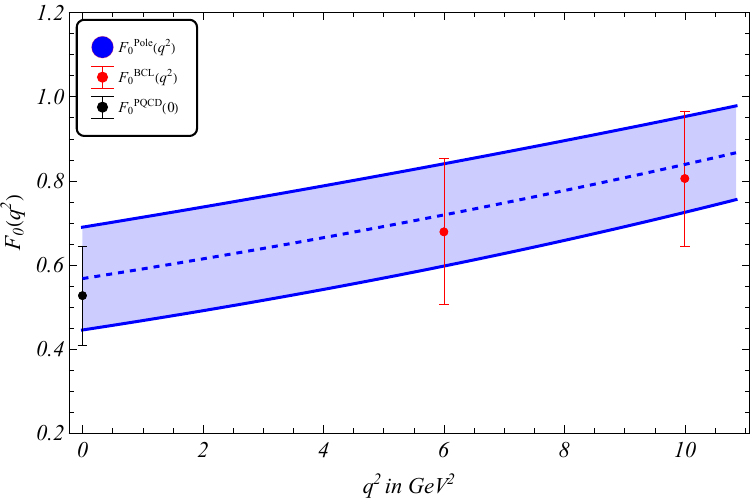}}
	\qquad
	\subfloat[\centering]{\includegraphics[width=6.1cm]{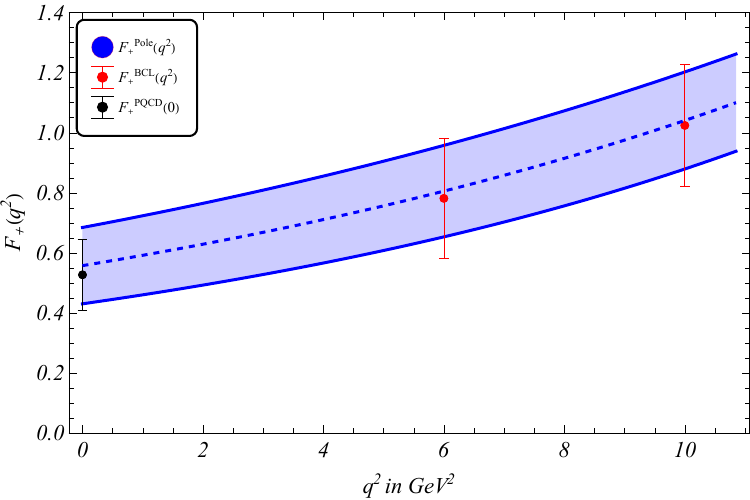}}
	\qquad
	\subfloat[\centering]{\includegraphics[width=6.1cm]{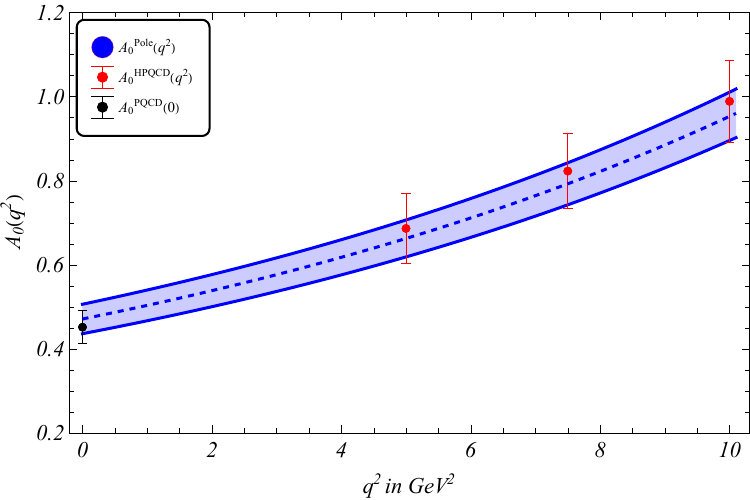}}
	\qquad
	\subfloat[\centering]{\includegraphics[width=6.1cm]{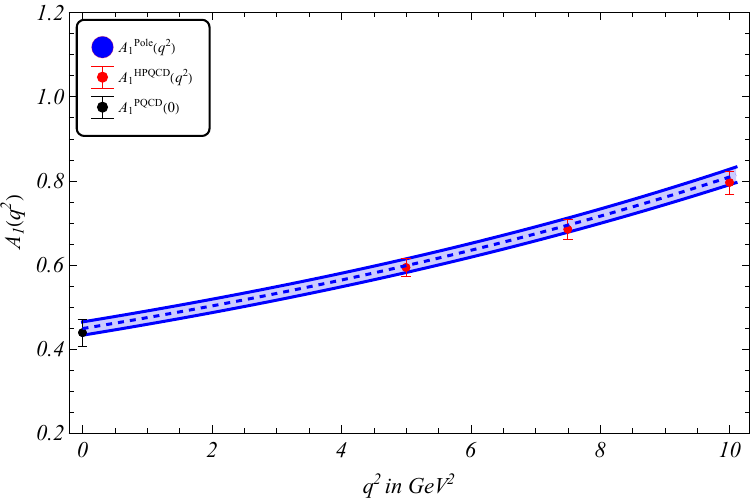}}
	\qquad
	\subfloat[\centering]{\includegraphics[width=6.1cm]{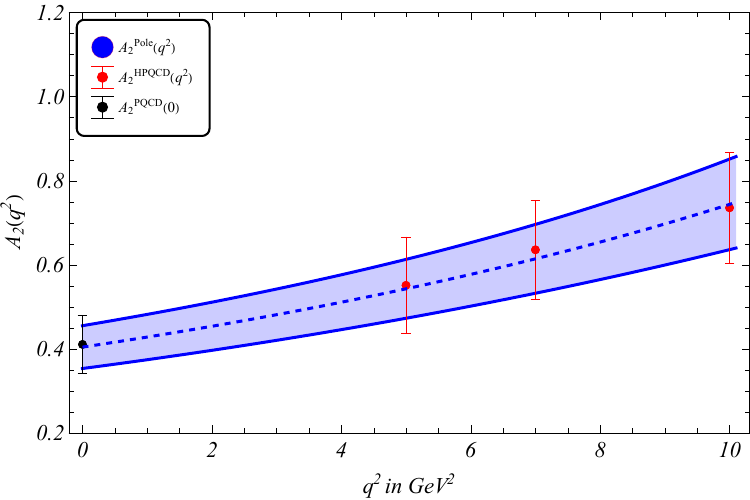}}
	\qquad
	\subfloat[\centering]{\includegraphics[width=6.1cm]{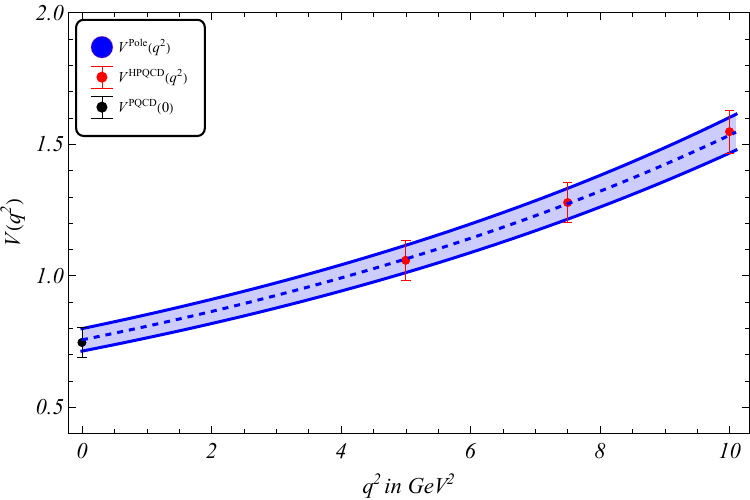}}
	\captionof{figure}{Plots showing the $q^{2}$ dependence of $B_{c}\rightarrow \eta_{c}$ and $B_{c}\rightarrow J/\psi$ semileptonic form factors. The blue curve denotes the distribution obtained using pole expansion parametrization, the red markers denote the synthetic data points from BCL and HPQCD, and the black marker is the pQCD value predicted in this work at $q^{2}=0.$}
	\label{fig:Pole Bc_JPsi}
\end{figure*}

In this appendix, we present $q^{2}$ distribution of the $B_{c}\rightarrow S$ wave semileptonic form factors obtained through pole expansion parametrization. The shape of the respective form factors are shown in Fig. \ref{fig:Pole Bc_JPsi} respectively. We can note that shapes obtained within the error bars could correctly accommodate the inputs used in the fit.

\section{$q^{2}$ distribution of $B_{c}\rightarrow P$ wave semileptonic decay widths:}
\label{section:q2 distribution of Bc to P decay width}

In this appendix, we have discussed the $q^2$ distributions of the $B_c \to J/\psi$ and $B_c\to \eta_c$ form factors, which we have obtained using the results of $q^2$ extrapolation parameters obtained in Table \ref{table:pole2}.  

	\begin{figure*}[htb!]
	
	\centering
	\subfloat[\centering]{\includegraphics[width=7.1cm]{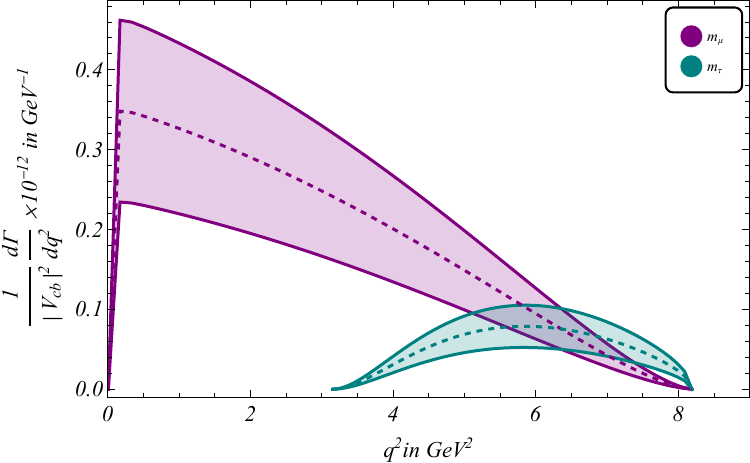}}
	\qquad
	\subfloat[\centering]{\includegraphics[width=7.1cm]{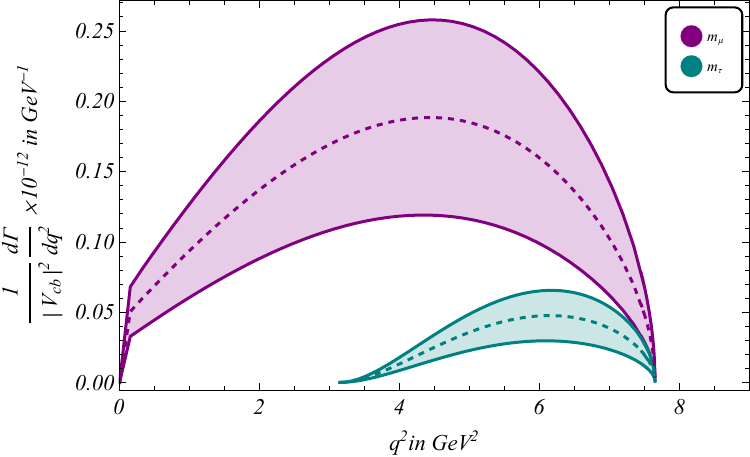}}
	\qquad
	\subfloat[\centering]{\includegraphics[width=7.1cm]{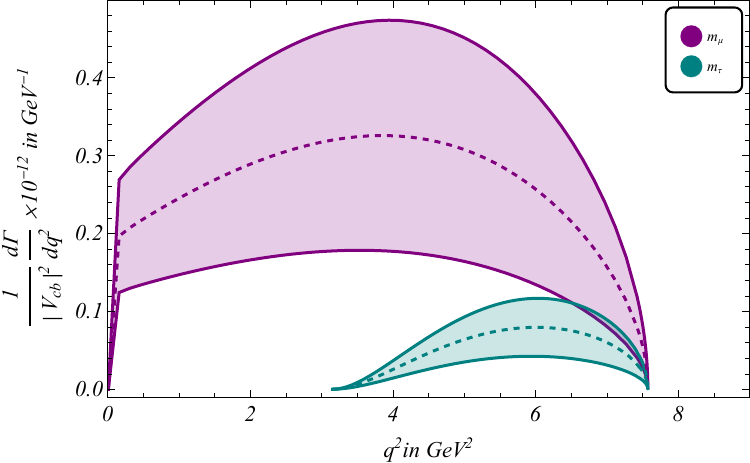}}
	\caption{$q^{2}$ distribution of (a) $d\Gamma(B_{c}\rightarrow\chi_{c0})/dq^{2}$, (b) $d\Gamma(B_{c}\rightarrow\chi_{c1})/dq^{2}$ and (c) $d\Gamma(B_{c}\rightarrow h_{c})/dq^{2}$ semileptonic channels. The violet and green plots denote the light lepton and heavy lepton cases respectively.}
	\label{fig:differential decay width}
\end{figure*}

\bibliographystyle{JHEP}
\bibliography{biblio}

\end{document}